\documentclass[iop,apj,numberedappendix,twocolappendix,revtex4]{emulateapj}
\usepackage{thumbpdf} 
\usepackage{amsmath,amssymb} 
\usepackage{graphicx,mathptmx} 
\usepackage{natbib} 
\usepackage[usenames]{color} 
\usepackage{ifpdf}  
\usepackage[protrusion=true,expansion=false]{microtype} 
\usepackage{lscape} 
\usepackage{xcolor}   
\usepackage{refcount} 
\usepackage{textcase} 
\usepackage{epstopdf} 
\usepackage{xstring}

\usepackage{etoolbox} 

\newtoggle{lowresversion} 
\togglefalse{lowresversion}

\ifpdf
\pdfoutput=1 
\else

\fi

\newcommand{\beq}{
\begin{equation}
}
\newcommand{\eeq}{
\end{equation}
}
\newcommand{\beqa}{
\begin{eqnarray}
}
\newcommand{\eeqa}{
\end{eqnarray}
}

\newcommand{\kgfigbeg}[1]{
\begin{figure}
\hypertarget{#1}{}%
}
\newcommand{\kgfigend}[2]{
\label{f:#1}
\end{figure}
\bookmarksetup{color=[rgb]{0.54,0,0}}
\bookmark[rellevel=1,keeplevel,dest=#1]{Fig \ref*{f:#1}: {#2}}
\bookmarksetup{color=black} 
}

\newcommand{\kgfigstarbeg}[1]{
\begin{figure*}
\hypertarget{#1}{}%
}
\newcommand{\kgfigstarend}[2]{
\label{f:#1}
\end{figure*}
\bookmarksetup{color=[rgb]{0.54,0,0}}
\bookmark[rellevel=1,keeplevel,dest=#1]{Fig \ref*{f:#1}: {#2}}
\bookmarksetup{color=black} 
}

\newcommand{\kgtabbeg}[1]{
\begin{deluxetable}{#1}
}
\newcommand{\kgtabend}[2]{
\label{t:#1}
\end{deluxetable}
\bookmarksetup{color=[rgb]{0,0,0.54}}
\bookmark[
rellevel=1,
keeplevel,
dest=table.\getrefnumber{t:#1}
]{Table \ref*{t:#1}: #2}
\bookmarksetup{color=[rgb]{0,0,0}}
}

\newcommand{\kgtabstarbeg}[1]{
\begin{deluxetable*}{#1}
}
\newcommand{\kgtabstarend}[2]{
\label{t:#1}
\end{deluxetable*}
\bookmarksetup{color=[rgb]{0,0,0.54}}
\bookmark[
rellevel=1,
keeplevel,
dest=table.\getrefnumber{t:#1}
]{Table \ref*{t:#1}: #2}
\bookmarksetup{color=[rgb]{0,0,0}}
}

\newcommand{\units}[1]  {\ensuremath{\mathrm{{#1}}}}
\newcommand{\msun}     {\ensuremath{{{M}}_{\scriptscriptstyle \odot}}}

\providecommand{\ion}[2]{#1$\;$\textsmaller{\@Roman{#2}}}

\def\spose#1{\hbox to 0pt{#1\hss}}
\newcommand{\lta}{\mathrel{\spose{\lower 3pt\hbox{$\mathchar"218$}}
      \raise 2.0pt\hbox{$\mathchar"13C$}}}
\newcommand{\gta}{\mathrel{\spose{\lower 3pt\hbox{$\mathchar"218$}}
      \raise 2.0pt\hbox{$\mathchar"13E$}}}
\def\simlt{\mathrel{\rlap{\lower 3pt\hbox{$\sim$}}\raise 2.0pt\hbox{$<$}}}
\def\simgt{\mathrel{\rlap{\lower 3pt\hbox{$\sim$}} \raise 2.0pt\hbox{$>$}}}

\definecolor{KayhanCiteColor}{rgb}{0,0.08,0.35}
\definecolor{KayhanURLColor}{rgb}{0,0.08,0.35}
\definecolor{KayhanLinkColor}{rgb}{0,0.08,0.35}
\definecolor{KayhanPageColor}{rgb}{0,0.08,0.35}
\definecolor{medred}{rgb}{0.75,0.0,0.0}

\shorttitle{Fundamental Plane of BH Accretion as Mass Estimator}
\shortauthors{G\"{u}ltekin et al.}

\makeatletter
\ifx\undefined\hyperref
\usepackage[pdftitle={The Fundamental Plane of Black Hole Accretion and its Use as a Black-Hole-Mass Estimator},pdfauthor={Kayhan Gultekin},pdfsubject={Astrophysics},pdfkeywords={black hole physics --- methods: statistical},letterpaper,linktocpage,colorlinks=true,linkcolor={KayhanLinkColor},urlcolor={KayhanURLColor},citecolor={KayhanCiteColor},hyperfootnotes=false]{hyperref}
\else\relax\fi
\makeatother
\usepackage[figure,figure*]{hypcap} 
\usepackage{atveryend} 
\usepackage[
  open,
  openlevel=3,
  atend
]{bookmark}[2010/04/08]  
\begin{document}

\label{firstpage}
 
\title{The Fundamental Plane of Black Hole Accretion and its Use as a Black Hole-Mass Estimator}

\author{Kayhan G\"{u}ltekin\altaffilmark{1}}
\author{Ashley L.\ King\altaffilmark{2}}
\author{Edward M.\ Cackett\altaffilmark{3}}
\author{Kristina Nyland\altaffilmark{4}}
\author{Jon M.\ Miller\altaffilmark{1}}
\author{Tiziana Di Matteo\altaffilmark{5}}
\author{Sera Markoff\altaffilmark{6,7}}
\author{Michael P.\ Rupen\altaffilmark{8}}
 \affil{\altaffilmark{1}Department of Astronomy, University of Michigan, 500 Church Street, Ann Arbor, MI 48109, USA; send correspondence to
\href{mailto:kayhan@umich.edu}{kayhan@umich.edu};}
\affil{\altaffilmark{2}{KIPAC, Stanford University, 452 Lomita Mall, Stanford, CA 94305 USA}}
\affil{\altaffilmark{3}Department of Physics and Astronomy, Wayne State University, 666 W. Hancock Street, Detroit, MI 48201, USA}
\affil{\altaffilmark{4}National Radio Astronomy Observatory, Charlottesville, VA 22903, USA}
\affil{\altaffilmark{5}McWilliams Center for Cosmology, Physics Department, Carnegie Mellon University, Pittsburgh, PA 15213, USA}
\affil{\altaffilmark{6}{Anton Pannekoek Institute for Astronomy, University of Amsterdam, Science Park 904, 1098 XH Amsterdam, The Netherlands}
\altaffilmark{7}{Gravitation Astroparticle Physics Amsterdam (GRAPPA) Institute, University of Amsterdam, Science Park 904, 1098 XH Amsterdam, The Netherlands };}
\affil{\altaffilmark{8}{National Research Council of Canada, Herzberg Astronomy and Astrophysics, Dominion Radio Astrophysical Observatory, P.O. Box 248, Penticton, BC V2A 6J9, Canada}.}

\begin{abstract}
\hypertarget{abstract}{} 
We present an analysis of the fundamental plane of black hole accretion, an empirical correlation of the mass of a black hole ($M$), its 5 GHz radio continuum luminosity ($\nu L_{\nu}$), and its 2--10 keV X-ray power-law continuum luminosity ($L_X$).  We compile a sample of black holes with primary, direct black hole-mass measurements that also have sensitive, high-spatial-resolution radio and X-ray data.  Taking into account a number of systematic sources of uncertainty and their correlations with the measurements, we use Markov chain Monte Carlo methods to fit a mass-predictor function of the form $\log(M/10^{8}\,\msun) = \mu_0 + \xi_{\mu R} \log(L_R / 10^{38}\,\units{erg\,s^{-1}}) + \xi_{\mu X} \log(L_X / 10^{40}\,\units{erg\,s^{-1}})$.  Our best-fit results are $\mu_0 = 0.55 \pm 0.22$, $\xi_{\mu R} = 1.09 \pm 0.10$, and $\xi_{\mu X} = -0.59^{+0.16}_{-0.15}$ with the natural logarithm of the Gaussian intrinsic scatter in the log-mass direction $\ln\epsilon_\mu = -0.04^{+0.14}_{-0.13}$.  { This result is a significant improvement over our earlier mass scaling result because of the increase in active galactic nuclei sample size (from 18 to 30), improvement in our X-ray binary sample selection, better identification of Seyferts, and {improvements in our} analysis that takes into account systematic uncertainties and correlated uncertainties.  Because of these significant improvements, we are able to} consider potential influences on our sample by including all sources with compact radio and X-ray emission but ultimately conclude that the fundamental plane can empirically describe all such sources.  We end with advice for how to use this as a tool for estimating black hole masses.

\bookmark[ rellevel=1, keeplevel,
dest=abstract
]{Abstract}
\end{abstract}
\keywords{galaxies:nuclei --- galaxies:active --- galaxies:jets --- accretion, accretion disk --- black hole physics}

\section{Introduction}
\label{intro}

Accretion and outflows, including jets, are seen in many astrophysical objects and are thought to have an intimate, physical connection to each other.  Among all of the objects seen to have accretion and outflows (e.g., protostars, white dwarfs, and neutron stars), black holes have the greatest range in mass so correlations with mass can be tested.  

An important observational connection related to accretion--jet phenomena was first noted by \citet{1998A&A...337..460H} in a power-law relation between the radio and X-ray fluxes of black hole candidate GX 339$-$4 at various levels of low/hard states.  Compiling many such black hole X-ray binaries (XRBs), \citet{2003MNRAS.344...60G} found that many black hole XRBs followed a similar trend such that the radio luminosity ($L_R$) of a low/hard state XRB scales with the X-ray luminosity ($L_X$) as $L_R \sim L_X^{0.7}$.  This same observational trend was examined in active galactic nuclei (AGNs) but found to depend on the mass of the black hole ($M$) as well \citep{merlonietal03, fkm04}.  Because the masses of XRB black holes are all within a factor of a few of each other, such a scaling is only apparent when it is examined across several orders of magnitude.  This $M$--$L_R$--$L_X$ relation is often called the fundamental plane of black hole activity because it occupies a two-dimensional manifold in the three-dimensional space.

The empirical relation of the fundamental plane has been used to generate insights about the physics of accretion onto black holes.  \citet{merlonietal03} interpreted the fundamental plane as the result of scale invariant disk--jet coupling relating jet power probed by radio and mass accretion rate probed by X-rays.  This interpretation was based on work by \citet{2003A&A...397..645M}, which was later generalized by \citet{2003MNRAS.343L..59H} and then extended by \citet{2004MNRAS.355..835H} to include the effects of synchrotron cooling.  \citet{fkm04} interpreted the fundamental plane as arising from sub-Eddington jet-dominated systems in which the emission arises from the jet as optically thick radio synchrotron and optically thin X-ray synchrotron.  For a description of the similarities and differences among these interpretations, see the discussion by \citet{2012MNRAS.419..267P}.  
An additional interpretation provided by \citet{2005ApJ...629..408Y} argues for a critical X-ray Eddington ratio below which the fundamental plane switches from accretion-flow-dominated X-ray to jet-dominated emission.

After the initial discovery studies, the empirical relation was pushed to a wider variety of accreting black hole sources.
\citet{2006ApJ...645..890W} considered the fundamental plane in radio-active Type 1 active galactic nucleus AGN and found only a weak dependence on mass, opening the possibility that AGN with $L_X/L_\mathrm{Edd} > 10^{-3}$ follow a completely different relation.  Their work was followed up by \citet{2008ApJ...688..826L} with a larger sample and showed a significant difference in fundamental plane fits for radio-quiet and radio-loud samples of Type 1 AGN.  \citet{2007A&A...467..519P} found a significant correlation between $L_R$ and $L_X$ in a sample of Seyfert galaxies and low-luminosity radio galaxies.  Although there were concerns that the fundamental plane was a manifestation of distance creating the illusion of a luminosity--luminosity relation, partial correlation analyses showed that the fundamental plane relations were not driven by distance \citep{merlonietal03, 2006ApJ...645..890W, merlonietal06}.

Work from the last five years has used even larger samples to further promote the understanding of the underlying physics of the fundamental plane.  \citet{2012MNRAS.419..267P} used a carefully selected sample of BL Lac objects to supplement a selection of low-luminosity AGNs and XRBs, arguing that when the radio spectrum is flat/inverted, the X-ray emission comes from jet synchrotron.  Further examination of the $L_R$--$L_X$ relation in XRBs has uncovered multiple tracks rather than a universal relation \citep{2012MNRAS.423..590G}.  The existence of multiple tracks raises the possibility of more complex physical underpinnings of the fundamental plane and led to work by \citet{2014ApJ...787L..20D} to observe that radiatively efficient XRBs and AGNs follow their own, separate fundamental plane.
\citet{2016ApJ...818..185F} demonstrated that the fundamental plane for compact steep-spectrum radio sources was best explained by a hot corona origin for X-ray emission.  In the faintest objects, with $L_X/L_\mathrm{Edd} < 10^{-6}$, the fundamental plane observed by \citet{2017ApJ...836..104X} argue for X-ray emission coming from hot thermal gas in the accretion flow as predicted by \citet{2005ApJ...629..408Y}, though this idea was ruled out by \citet{2013ApJ...773...59P}.


In addition to the insights it may provide regarding accretion physics, the fundamental plane is also interesting because it relates two relatively simple electromagnetic observations, $L_R$ and $L_X$, to a notoriously difficult one, $M$.  A black hole's mass is of paramount interest, as mass and spin are the only two parameters intrinsic to an astrophysical black hole.  It also sets the scale for accretion properties for such things as the Eddington luminosity, $L_\mathrm{Edd}$.  The Eddington fraction $f_\mathrm{Edd} \equiv L_\mathrm{bol} / L_\mathrm{Edd}$, in fact, may be the most important parameter of an AGN \citep{1992ApJS...80..109B, 2000ApJ...536L...5S, 2002ApJ...565...78B, 2014Natur.513..210S}.  Black hole mass also sets relative size scales, including the Schwarzschild radius, the innermost stable circular orbit radius, and the the black hole shadow size \citep{2008Natur.455...78D}.  

Although black hole-mass estimation methods exist at varying levels of resource intensiveness (e.g., stellar dynamics, \citealt{2009ApJ...695.1577G}; reverberation mapping, \citealt{2014SSRv..183..253P}; and host scaling relations, \citealt{2009ApJ...698..198G}), the ability to use radio and X-ray observations to estimate black hole mass would be a useful additional method in cases where other methods do not work well.  It may be useful, for example, for (i) distinguishing between XRBs, accreting intermediate-mass black holes (IMBHs), and AGNs; (ii) determining the mass of a black hole in a Type 2 AGN in a host galaxy with disturbed morphology making host-galaxy scaling relations unusable; or (iii) investigating the evolution of host-galaxy scaling relations with redshift.

The fundamental plane has been used to estimate black hole masses in a number of cases  where other methods were not viable.  For instance, {this} plane was used to claim the presence of a $10^{6}\,\msun$ black hole in the dwarf starburst galaxy Henize 2-10, which had no obvious spheroidal component from which one could use host scaling relations \citep{2011Natur.470...66R}.  \citet{2011ApJ...738L..13M} used the fundamental plane to estimate the mass of the black hole in the tidal disruption event Swift J164449.3+573451.  The latter case assumes that the fundamental plane is appropriate at high\,---\,probably super-Eddington\,---\,accretion rates expected from a tidal disruption event.




In \citet{2009ApJ...706..404G}, we looked at the fundamental plane for a sample of AGNs that have direct, primary measurements of black hole mass to eliminate this source of systematic uncertainty.  We found a number of potentially interesting empirical results, but it was not clear whether these arose because of actual correlations or because of the relatively small number {(18)} of AGNs with direct, primary $M$ measurements that also have suitable X-ray and radio data.  In this current paper, we use new data to continue our study of the fundamental plane with black holes with direct, primary mass measurements.  The use of such mass measurements enables us (i) to eliminate the systematic uncertainty of using secondary measurements and (ii) to calibrate the relation for use as a mass estimator.  We use the X-ray data from \citet{2012ApJ...749..129G} as well as hitherto unpublished radio data plus archival and data from the literature to make the largest $M$--$L_R$--$L_X$ analysis for black holes with dynamical mass measurements to date now including 30 AGN sources and 6 XRB sources.  We discuss our sample selection and standards for mass, radio, and X-ray data inclusion in section \ref{sample}.  In section \ref{fitting} we describe our fitting methodology and results from the fundamental plane fit to the data.  We discuss our results in section \ref{discussion} and conclude with advice for how to use the fundamental plane to estimate black hole mass in section \ref{massestimation}.  Appendices \ref{archivalxray}--\ref{xrbdata} contain details of the new and archival data analysis used in this paper.


\section{Sample}
\label{sample}

In this section we describe our sample.  Because of the differences in the two, we discuss the AGN and XRB samples separately.  The data are summarized in Figures \ref{f:mbhhist}--\ref{f:lrlxhist} and listed completely in Tables \ref{t:arcombhfits}--\ref{t:xrbdata}.  { We note here that there is significant improvement in the sample here over our earlier sample of \citet{2009ApJ...706..404G}.  We have increased the AGN sample from 18 to 30, and inclusion in the current {expanded} XRB sample required stricter mass and distance determinations {changing the number from 3 to 6}.  We also have improved AGN classification (section \ref{sample:smbhclass}).}

\subsection{AGNs}
\label{sample:smbhs}

The ideal data set of AGNs will have a large number of sources with multiple, independent measurements of mass, strictly simultaneous measurements of the X-ray and radio fluxes at the highest possible spatial resolution in multiple bands.  Such ideal data do not exist, and thus we have made a number of compromises in terms of simultaneity and multiple bands.  To mitigate the impact of these comprimises, we have implemented a number of measures, which we describe below.  Overall, these measures help ensure that our final results {are} robust and meaningful.

\subsubsection{Mass estimates}
\label{sample:smbhmass}

We require our AGN sources to have primary, direct dynamical mass measurements.  This requirement implies that all black hole-mass measurements were done using stellar dynamical \citep[e.g.,][]{2009ApJ...695.1577G}, gas dynamical \citep[e.g.,][]{
2010ApJ...721..762W}, or megamaser techniques \citep[e.g.,][]{2011ApJ...727...20K}.  Because reverberation mapping techniques \citep{1982ApJ...255..419B} mostly rely on normalizing the mass estimates to the primary measurements and thus are not independent, we do not include them.  Requiring that mass measurements be independent is essential for using the fundamental plane as a mass estimator.

Our sample of AGNs begins with the compilation of primary, direct dynamical mass measurements in \citet{2013ARA&A..51..511K}.  We supplement their compilation with upper limits compiled by \citet{2009ApJ...698..198G}.  We also adopt the distances determined by \citet{2013ARA&A..51..511K}, unless they are unavailable, in which case we use the value determined by \citet{2009ApJ...698..198G}.  We note that NGC 1399 has two stellar dynamical mass measurements \citep{2007ApJ...671.1321G, 2006MNRAS.367....2H} that are independently reliable but inconsistent with each other.  \citet{2009ApJ...698..198G} and \citet{2013ARA&A..51..511K} took both results into consideration, but given the larger number of tests done by the code used in the \citet{2007ApJ...671.1321G} result, we use only their mass value.  All AGN sources are listed by host galaxy name in Table \ref{t:agndata} including black hole masses ($M$) along with the references for the measurement.

\subsubsection{Radio data}
\label{sample:smbhradio}

The ideal radio data for this project requires good spatial resolution to isolate the nuclear core flux with good point-source sensitivity to reach as deep as possible to avoid having only an upper limit.  This effectively requires sensitive radio interferometry, ideally with the Karl Jansky Very Large Array (VLA), which has a resolution of about $0\farcs4$ at 5 GHz, corresponding to 40 pc at a distance of 20 Mpc.  Whenever possible, we have used VLA data either from the modern epoch or from earlier epochs, but we have also used data from other radio interferometers and even single-dish data for a few sources.  We use the 5 GHz band as the frequency of reference to be compatible with other fundamental plane studies, but we do not require that the observations be taken at this frequency.  We do not make such a requirement because (i) the radio spectra near 5 GHz are almost always power-law spectra that can be relatively easily translated to 5 GHz, (ii) at higher frequencies the spatial resolution is superior, allowing better isolation of nuclear core flux from contaminants, and (iii) limiting ourselves to 5 GHz would severely limit the amount of archival and literature data at our disposal.  We discuss how we translate to 5 GHz in further detail in section \ref{radioluminosities}.  The data in this paper come from (i) our previous analysis of archival data \citep{2009ApJ...706..404G}, (ii) results published in the literature, (iii) our analysis of previously unpublished data obtained for this project (Appendix \ref{newradio}), and (iv) our analysis of archival data (Appendix \ref{archivalradio}).

\subsubsection{X-Ray data}
\label{sample:smbhxray}

The ideal X-ray data for our project requires good spatial resolution to isolate the nuclear flux as best as possible and good sensitivity to reach to low Eddington fractions.  This effectively requires moderate to long exposures with the \emph{Chandra X-ray Observatory} (\emph{Chandra}), which has a 90\% encircled energy radius of about 0\farcs8 at 1.5 keV.  Therefore, we only use \emph{Chandra} observations of our selected AGN.  As with other fundamental plane studies, we use the 2--10 keV flux arising from a power-law spectral component of the continuum emission from the AGN.  If present, we exclude any emission lines in the bandpass.  By using the 2--10 keV bandpass, we avoid most problems arising from inferring intrinsic flux in the presence of absorption.  The AGN X-ray data in this paper come from (i) our previous analysis of archival data \citep{2009ApJ...706..404G}, (ii) our previous analysis of \emph{Chandra} data obtained for this project \citep{2012ApJ...749..129G}, and (iii) an analysis of archival X-ray 
data presented in this paper in Appendix \ref{archivalxray}.

\subsubsection{AGN Classification}
\label{sample:smbhclass}

One of the goals of this project is to determine whether or not the X-ray and radio emission by Seyferts are explained by the same fundamental plane as low-luminosity AGNs (LLAGNs) and low-hard XRBs.  Therefore we include data from both Seyferts and LLAGNs.  We return to this topic in section \ref{discussion:seyferts} in which we discuss how we identify Seyferts {based on classification by \citet{1997ApJ...487..568H} and inspection of the X-ray spectrum}.

\subsection{XRBs}
\label{sample:xrbs}

We choose XRBs with well measured black hole masses and distances for which there are strictly or nearly simultaneous radio and X-ray data.  It is obvious that an analysis of a relation among mass, radio luminosity, and X-ray luminosity requires mass measurements to be accurate and to have reliable measurement uncertainties.  The distances to the XRBs are essential for turning the measured radio and X-ray fluxes into luminosities.  The need for simultaneity arises from the variability of XRBs on timescales as short as hours.  These selection requirements leave us with six black hole XRBs, which we list in Table \ref{t:xrbsample}.  While most of the XRB data come from accreting systems in the low/hard state, we also include a few data from XRBs that are in an intermediate, high/soft state, or very high state.  Although XRBs are generally considered to have had their jets quenched when leaving the low/hard state and thus should not go on the fundamental plane, we include them even if they are in a state other than low/hard so long as they have measurable radio flux densities.  We return to this topic in section \ref{discussion}.  The XRB data come from a combination of literature mass, radio, and X-ray measurements with archival analysis of \emph{Rossi X-ray Timing Explorer} (\emph{RXTE}) data for a few sources (Appendix \ref{xrbdata}).  The requirement for \emph{Chandra} X-ray data for AGNs does not extend to XRBs as they do not suffer from the same level of contamination issues.

{
We note that the sample here differs from \citet{2009ApJ...706..404G} in that we have increased our sample size from 3 XRBs to 6, keeping GRS 1915+105 and Cygnus X-1, dropping V404 Cyg because of questions raised about its nature during its outburst in 2015 \citep{2015GCN.17929....1B, 2015ATel.7647....1K, 2015ApJ...813L..37K}, and adding XTE J1118+480, 4U 1543$-$47, XTE J1550$-$564, and GRO J1655$-$40.  The \citet{2009ApJ...706..404G} sample had a total of 5 observations (2 of GRS 1915+105, 2 of Cygnus X-1, and 1 of V404 Cyg), whereas in this work we have 69 total observations split as shown in Table \ref{t:xrbdata} with at least two observations of each.  
}

\kgfigbeg{mbhhist}
\includegraphics[width=\columnwidth]{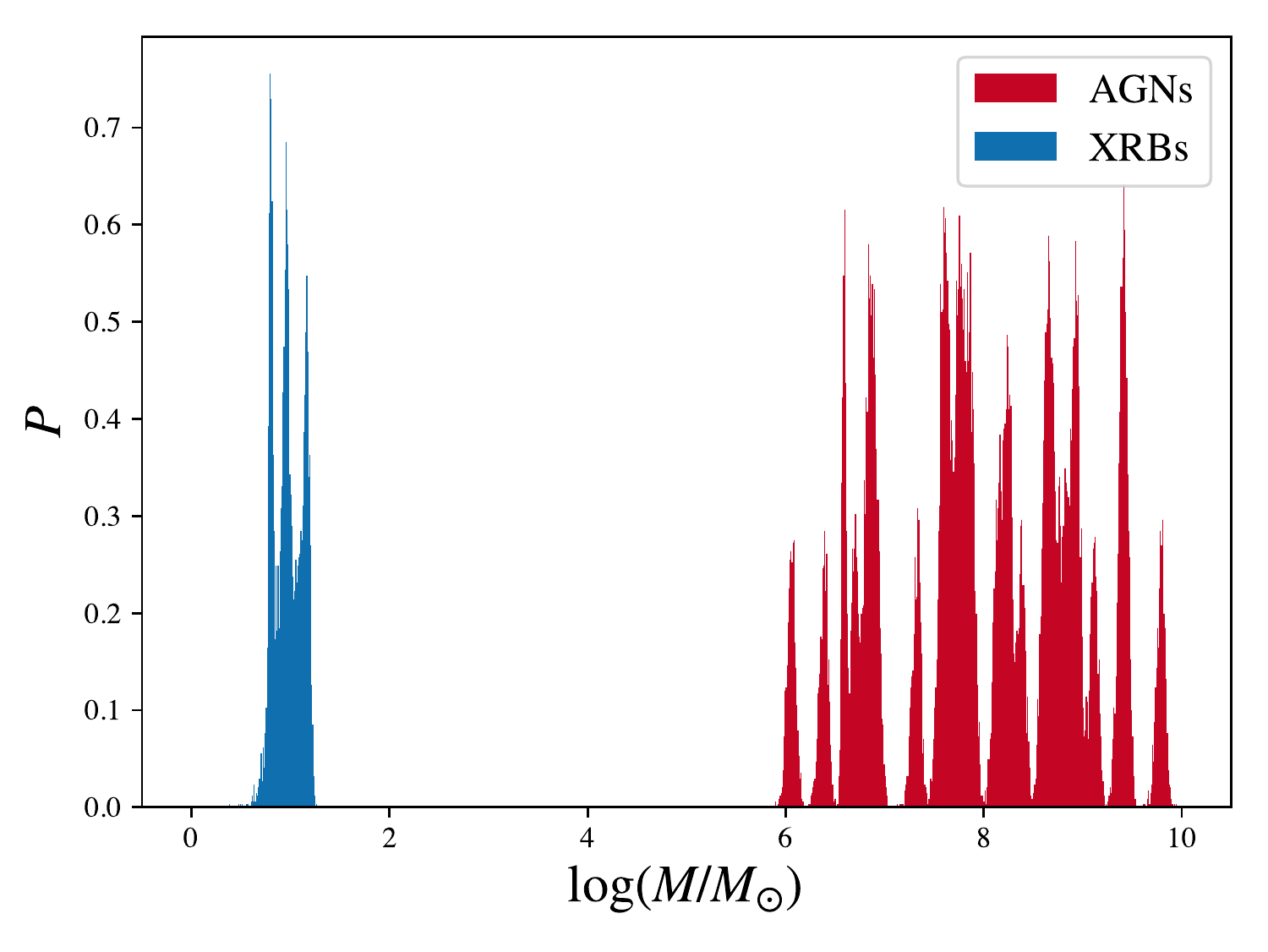}
\caption{Probability density of black hole masses for the entire sample. We plot multiple realizations of our sample from the masses with uncertainties and their dependence on distance for the AGN sample.  Thus, the distribution of the probability density of an individual black hole's mass combines the statistical measurement uncertainties of both the mass and the distance.  We do not include distance uncertainty in the XRB mass probability density because it generally does not affect it.  The range of masses included in our sample is illustrated in this figure.}
\kgfigend{mbhhist}{BH mass histogram.}

\kgfigbeg{feddhist}
\includegraphics[width=\columnwidth]{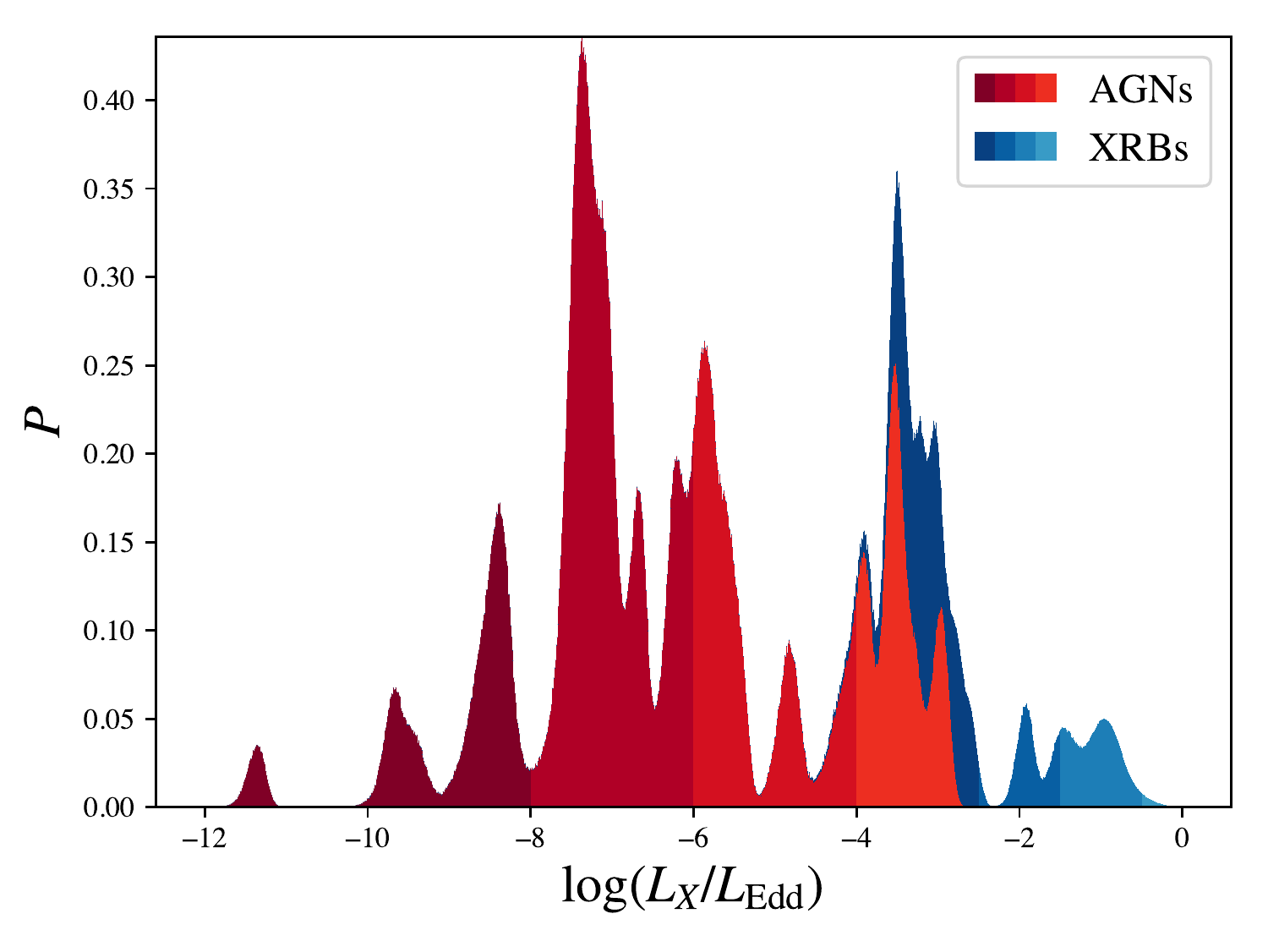}
\caption{Probability density of logarithmic 2--10 keV X-ray Eddington fraction.  We plot multiple realizations of our sample using statistical uncertainties of X-ray flux, distance to source, and black hole mass.  For AGN sources, we take into account the correlation between distance in the luminosity calculation and distance in the mass estimate.  The multiple observations of individual XRBs are incorporated by weighting each as $1/N$, where $N$ is the number of observations of a given XRB.  The probability density curves are colored according to mass category (red for AGNs and blue for XRBs) and the shade is given by the value of $\log(L_X/L_\mathrm{Edd})$. AGNs are grouped into the following discrete bins: $(-\infty, -8)$, $[-8, -6)$, $[-6, -4)$, and $[-4, +\infty)$.  XRBs are grouped into the following discrete bins: $(-\infty, -2.5)$, $[-2.5, -1.5)$, $[-1.5, -0.5)$, and $[-0.5, +\infty)$.  We use this color scheme in figures throughout this paper.  Note that the probability density plotted is the total probability density of all sources so that at, e.g., $\log(L_X/L_{\mathrm{Edd}}) = -4$, the probability density is dominated by AGNs with a small contribution coming from XRBs.
  The nearly 10 orders of magnitude in X-ray Eddington fraction covered by our sample is illustrated in this figure.}
\kgfigend{feddhist}{Logarithmic X-ray Eddington fraction histogram.}

\kgfigbeg{lrlxhist}
\includegraphics[width=\columnwidth]{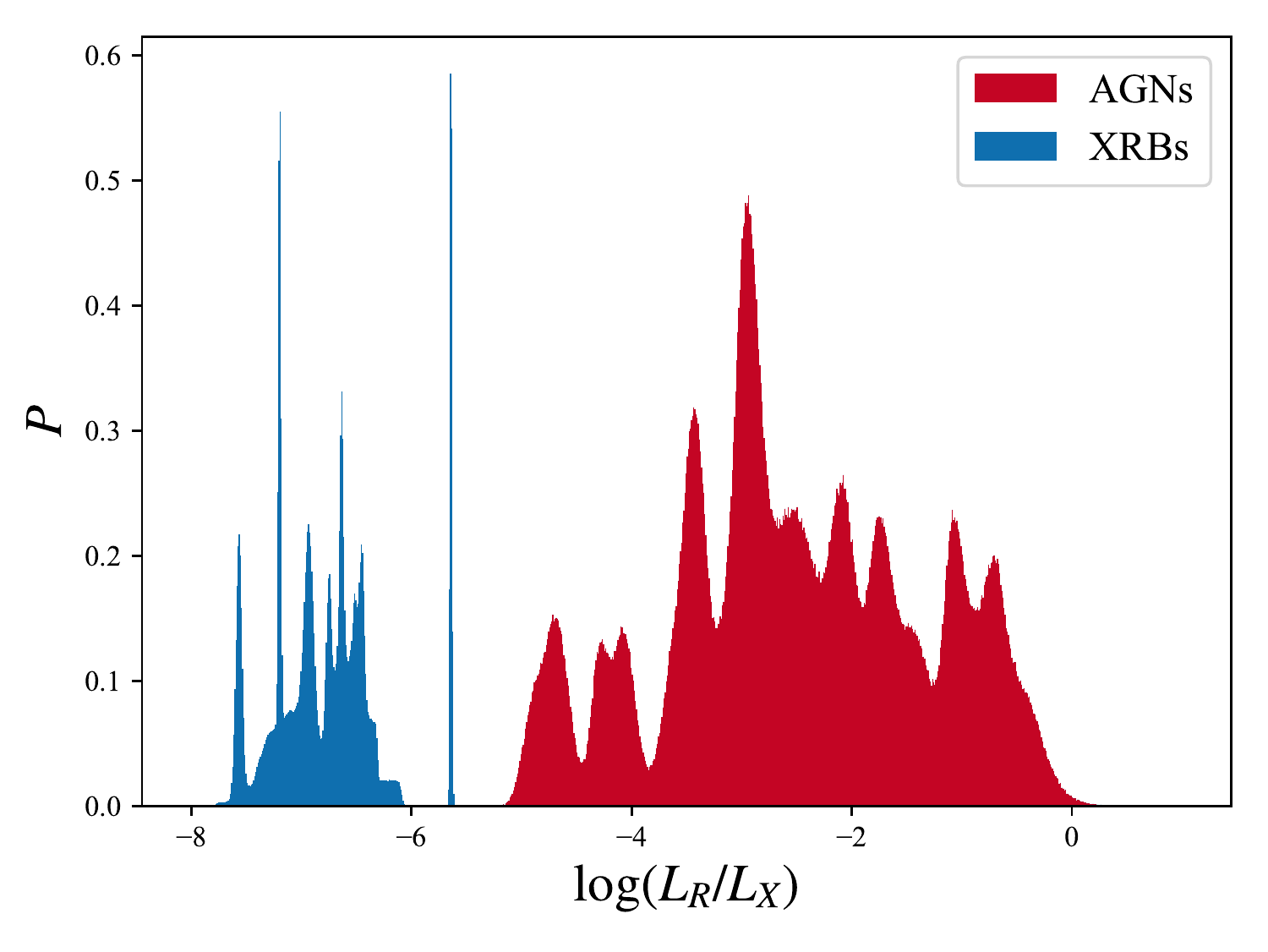}
\caption{Probability density of logarithmic ratio of radio to X-ray luminosity.  We plot multiple realizations of our sample using statistical uncertainties of radio flux and X-ray flux.  As in the rest of this paper, radio ``luminosity'' is defined as $L_R \equiv \nu L_\nu$ for $\nu = 5\,\mathrm{GHz}$, whereas the X-ray luminosity is a true 2--10 keV bandpass luminosity.  The uncertainty in the radio luminosity also incorporates our estimate of systematic uncertainty in converting from other frequencies to 5 GHz.  The nearly 8 orders of magnitude probed by our sample is evident in this figure as is the larger fractional uncertainties in the AGNs.}
\kgfigend{lrlxhist}{Rado to X-ray flux.}


\section{Fitting the fundamental plane}
\label{fitting}

The primary analysis of this paper is to fit the fundamental plane to the data gathered in section \ref{sample} with special consideration of uncertainties.  In this section, we first describe our Markov chain Monte Carlo (MCMC) fitting methodology and our treatment of statistical and systematic uncertainties.  Then, we present the results of the fits.  { The fitting and handling of uncertainties in this work is a significant improvement over our earlier analysis in \citet{2009ApJ...706..404G}.}  {Here, we include a treatment of correlated uncertainties, which avoids a source of systematic uncertainty.  We also handle multiple observations of individual objects, which increases the information provided without unduly biasing results to the particulars of an individual sources.  We now mitigate the effects of contamination by XRBs on the AGN X-ray flux measurements statistically, which would produce a systematic uncertainty if we did not.  We also handle non-simultaneous observations of X-ray and radio fluxes statistically to account for another source of systematic uncertainty.}

\subsection{Statistical treatment}
\label{fitting:statistics}

We employ emcee, the \citet{ForemanMackey2016} Python implementation of MCMC, with random sampling of the data from the measurement uncertainties self-consistently for each realization.  First, we describe our treatment of statistical and systematic uncertainties in our realization of the data set, and then we describe the fitting of the realized data sets.

Our strategy for treating measurement uncertainties in our fits is to use Monte Carlo methods to randomly sample each of the data points from the measurement uncertainties, keeping track of correlated uncertainties.  Thus, in the maximum likelihood or MCMC methods described below, every time that we compare the data to a model, we generate a full realization of the data set.  Because of the large number of data--model comparisons done in MCMC methods, we fully sample the measurement uncertainties and their covariances.  We outline our method of Monte Carlo data realization in the following subsections.

\subsubsection{Distances}
\label{distances}
For each source we assign a distance, $D$, drawn from the measured distance and assumed normal distribution from the given 1$\sigma$ measurement uncertainty.  We assume a 10\% uncertainty for all extragalactic distances.  Note that we use this distance for all subsequent calculations in a given realization.

\subsubsection{Masses}
\label{masses}
For each source we assign a black hole mass, $M$, drawn from the measured black hole mass and assumed normal distribution from the given 1$\sigma$ measurement uncertainty.  If the high and low measurement uncertainties are asymmetric, we approximate it by averaging the two and using as the 1$\sigma$ measurement uncertainty.  For AGNs, mass estimates depend on the assumed distances to the sources.  To take this into account, we scale the mass to the  distance realized in section \ref{distances} linearly, except for Sgr A*, which scales as $D^2$.  As an example, consider NGC 3607, which has a distance $D = 22.65\,\mathrm{Mpc}$ and mass $M = 1.37^{+0.45}_{-0.47} \times 10^{8}\,\msun$.  First we symmetrize the measurement uncertainty to be $M = (1.37 \pm 0.46) \times 10^8\,\msun$.  We simulate the mass by drawing normal deviate to be, e.g., $1.53 \times 10^8\,\msun$.  If for a given realization, the realized distance is $D = 19.15\,\mathrm{Mpc}$, then we scale the mass to $M = 1.29 \times 10^{8}\,\msun$.

\subsubsection{Radio luminosities}
\label{radioluminosities}
For each source, we calculate a 5 GHz radio ``luminosity'' as $L_R \equiv \nu L_\nu = 5\,\mathrm{GHz}\ L_5 = (5\,\mathrm{GHz}) 4\pi D^2 F_5$, where $L_5$ and $F_5$ are the 5 GHz luminosity density and flux density, respectively, and $D$ is the distance simulated in section \ref{distances}.  Because we do not always have 5 GHz radio data, we use the following procedure.  First, if we have 5 GHz data {(as we do for eight sources)}, we use it.  If not, we convert the $\nu \ne 5\,\mathrm{GHz}$ data using a literature value for the {five available} measured spectral index $\alpha$ (using the objectively superior sign convention that $S_\nu \propto \nu^{-\alpha}$).  { The literature radio spectral index measurements are listed in Table\ \ref{t:agndata} in the table notes.}  We simulate a value of $\alpha$ based on the 1$\sigma$ measurement uncertainties, and use that to convert the observed $F_\nu$ data to $F_5$.  {For the 13 sources for which} no literature $\alpha$ value is available, we use additional $\nu \ne 5\,\mathrm{GHz}$ data to calculate $F_5$ by interpolating between or extrapolating from the two $F_\nu$ observations, each simulated from their corresponding 1$\sigma$ measurement uncertainties and assuming a power-law spectral form.  { The data used for these {simple} spectral energy distributions in the 5 GHz region are listed in Table\ \ref{t:agndata} in the table notes for the sources for which it was done.}  Finally, if no other usable data exist {(four sources)}, we simulate $\alpha$ by drawing from a uniform distribution of $[-0.5, +0.5)$ to calculate $F_5$.  We tried several different ranges of the uniform distribution of $\alpha$ and found that it had very little impact.  Because we use high-angular resolution data as much as possible, we are generally insensitive to contamination from star formation at the galaxy nucleus.  We note that there are two reasons to calculate $\alpha$.  First is to get 5 GHz flux density from measurements at other frequencies as mentioned above.  Second is to determine whether the radio emission is due to core AGN activity, which would have a flat radio spectrum, or from extended jets, which would have a steep radio spectrum.  
We use the $\alpha$ {estimates} above for core--jet determination in section \ref{discussion:flatsteep} {so that we need an estimate of $\alpha$ even if we have $\nu = 5\,\mathrm{GHz}$ data in the same method as described above}.  {The uncertainty of an $\alpha$ estimate when using two radio measurements is a combination of the uncertainties in flux densities and in the distance between the two frequencies.  We assume that the uncertainty in frequency is negligibly small.  While we could improve the precision of the $\alpha$ estimates by acquiring additional observations at other frequencies, the precision and accuracy of our estimates is sufficient for determining $F_5$ and for determining whether the emission is flat or steep.  With two data points with arbitrarily small uncertainties in the flux density measurements, our $\alpha$ estimates will be very precise.  As detailed in Table\ \ref{t:agndata} and as can be seen in Figure \ref{f:alphased}, most of our $\alpha$ estimates come from a measurement near 5 GHz and another near 8.5 or 15 GHz.  This factor of $\sim2$--3 in frequency range is sufficient for our purposes.}

\kgfigbeg{alphased}
\centering
\includegraphics[width=1.0\columnwidth]{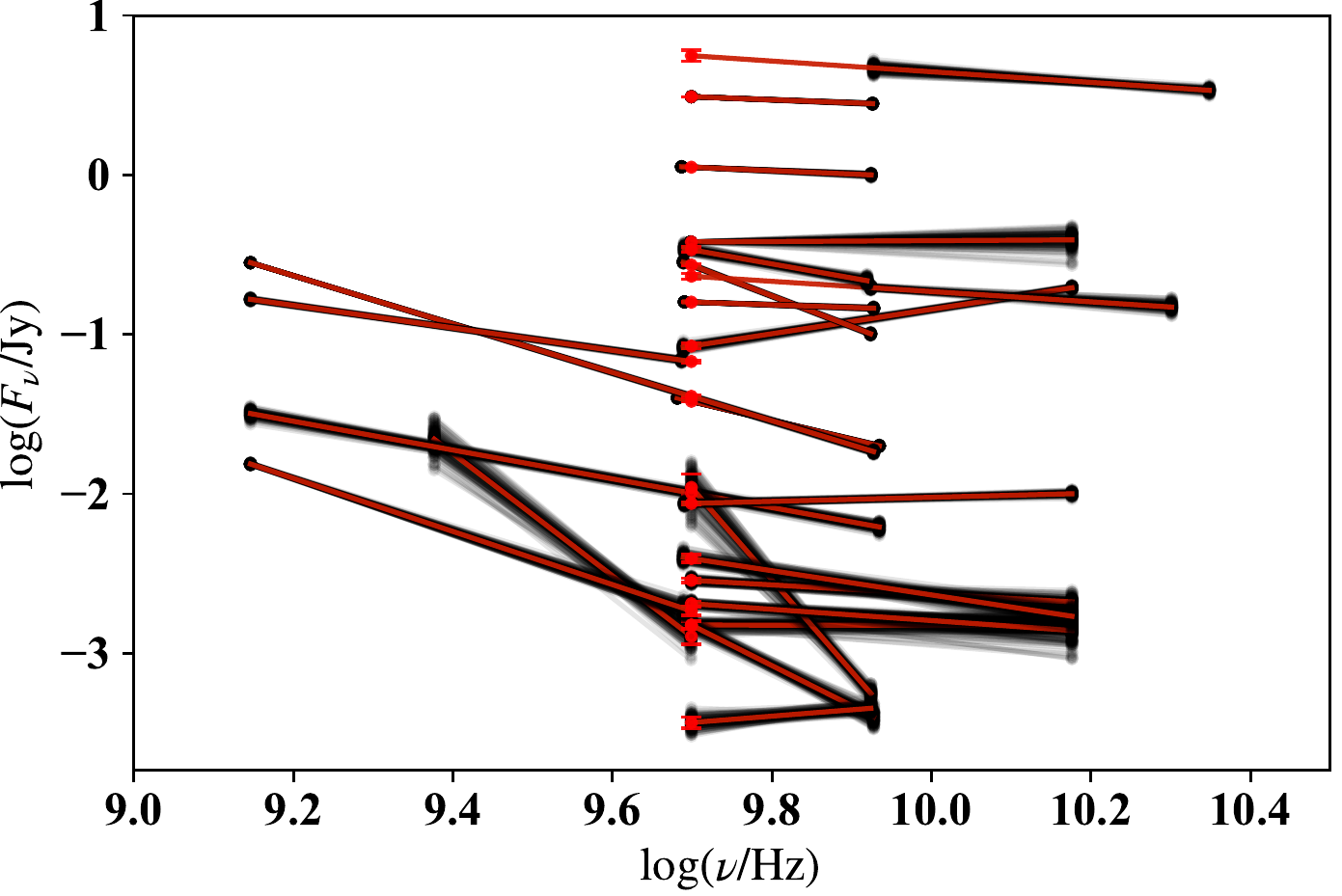}
\caption{\bf
A summary plot of radio data for all AGN for which we have calculated radio spectral indexes ($\alpha$).  Each set of black lines was constructed by doing Monte Carlo simulations of the measurements of the source at each frequency assuming a Gaussian distribution with dispersion equal to their $1\sigma$ uncertainties.  Because the sources are all at low redshift, the correlated uncertainties in distance do not come into effect here, and we just plot flux density.  Then we calculate $\alpha$ for the two values.  We do this for $10^5$ realizations, but we only plot a random subset of 100 of them.  For using a single value of alpha for our tables, we use the median value of alpha with the 68\% intervals for our $1\sigma$ uncertainties. We plot, in red, the median value and the median interpolated/extrapolated value for the 5 GHz $F_{\nu}$ value with $1\sigma$ error.  Although we plot the medians for reference, we  use the Monte Carlo realizations for each of our fit realizations. In this way, e.g., a source with $\alpha = 0.3 \pm 0.1$, will be classified and treated as a flat source for the roughly 84\% of the time that $\alpha < 0.4$. 
}
\kgfigend{alphased}{Simple radio SED to determine $\alpha$.}

\subsubsection{X-Ray luminosities}
\label{xrayluminosities}
For each source, we calculate a 2--10 keV X-ray luminosity from the measured 2--10 keV X-ray flux, simulated from symmetrized uncertainties and converted to a luminosity using the distance simulated in section \ref{distances}.  We incorporate an additional systematic uncertainty based on the non-simultaneity of radio and X-ray observations.
Because we only consider XRBs with strictly or very nearly simultaneous radio and X-ray observations, we only consider this systematic uncertainty for AGNs.  For radio and X-ray observations of AGNs that were taken more than 60 days apart, we include an additional 20\% uncertainty.  This accounts for the typical variability seen in AGNs with relatively low values of $L_X / L_\mathrm{Edd}$ as in our data set.  

\subsubsection{Background AGN contamination}
\label{bgagncontam}
For extragalactic sources, we must consider the possibility that a background AGN anti-serendipitously appears at the location of our source's nucleus.  We calculate the background contamination probability as 
\begin{equation}
P_\mathrm{BG} = A_\mathrm{PSF} \times
 \begin{cases} 
      5.93 \times 10^{-2} F_X^{-0.32} - 1639.51 & F_X < F_\mathrm{br}\\
4.26 \times 10^{-20} F_X^{-1.55} & F_X \ge F_\mathrm{br} 
   \end{cases},
\label{e:bgagn}
\end{equation}
where $A_\mathrm{PSF} = 2.424 \times 10^{-7}$ is the area of \emph{Chandra}'s point-spread function (PSF) in square degrees, $F_X$ is measured in units of $\mathrm{erg\,s^{-1}\,cm^{-2}}$, and $F_\mathrm{br} = 6.4\times 10^{-15}\,\mathrm{erg\,s^{-1}\,cm^{-2}}$ is the location of the break.  This comes from $\log N$--$\log S$ cumulative number density of cosmic X-ray background sources from deep field surveys \citep[e.g.,][]{2001AJ....122....1B, 2001A&A...365L..45H, 2002ApJ...566L...5C, 2002ApJ...566..667R, 2003ApJ...588..696M, 2004AJ....128.2048B, 2005ARA&A..43..827B}.  Because of the low probabilities associated with the small value of $A_\mathrm{PSF}$, the exact form of Equation (\ref{e:bgagn}) does not significantly change our results.
 
\subsubsection{XRB contamination}
\label{xrbcontam}
For extragalactic sources, we also need to consider contamination from low-mass X-ray binaries (LMXBs) and high-mass X-ray binaries (HMXBs) intrinsic to the galaxy near the nucleus.  LMXBs tend to scale with stellar mass ($M_*$), and HMXBs tend to scale with star formation rate ($\mathrm{SFR}$) so that we consider them separately.  We use the stellar masses calculated by \citet{2016ApJ...830L..12T}, if available; otherwise we use that calculated by \citet{2016ApJ...831..134V}.  We also calculate $f_\mathrm{nuc}$, the fraction of total stellar light from the galaxy at the nucleus, defined by a circular radius equal to 1\arcsec, the combined astrometric uncertainty when registering optical images usually used to determine the galaxy center, radio observations of the nucleus, and \emph{Chandra} observations of the nucleus.  We calculate this with simple photometry of Two Micron All Sky Survey or DSS images of the galaxies.  The expected number of LMXBs is then given by \citet{2004ApJ...611..846K} to be 
\beq
N_\mathrm{LMXB} = 25.4 f_\mathrm{nuc} (M_* / 10^{11}\,\msun) (L_X / 10^{38}\,\mathrm{erg\,s^{-1}})^{-1}.
\eeq

To calculate HMXB contamination, we need the SFR of each host galaxy.  We use the values calculated by \citet{2016ApJ...830L..12T}, if available.  If the source is not in \citet{2016ApJ...830L..12T}, we calculate SFR using the same methods from similar available far infrared data.  In particular, we estimate the SFR to be $\log(\mathrm{SFR} / \msun\,\mathrm{yr^{-1}}) = \log(2 L_\mathrm{FIR} / \mathrm{erg\,s^{-1}}) -43.41$, where $L_\mathrm{FIR}$ is the far infrared luminosity estimated from the 60 and 100 $\mu$m data as 
\beq
\frac{F_\mathrm{FIR}}{\mathrm{erg\,s^{-1}\,cm^{-2}}}  = 1.26 \times 10^{-11} \left(2.58 \frac{S_{60}}{Jy} + \frac{S_{100}}{Jy}\right),
\eeq
where $F_\mathrm{FIR}$ is the FIR flux, and $S_{60}$ and $S_{100}$ are the 60 and 100 $\mu$m flux densities as found in \emph{IRAS} and other data.  Because of the approximate nature involved in estimating SFR this way, we do not fold into it the distance uncertainty as the systematics of using this correlation dominate.  

With the SFR estimated, we may calculate the expected number of HMXBs based on \citet{2003MNRAS.339..793G} to be
\beq
N_\mathrm{HMXB} = 5.4 f_\mathrm{nuc} (\mathrm{SFR} / \msun\,\mathrm{yr^{-1}}) (L_X / 10^{38}\,\mathrm{erg\,s^{-1}})^{-0.61}.
\eeq
In our data set, only one of $N_\mathrm{LMXB}$ and $N_\mathrm{HMXB}$ is ever very large, we only consider the larger of the two to calculate the probability of XRB contamination  as $P_\mathrm{XRB} = 1 - \exp(-\mathrm{max}(N_\mathrm{LMXB}, N_\mathrm{HMXB})]$.

\subsubsection{Censoring data}
When realizing each data set, we draw a pair of uniform random deviates from the semi-open interval $[0,1)$ for each source.  If the first random deviate is smaller than $P_\mathrm{BG}$ or the second random deviate is smaller than $P_\mathrm{XRB}$, then we discard the given source from that realization. Thus, a source that is very unlikely to be contaminated by either XRBs or background AGNs will be represented in almost all realizations, whereas a source that has a higher chance of contamination will be represented in a smaller fraction of realizations.  This allows us to consider contamination in a probabilistic way.  We also discard data in which the random normal deviates used in measurement uncertainty Monte Carlo simulations result in a negative mass, radio luminosity, or X-ray luminosity.  Because most of the data are detected at 3$\sigma$ or better, this is not a frequent occurrence.

\subsubsection{Multiple observations of individual sources}
\label{multobs}
As a final consideration of sources with multiple observations, primarily XRBs but also Sgr A*, we assign a weight equal to the reciprocal of the number of observations of that source that were not censored.  This gives us the ability to use multiple observations as a probe of filling out the fundamental plane if, e.g., an XRB is observed at vastly different regions of the $L_R$--$L_X$ plane without over-weighting a source that happens to have many observations.

\subsection{MCMC Fitting}
\label{mcmcfitting}
We use the following model, describing the plane
\beq
\mu = \mu_0 + \xi_{\mu R} R + \xi_{\mu X} X,
\eeq
where $\mu \equiv \log(M/10^8\,\msun)$, $R \equiv \log(L_R/10^{38}\,\mathrm{erg\,s^{-1}})$, and $X \equiv \log(L_X / 10^{40}\,\mathrm{erg\,s^{-1}})$. The model parameters  are the mass intercept ($\mu_0$), radio slope ($\xi_{\mu R}$), X-ray slope ($\xi_{\mu X}$), and a Gaussian intrinsic scatter in the log-mass direction ($\epsilon_\mu$). In our fitting, we first find the maximum likelihood model by minimizing the negative of the following likelihood:
\beq
\log{\cal{L}} = - \frac{1}{2} \sum_{i=1}^{N} w_i {\left(\mu_i - \mu(R_i, X_i, \theta)\right)^{2}}{\epsilon_\mu^{-2}} + \log\epsilon_\mu^{2},
\eeq
where $w_i$ is a weight for each of the $N$ data points ($\mu_i$, $R_i$, $X_i$) and $\theta$ is a vector of the model parameters.  In practice, we use $\ln\epsilon_\mu$ as a fit parameter to avoid numerical problems associated with negative values of $\epsilon_\mu$ in minimization techniques.

We use the results of the maximum likelihood finding as a starting location with the \citet{ForemanMackey2016} implementation of the \citet{2010CAMCS...5...65G}
affine-invariant MCMC ensemble sampler.  We start 100 walkers in a small region centered on the maximum likelihood results randomized in each parameter with a small ($10^{-4}$) deviation for each of the walkers.  For all model parameters ($\mu_0$, $\xi_{\mu R}$, $\xi_{\mu X}$, and $\ln\epsilon_\mu$) we use an uninformative uniform prior of $(-5, +5)$.  Tests with different priors showed no differences to the results.  We ran the sampler for different numbers of steps, inspecting the chains visually to determine how many steps should be used for burn-in.  We found that 200 steps was always sufficient for burn-in. We tried various numbers of steps, up to $10^6$ to ensure robust results.  These experiments showed that any number of steps above 10 times the autocorrelation time gave essentially the same median and 68\% interval, and the only merit in increasing the number of steps above was smoother posterior figures.  Thus, we present results with $10^6$ steps to show the smoothest figures.

\subsection{Fitting Results}
\label{fitting:results}

The results of our MCMC fitting are summarized in Figure \ref{f:cornerplot}.  We take the median and 68\% interval of the posterior distribution as our final results.  Our best fitting correlation parameters are 
\beqa
\mu_0 &=& 0.55 \pm 0.22,\nonumber\\
\xi_{\mu R} &=& 1.09 \pm 0.10,\nonumber\\
\xi_{\mu X} &=& -0.59^{+0.16}_{-0.15}, \text{ and}\nonumber\\
\ln\epsilon_\mu &=& -0.04^{+0.14}_{-0.13}.
\eeqa  The posterior probability distributions are singly peaked with a roughly normal distribution, indicating robust results.  

\kgfigstarbeg{cornerplot}
\centering
\includegraphics[width=0.8\textwidth]{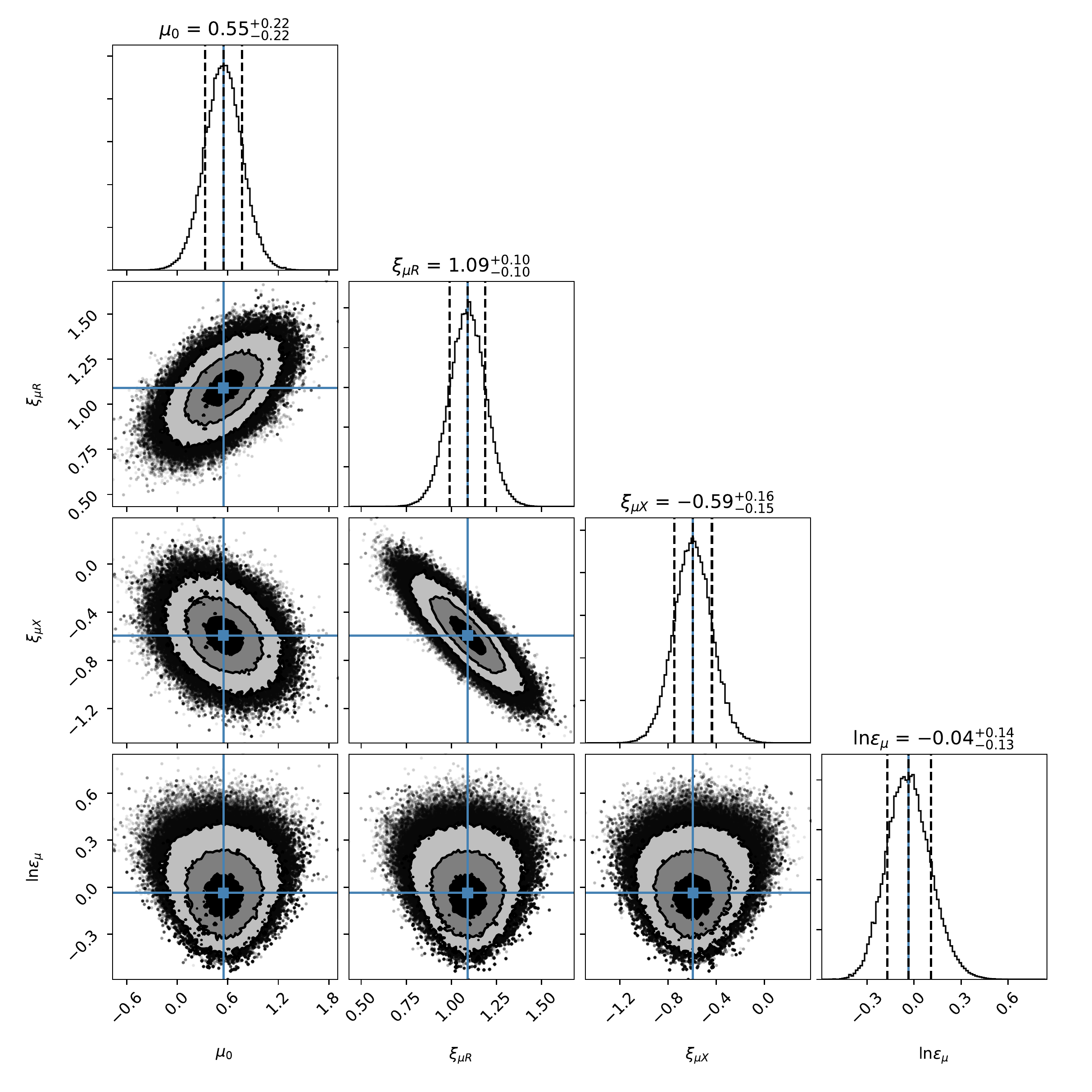}
\caption{A \citet{ForemanMackey2016} corner plot of MCMC results. Each panel in this corner plot of our MCMC results shows either the posterior probability distribution of an individual parameter in our fits (histograms) or the joint posterior probability distribution of pairs of parameters (scatter plots).  The equations at the top of each column show the median and 68\% interval of each parameter, $\mu_0 = 0.55 \pm 0.22$, $\xi_{\mu R} = 1.09 \pm 0.10$, $\xi_{\mu X} = -0.59^{+0.16}_{-0.15}$, and $\ln\epsilon_\mu = -0.04^{+0.14}_{-0.13}$.  The posterior distributions show well behaved, mono-modal distributions.  The joint posterior distributions show some covariance between $\xi_{\mu R}$ and $\xi_{\mu X}$ as well as between $\mu_0$ and either of $\xi_{\mu R}$ and $\xi_{\mu X}$.  The asymmetry in the joint posterior distributions that include $\ln\epsilon_\mu$ is typical when using a logarithmic instinsic scatter term.  {For comparison, the corresponding fits from \citet{2009ApJ...706..404G} are $\mu_0 = 0.19 \pm 0.19$, $\xi_{\mu R} = 0.48 \pm 0.16$, $\xi_{\mu X} = -0.24 \pm 0.15$, and $\ln\epsilon_\mu = -0.26$. } }
\kgfigstarend{cornerplot}{MCMC results}

Our results are also summarized in Figure \ref{f:edgeonfp}, which shows the edge-on projection of the fundamental plane with $\log M$ as the dependent variable.  The figure shows no apparent residual trend with Eddington fraction, nor do the non-low/hard state XRBs appear to be outliers.  There is, however, substantial intrinsic scatter of  $\epsilon_\mu \approx 1\ \mathrm{dex}$, indicating a large amount of unexplained variance.  The edge-on projection shows XRBs to be low compared to the projected median relation.  As seen in the three-dimensional views in Figure \ref{f:3dfp}, this offset cannot be fixed by a simple adjustment of a slope in the edge-on projection.

\kgfigstarbeg{edgeonfp}
\iftoggle{lowresversion}{%
\includegraphics[width=\textwidth]{./edgeonfp.png}%
}{%
\includegraphics[width=\textwidth]{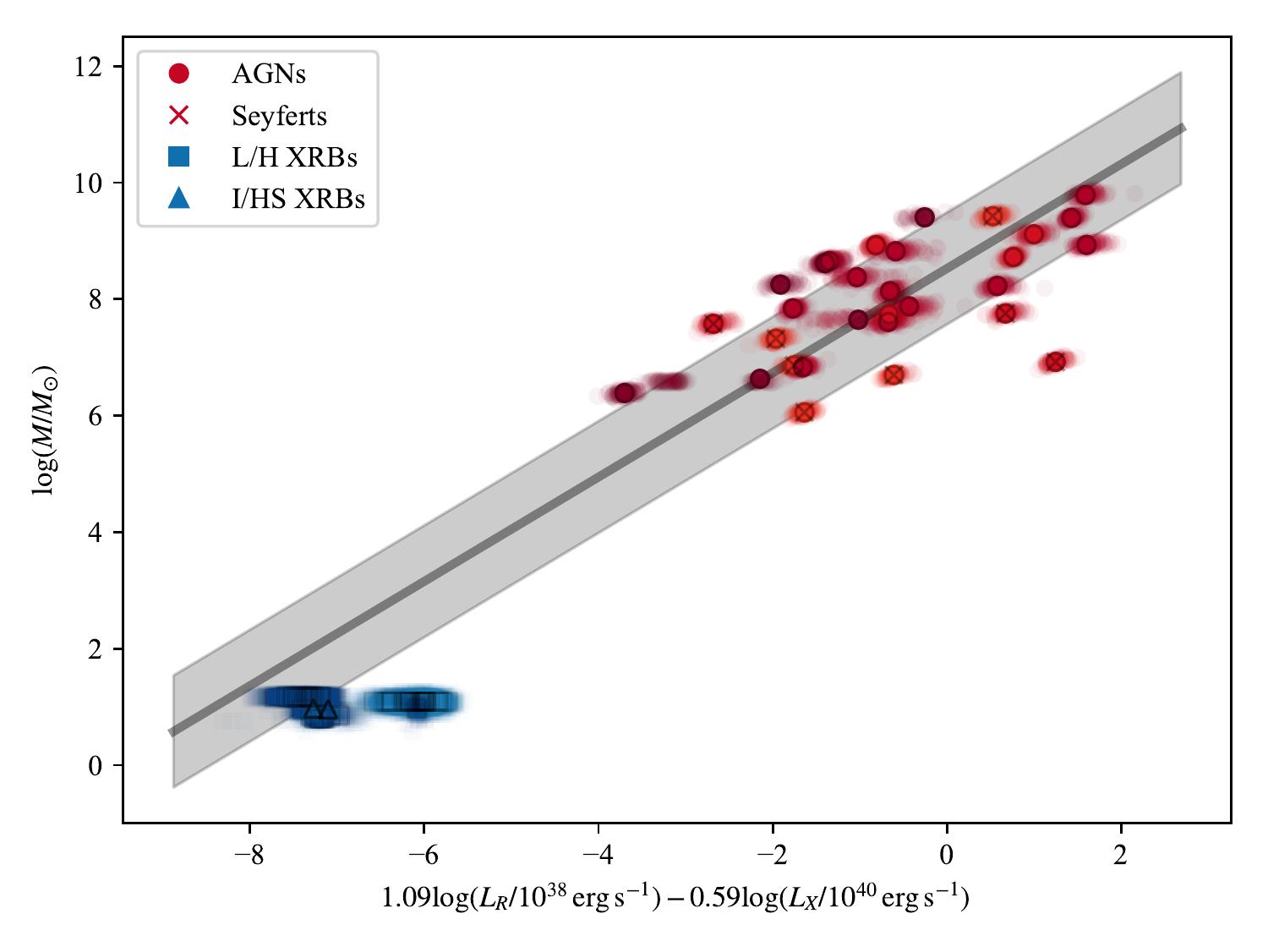}%
}
\caption{Edge-on view of the fundamental plane with mass as dependent variable.  Here we plot all data realized $N$ times to show the correlated uncertainties.  Colors are as in Fig.\ \ref{f:feddhist}, and symbols indicate whether the source is an AGN (red circles), a Seyfert AGN (red circle with cross), an XRB in a low/hard state (blue squares), or an XRB in an intermediate or high/soft state (blue triangles). Each source is sampled from its measurement uncertainties as is done in the fitting procedure and is plotted with a partially transparent symbol plus a dark outline symbol on top at the nominal values.   We plot the best fit relation as a dark gray line with a light gray shaded region to indicate the 1$\sigma$ region of the Gaussian intrinsic scatter, which has magnitude of 1 dex.  This figure summarizes the results of the fits as well as indicates the fidelity with which one can use the fundamental plane to estimate black hole mass.}
\kgfigstarend{edgeonfp}{Edge-on view of the fundamental plane.}

\kgfigstarbeg{3dfp}
\centering
\includegraphics[width=0.95\textwidth]{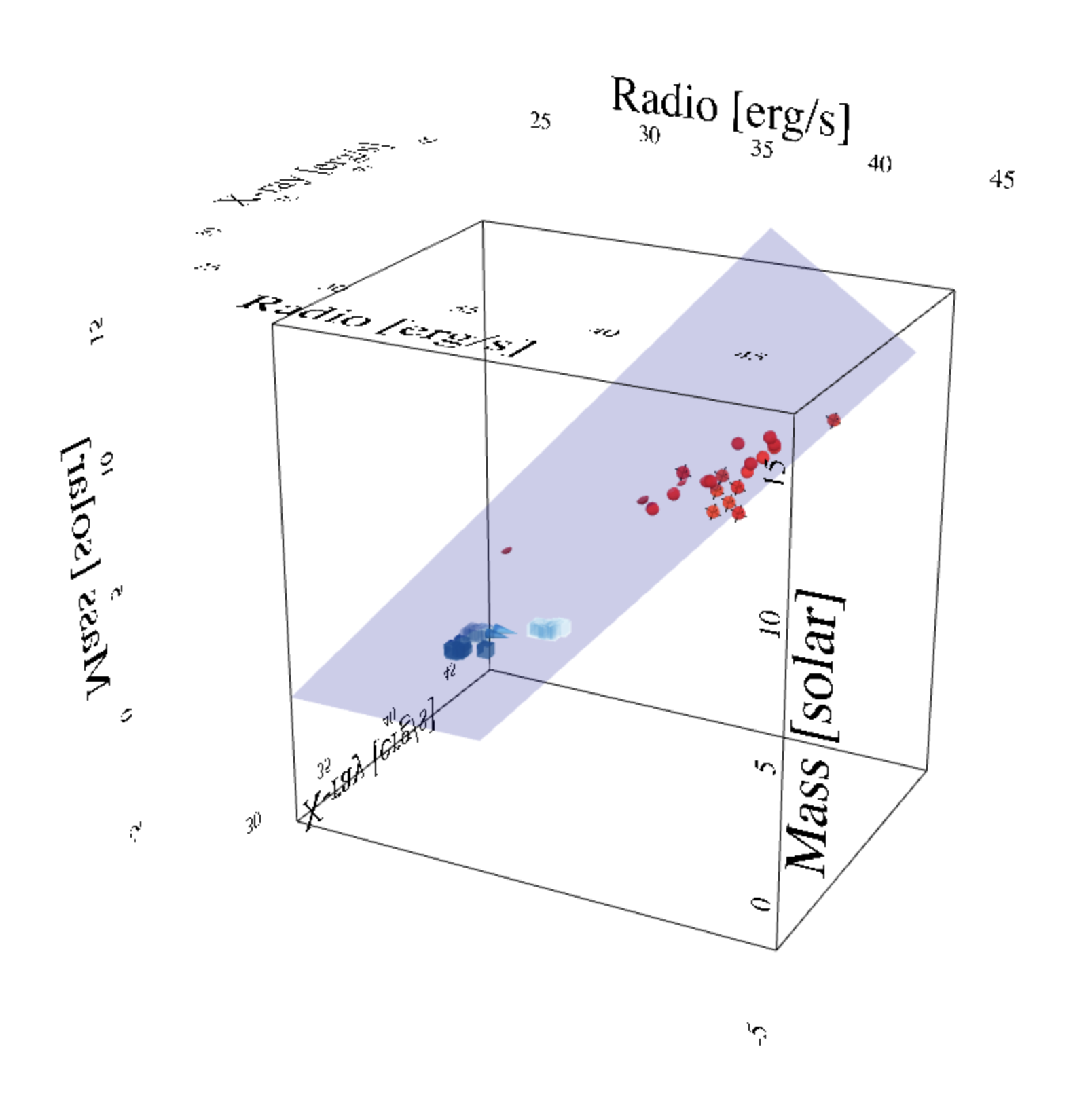}
\caption{Three-dimensional view of the fundamental plane. The online version of this figure (\url{http://kayhan.astro.lsa.umich.edu/supplementary_material/fp/}) has an interactive three-dimensional viewer to explore the data.  We plot the median plane from the MCMC fits.  Colors are as in Fig.\ \ref{f:edgeonfp}.  Spheres are AGN, and spheres with axes drawn in them are Seyferts.  Cubes are low/hard-state XRBs, and cones are intermediate or high/soft-state XRBs.  The three-dimensional view of the data reveals that the data encompass a primarily planar region rather than either a line or a three-dimensional solid within the $M$--$L_R$--$L_X$ space.  This indicates that it can, in fact, be used to estimate black hole mass.}
\kgfigstarend{3dfp}{Three-dimensional view of the fundamental plane from multiple angles.}

\section{Discussion}
\label{discussion}

In section \ref{fitting:results} we found the best-fit mass-predictor relation for our full sample to be 
\beq
\mu = 0.55 \pm 0.22 + (1.09 \pm 0.10) R + (-0.59 ^{+0.16}_{-0.15}) X,
\label{bestfit}
\eeq
with an intrinsic scatter of $\ln\epsilon_\mu = -0.04^{+0.14}_{-0.13}$.  Here we 
consider alternative samples from our parent full sample { and comment on the implications}.  { In section \ref{differences}, we compare the differences between this work and our earlier results in \citet{2009ApJ...706..404G}.}


\subsection{Sgr A*}
\label{discussion:sgrastar}
As the source with the lowest $L_X/L_\mathrm{Edd}$ value, Sgr A* deserves special consideration that its accretion properties may be different from the rest as it would be undetectable outside the Local Group.  Sgr A* is listed in Table \ref{t:agndata} three times for three different X-ray states.  It has been argued previously that Sgr A* only approaches the fundamental plane during its X-ray flare state \citep{2012MNRAS.419..267P}.  To investigate this, we try fitting the fundamental plane using only the brightest X-ray flux from $\mathrm{MJD} = 51843$.  When doing so, the best-fit relation becomes
\beq
\mu = 0.53^{+0.22}_{-0.23} + (1.08 \pm 0.11) R + (-0.56 \pm 0.18) X,
\eeq
with an intrinsic scatter of $\ln\epsilon_\mu = -0.02^{+0.15}_{-0.14}$.  This result is very close to the result in Eq.\ (\ref{bestfit}), which is unsurprising giving that we weight multiply observed sources as the reciprocal of the number of observations.  Thus, it is difficult to tell if the degree to which Sgr A* is an outlier is a result of the relatively large intrinsic scatter or a result of substantially different accretion physics at low Eddington fractions.  We also try fitting without Sgr A* represented at all, and the best-fit relation becomes
\beq
\mu = 0.49 \pm 0.23 + (1.05 \pm 0.11) R + (-0.50 \pm 0.19) X,
\label{nosgrastar}
\eeq
with an intrinsic scatter of $\ln\epsilon_\mu = -0.04^{+0.15}_{-0.13}$.  Again, there is no significant difference with the result in Eq.\ (\ref{bestfit}).

\subsection{Radio-active high/soft state XRBs}
\label{discussion:highsoft}
The fundamental plane has generally been applied to low/hard state (and similar states) XRBs because in the high/soft states, jets are usually quenched \citep[e.g.,][]{1972ApJ...177L...5T}.  In our XRB sample selection, however, we have taken an inclusive approach by including all detections that meet our criteria set forth in section \ref{sample:xrbs}.  This has resulted in two sources (4U 1543$-$47 and XTE J1550$-$−564) with radio detections at nearly simultaneous epochs with X-ray spectra classifiable as very high or intermediate states.  In both cases, the radio observations are less than 24 hr apart from the X-ray observations, and the radio emission is compact.  As can be seen in Figs.\ \ref{f:edgeonfp} and \ref{f:3dfp} the sources do not appear to be substantial outliers compared to the other XRBs.  Nevertheless, we may exclude them and Sgr A* from the sample, and the best-fit relation becomes
\beq
\mu = 0.51^{+0.25}_{-0.24} + (1.06^{+0.11}_{-0.12}) R + (-0.51^{+0.19}_{-0.18}) X,
\eeq
with an intrinsic scatter of $\ln\epsilon_\mu = -0.02^{+0.14}_{-0.13}$.  The small change from Eq.\ \ref{nosgrastar} is likely a result of a combination of the fact that (i) only two sources have a single observation each in non-low/hard state and (ii) the radio-active intermediate- and high-state observations in our sample do not deviate strongly from low/hard state XRBs.

\subsection{Optically Thick versus Optically Thin Radio Emission}
\label{discussion:flatsteep}
We may expand on our examination of different states in XRBs by considering the influence of including AGNs with optically thin radio emission, which may arise from substantially older synchrotron emission than the currently observable X-ray emission.  In all cases, we only use the compact, { unresolved} emission from any AGN, but with the exception of this subsection, we include the sources regardless of radio optical depth { as determined by the spectral index}.  We use the radio spectral index of $\alpha = 0.4$ to delineate between optically thick $\alpha < 0.4$ and optically thin $\alpha \ge 0.4$ radio emission, recalling that we use the $S_{\nu} \propto \nu^{-\alpha}$ convention. {The choice of $\alpha = 0.4$ as the division is conservative compared to more widely used value of 0.5, but makes no difference as there are no sources in our sample with radio spectra index in the range $0.35 < \alpha < 0.71$.}  {As discussed in section \ref{radioluminosities}, some of our estimates of $\alpha$ come from two radio flux density measurements.  The uncertainties in $\alpha$ (listed in Table \ref{t:agndata}) are generally small enough that we can determine whether a given source is flat or steep.  There are, however, four sources with $\alpha$ estimates within 3$\sigma$ of our flat/steep boundary (IC 4296, $\alpha = 0.1 \pm 0.24$; NGC 3607, $\alpha = 0.35 \pm 0.10$; NGC 5077, $\alpha = 0.32 \pm 0.07$; and NGC 5128, $\alpha = 0.33 \pm 0.06$).  To account for these, we adopt the following method for estimating $\alpha$ from data for all AGNs.  For each realization, we use flux densities sampled from the measurement with uncertainties assumed to be normally distributed with mean equal to the reported flux density and $\sigma$ equal to the reported flux density uncertainty.  Then, for each pair of flux density measurements, there is an implied $\alpha = \log(S_{\nu_1}/S_{\nu_2})/\log(\nu_2 / \nu_1)$.  For each realization, we treat the object according to the realized $\alpha$ as described below.  Finally, we note that identifying objects as cores or jets based on measurements of $\alpha$ alone is still a heuristic as there are a non-negligible fraction of AGN cores identified at Very Long Baseline Array resolution that have $\alpha = 0.5$--1.6 \citep{2014AJ....147..143H}.}

First, we remove all sources with $\alpha \ge 0.4$ as determined by the method described {above}.  When doing so, our best fit is
\beq
\mu = 0.70^{+0.28}_{-0.29} + (1.10 \pm 0.12) R + (-0.53^{+0.20}_{-0.21}) X,
\label{eq:nosteep}
\eeq
with the natural logarithm of the intrinsic scatter $\ln\epsilon_\mu = 0.01^{+0.17}_{-0.16}$.  This result is very close to our full sample results.  {Eight} of the sources in our sample, however, do not have multi-frequency radio data sufficient for a robust measurement of $\alpha$.  In those cases, as described in section \ref{radioluminosities}, we have assumed an optically thick value $\alpha = 0 \pm 0.5$.  If we remove those sources and only include AGNs with positive evidence for having a flat spectrum ($\alpha < 0.4$), the best-fit result is
\beq
\mu = 0.28^{+0.32}_{-0.36} + (0.92 \pm 0.15) R + (-0.18 \pm 0.29) X,
\label{eq:onlyflat}
\eeq
with the natural logarithm of the intrinsic scatter $\ln\epsilon_\mu = -0.04^{+0.22}_{-0.21}$.  The difference between the above results and our full sample results is at slightly more than $1\sigma$ in the joint $\xi_{\mu R}$--$\xi_{\mu X}$ posterior and is most likely a result of the decreased sample size when requiring positive evidence for optically thick radio emission.  Our conclusion is that if the emission is sufficiently compact, as it is for our nearby sources at VLA resolution, then there is either unlikely to be substantial contamination from optically thin radio emission, or it makes little difference.

\subsection{Seyferts on the Fundamental Plane}
\label{discussion:seyferts}
We consider whether Seyferts belong on the fundamental plane of black hole accretion.  Some previous fundamental plane studies have restricted samples not to include Seyferts on the grounds that Seyferts are dominated by a radiatively efficient disk, which presumably will have a different correlation from low/hard state systems.  In addition to this, at accretion rates high enough for an accreting black hole to be a Seyfert, the AGN should have quenched its jet and thus any radio emission seen is likely relic emission from an earlier epoch of low/hard state-like accretion.  There are, however, Seyferts known to have compact flat or inverted-spectrum continuum radio emission at very long baseline interferometry resolution such as NGC 5033 \citep{2009ApJ...706L.260G}.  Thus, we have continued our empirical approach by including Seyferts as long as they met our criteria discussed in section \ref{sample}.  It is, however, possible that at different accretion rates, different physics manifests itself in the radio and X-ray accretion.  For this reason we try fitting the fundamental plane without Seyferts and only with Seyfert AGNs to see if any differences arise.

To cull Seyferts from our sample, we must first identify them.  This work improves on our previous Seyfert identification method \citep{2009ApJ...706..404G}, which relied solely on optical line ratios from \citet{2006A&A...455..773V}.  The \citet{2006A&A...455..773V} catalog and related material \citep[e.g.,][]{2010A&A...518A..10V} is generally reliable for bright AGNs, but at low Eddington ratios the contamination from starlight in the host galaxies is substantial.  This leads to, e.g., NGC 3607's being identified as a Seyfert despite the fact that (i) there is no obvious optical AGN and (ii) it has a 2--10 keV X-ray luminosity of $7.7 \times 10^{38}\,\mathrm{erg\,s^{-2}}$, corresponding to an X-ray Eddington fraction below $10^{-7}$.

In this work we use the \citet{1997ApJ...487..568H} classification scheme if available for our sources and inspection of the X-ray spectrum if not.  \citet{1997ApJ...487..568H} use information about the luminosity to inform whether to identify a source as a Seyfert.  The approach of using the X-ray spectrum has the advantage of probing the bands more closely associated with the accretion inflow and jet production physics involved and allows one to inspect the data at hand to see if it is appropriate for inclusion.  We classify objects as Seyferts if their X-ray spectrum requires a soft-excess component, a warm absorber, a Seyfert-like Fe line, or a pexmon reflection spectrum \citep{2007MNRAS.382..194N}.  For nearly all cases where the source was in the \citet{1997ApJ...487..568H} catalog, our X-ray spectrum classification agreed.  We identify Seyferts in Table \ref{t:agndata}.

When we remove Seyferts from our sample, the best-fit relation is
\beq
\mu = 0.53^{+0.24}_{-0.23} + (1.16 \pm 0.11) R + (-0.74^{+0.20}_{-0.21}) X,
\label{noseyfs}
\eeq
with an intrinsic scatter of $\ln\epsilon_\mu = -0.06^{+0.16}_{-0.15}$.  This is very close to the results obtained from our full sample in Eq.\ \ref{bestfit}.  Because this may be a result of the fact that there are relatively few Seyferts in our sample, we also fit to a sample in which the only AGN we include are Seyferts.  In this case, the best-fit relation is
\beq
\mu = -1.15^{+0.72}_{-0.60} + (0.58^{+0.24}_{-0.20}) R - (0.26^{+0.30}_{-0.35}) X, 
\label{onlyseyfs}
\eeq
with an intrinsic scatter of $\ln\epsilon_\mu = -0.18^{+0.26}_{-0.22}$.  Although the difference between this relation and the previous is large in absolute terms, it is still consistent at about the $2\sigma$ level in the joint posterior distribution of $\xi_{\mu R}$ and $\xi_{\mu X}$.  The differences most likely show the result of having only seven Seyferts in our sample.  Thus, we do not have sufficient data to state that radio-active Seyferts decidedly do or do not belong on the same fundamental plane relation.

\subsection{Just AGN}
\label{discussion:justagn}
In \citet{2009ApJ...706..404G} we reported a difference between fundamental plane fits to an AGN-only sample and fits to samples with both AGNs and XRBs.  Given the relatively few sources, it was not clear if AGNs and XRBs actually did not belong on the same relation or if small number statistics and the reduced dynamic range in mass can lead to fits of a 2D manifold in 3D space with an intrinsic scatter to a spurious result.  \citet{2014ApJ...788L..22G} tested for this by using some of the lowest-mass AGN available to put on the two different fundamental plane relations.  These low-mass AGN were all Seyferts with masses determined from single-epoch H$\alpha$ line widths.  We discussed above in section \ref{discussion:seyferts} that we cannot definitively conclude that Seyferts belong on the same fundamental plane as the rest of the sample, but assuming that they do, the low-mass AGNs better followed the all black hole fundamental plane than the AGN-only fundamental plane.  When limiting our current sample to only AGN, the best-fit relation is
\beq
\mu = 0.37 \pm 0.18 + (0.56^{+0.13}_{-0.14}) R + (-0.29^{+0.14}_{-0.13}) X,
\eeq
with an intrinsic scatter of $\ln\epsilon_\mu = -0.30^{+0.16}_{-0.15}$.  This is noticeably different from the fit in Eq.\ (\ref{bestfit}) and very close to that found in \citet{2009ApJ...706..404G}.  This reintroduces the possibility that there are real differences in the coupled radio and X-ray emission in AGNs and XRBs.  Note that while the AGN-only fit parameters fall outside the 1$\sigma$ intervals of the posterior probability distributions seen in Fig.\ \ref{f:cornerplot}, the covariance between $\xi_{\mu R}$ and $\xi_{\mu X}$ does allow a decrease in $\xi_{\mu R}$ coupled with an increase in $\xi_{\mu X}$.  Thus, the difference between the AGN-only fit and the full sample fit is at roughly the $2.4\sigma$ level.  
Given such a small difference, we cannot claim that the difference is significant as it is just as likely a result of the reduced dynamic range from limiting the sample to just AGNs.  At $2.4\sigma$, however, it is worth further investigation with a sample that expands the range of AGN masses.

\subsection{Low Eddington rates}
\label{discussion:lowfedd}
A series of works \citep{2008ApJ...688..826L, 2016ApJ...818..185F, 2016MNRAS.456.4377X, 2017ApJ...836..104X, 2017arXiv170704029Q} suggests that at the lowest accretion rates, the fundamental plane will take a different form.  To investigate this, we restrict our sample to only sources with $L_X / L_\mathrm{Edd} < 10^{-6}$.  This has the effect of limiting the sample to only AGNs (see Fig.\ \ref{f:feddhist}) so that the issues raised in section \ref{discussion:justagn} apply here as well.  The best-fit relation is
\beq
\mu = 0.69 \pm 0.18 + (0.35^{+0.17}_{-0.16}) R + (0.06^{+0.21}_{-0.22}) X,
\label{lowfedd}
\eeq
with an intrinsic scatter of $\ln\epsilon_\mu = -0.67^{+0.23}_{-0.21}$.  Here we note that Sgr A* may have a substantial influence on the overall fit.  If we also exclude Sgr A* from the sample, the best fit is
\beq
\mu = 0.74^{+0.20}_{-0.21} + (0.29 \pm 0.20) R + (0.21^{+0.33}_{-0.34}) X,
\label{lowfeddnosgra}
\eeq
with an intrinsic scatter of $\ln\epsilon_\mu = -0.67 \pm 0.25$.  The difference in results between Eqs.\ (\ref{lowfedd}) and (\ref{lowfeddnosgra}) is not significant.  

\subsection{Regression in other directions}
Because the focus of this work is to provide a mass estimator, we have until now only done a regression analysis with $R$ and $X$ as the independent variables.  This results in a mean value of $\mu$ for given values of $R$ and $X$.  Inverting the best-fit plane found in this method to predict $R$ or $X$ from two other measurements, however, is not appropriate.  In this section, we report results from regression with $R$ or $X$ as the dependent variable.  In the case of using radio as the dependent variable, the best-fit relation is
\beq
R = -0.62^{+0.15}_{-0.17} + (0.70^{+0.08}_{-0.09})X + (0.74 \pm 0.06) \mu,
\label{radioregression}
\eeq
with an intrinsic scatter in the log-radio direction of $\ln\epsilon_R = -0.23^{+0.14}_{-0.13}$.  In the case of using X-ray as the dependent variable, the best-fit relation is 
\beq
X = 0.58 \pm 0.23 + (-0.59 \pm 0.15) \mu + (0.99^{0.12}_{-0.13}) R,
\label{xrayregression}
\eeq
with an intrinsic scatter in the log-X-ray direction of $\ln\epsilon_X = -0.03^{+0.15}_{-0.14}$.  The intrinsic scatters measured by regressing in the log-radio and log-X-ray directions are not significantly smaller, though there is a suggestion that it is smaller in the $R$ direction.

{
\subsection{Summary of differences between current work and G{\"u}ltekin et al.\ (2009a)}
\label{differences}

As this work is an extension of the work started by \citet{2009ApJ...706..404G}, it is worth directly comparing and contrasting the work done here with the earlier work in terms of data, analysis, and conclusions.  In this subsection we summarize the advances made in this work.  First, and most importantly, we have increased the sample size of AGNs with primary direct mass measurements and requisite radio and X-ray data from 18 to 30, roughly doubling (Section \ref{sample:smbhs}).  Second, we have improved our selection of XRBs so that our analysis is not hindered by poor distance or mass estimates (Section \ref{sample:xrbs}).
Third, we improved our identification and handling of Seyferts (Section \ref{discussion:seyferts}).  Fourth, we have improved our analysis so that we now (i) include a treatment of correlated uncertainties (Section \ref{fitting:statistics}), (ii) handle multiple observations of individual objects to increase the information available (Section \ref{multobs}), (iii) have an improved  handling of contamination of AGN X-ray flux measurements from background AGN (Section \ref{bgagncontam}) and XRBs near the galaxy center (Section \ref{xrbcontam}), (iv) statistically treat the effects of non-simultaneous observations of X-ray and radio fluxes, and (iv) use MCMC methods (Section \ref{mcmcfitting}) rather than a merit function, which has some disadvantages \citep{2012MNRAS.419..267P}.  

The above improvements in data selection and analysis have allowed us to better consider two speculations raised in \citet{2009ApJ...706..404G}.  First, \citet{2009ApJ...706..404G} compared fits to two subsamples of the earlier, smaller parent sample: one subsample whose AGN consisted only of LLAGN and LINERs and another subsample whose AGN consisted only of Seyferts.  In doing so, they \citet{2009ApJ...706..404G} found a much smaller intrinsic scatter in the LLAGN/LINER subsample as well as a statistically different fit.  As mentioned above, we now have a larger sample and an improved identification of Seyferts.  With the improved sample and identification, we no longer find a significant difference in the intrinsic scatter, nor in the fundamental plane fit parameters (section \ref{discussion:seyferts}).  Second, \citet{2009ApJ...706..404G} compared an AGN-only sample to a sample with both XRBs and AGNs.  In the comparing these two samples, \citet{2009ApJ...706..404G} found a significant difference between the fits, suggesting the possibility that the fundamental plane was not fundamental to all black holes but that there were separate relations for XRBs and AGNs.  Based on our improved sample and in combination with results of \citet{2014ApJ...788L..22G}, we no longer have strong evidence supporting this speculation, though it is worth testing with as large a dynamic range in AGN mass as possible (section \ref{discussion:justagn}).

Finally, a major goal of this project is to give the best possible mass estimation tool.  The mass-predictor relation we present in this work (given the discussion below in section \ref{massestimation}) is {more robust than} the relation presented by \citet{2009ApJ...706..404G}.  The {robustness} comes from all of the above-mentioned improvements in data and analysis.

{
As an example of the significant improvements made in this paper, we consider how the mass estimation tool we present in this paper better predicts the mass of the black holes in M87 and GRS 1915+104.  The black hole in M87 has logarithmic mass in solar units of $\mu =9.79 \pm 0.03$ \citep[but see also \citealt{2013ApJ...770...86W}]{2011ApJ...729..119G}.  Based on the data provided in Table \ref{t:agndata}, our current mass estimator based on the fundamental plane of black hole accretion is $\mu = 10.14 \pm 0.96$, well within the measured scatter of the relation.  On the other hand, the mass predicted by equation 6 in \citet{2009ApJ...706..404G} is $\mu 8.90 \pm 0.77$.  The discrepancy is even more apparent at the low-mass end, such as for GRS 1915+104, which would be predicted to have a logarithmic mass of $\mu = 2.2 \pm 0.96$ with our current relation but a mass of $\mu = 5.4 \pm 0.96$ with the older version.  Thus, with the current relation, one would correctly identify it as consistent with a stellar mass black hole, while with the old relation one would conclude that it is either an IMBH or a low-mass AGN.  The underlying reason for this is that the mass-predictor regressions in \citet{2009ApJ...706..404G} only used AGN sources as they were only intended to be used for AGNs.  With the currently better measured intrinsic scatter of $\ln\epsilon_0 = -0.04$, it is clear that one of the best uses for this mass-predictor relation is for discerning between XRBs, IMBHs, and AGNs, and this requires a mass-predictor relation that uses XRB data as we have done here.  We also reported regressions for the prediction of radio luminosity in Eq.\ (\ref{radioregression}) and of X-ray luminosity in Eq.\ (\ref{xrayregression}), the latter of which has not been reported in the literature before.
}

{
\subsection{Future Work}
\label{futurework}

We finally note that we expect that further improvements of the fundamental plane can be made by including more XRBs and expanding the mass range of AGNs in the sample.  The mass range of AGNs in the sample can be improved by targeting known high-mass AGNs with current instrumentation as well as future instrumentation.  In particular, high-mass AGNs are difficult to measure black hole masses because of (i) their typical distances, which can be addressed with Atacama Large Millimeter Array and 30 m class infrared telescopes with adaptive optics and/or (ii) their low surface brightnesses, which can be addressed with 30 m class telescopes and the \emph{James Webb Space Telescope}.  With more high-mass AGNs, the lever arm of the fits will be better established.  Low-mass AGNs also require high-angular resolution instrumentation in optical or infrared to measure black hole masses as well as sensitive and high-angular X-ray and radio instruments, e.g., Lynx \citep{2017SPIE10397E..0SG} and Next Generation VLA \citep{2015arXiv151006438C}, to probe the typically fainter sources and rule out contamination from XRBs.

}

}

\section{How to Estimate Black Hole Masses with the Fundamental Plane}
\label{massestimation}

For those who wish to use the fundamental plane of black hole accretion to estimate the mass of a black hole, we provide the following guidelines.  First, the prospective mass estimator needs some assurances that the object in question is a black hole or should explicitly acknowledge that they are making such an assumption.  The fundamental plane we have studied here only uses known black holes\,---\,at least to the extent that any given XRB or AGN is known to be a black hole.  The fundamental plane does not, by itself, constitute a means for discriminating between black holes and other objects, though other means for such an exercise exist \citep{2012Natur.490...71S}.

Second, mass estimation from the $M$--$L_R$--$L_X$ relation requires good $L_R$ and $L_X$ data.  Because we do not use upper limits in our analysis, the data must be detections in both cases.  The best data will have high-angular resolution to avoid contamination from other sources of radio or X-ray emission.  This is especially important at low X-ray luminosities of AGNs, which could be confused with XRBs.  Radio data should be converted to 5 GHz in a manner similar to the one described in section \ref{radioluminosities} and to use 2--10 keV power-law continuum flux.  Obviously, one needs the distance to turn what assumes and hopes is isotropic flux into a luminosity.  The radio and X-ray data ought to be from a similar epoch, the closer in time the better.  As a very rough rule of thumb, we recommend they be observed within $\Delta t < (2 + M / 10^6\,\msun) \mathrm{day}$.  Obviously, without knowing the mass of the black hole one is trying to estimate, it is impossible to know how close to simultaneous one must schedule the observations, but one may see what masses they are sensitive to.  

Finally, recognize that there is substantial intrinsic scatter in the relation of an assumed normal distribution with $\epsilon_{\mu} = 1\,\mathrm{dex}$.  This means, for example, that for a large collection of black holes with masses estimated from the fundamental plane to be $10^8\,\msun$, 5\% of them will be below $10^6\,\msun$ or above $10^{10}\,\msun$, assuming that all logarithmic black hole masses are equally represented.  This makes it a relatively crude tool for black hole-mass estimation, but if it is the only tool available, it will be the best tool available.  

The fundamental plane is most useful in mass estimation when one wants to discriminate between an XRB and an IMBH or AGN.  The fundamental plane is also useful in estimating the mass of a Type 2 AGN (without broad lines) in a galaxy without a well-defined bulge from which to use host-galaxy scaling relations.  

Given the above considerations, then we recommend the use of the following mass estimator:
\beq
\mu = 0.55 \pm 0.22 + (1.09 \pm 0.10) R + (-0.59^{+0.16}_{-0.15}) X.
\label{mass estimator}
\eeq

One source of data for which the fundamental plane could prove especially useful is that from extended Roentgen Survey with an Imaging Telescope Array (eROSITA) on the
Spectrum-R{\"o}ntgen-Gamma satellite \citep{2010SPIE.7732E..0UP, 2012arXiv1209.3114M}.  Expected to detect $\sim3\times10^{6}$ AGNs in the 2--10\,keV band, eROSITA will provide half of the needed data to use the fundamental plane for black hole-mass estimation.  A radio survey of the detected AGNs would complete the necessary data to make a black hole-mass catalog.

\section{Summary}

In this paper we have analyzed the dependence of an accreting black hole's mass on its radio and X-ray emission.  Using only black holes with high-quality, direct, primary mass measurements and sensitive, high-spatial-resolution radio and X-ray data, we used MCMC methods to find the best mass-predictor relation to be
\beq
\mu = 0.55 \pm 0.22 + (1.09 \pm 0.10) R + (-0.59^{+0.16}_{-0.15}) X
\label{summarybestfit}
\eeq
with the natural logarithm of the intrinsic scatter $\ln\epsilon_\mu = -0.04^{+0.14}_{-0.13}$.  After considering a number of potential modifications to our original, inclusive sample, we conclude that the fundamental plane can be used to describe any accreting black hole with both X-ray and compact radio emission.  In particular, we cannot rule out that radio-active high/soft state XRBs and radio-active Seyferts are inconsistent with the fundamental plane made up of low/hard state XRBs and LLAGNs and LINERs.  The low numbers of radio-active high/soft state XRBs and radio-active Seyferts, however, make such conclusions tentative and warrant further study.  Given the wide variety of sources that are included in our sample and the substantial intrinsic scatter we found, the fundamental plane is a useful\,---\,though relatively low precision\,---\,tool for estimating black hole masses.

\hypertarget{ackbkmk}{}%
\acknowledgements 
\bookmark[level=0,dest=ackbkmk]{Acknowledgments}

K.G.\ acknowledges support provided by the NASA through \emph{Chandra} Award Number
GO0-11151X issued by the \emph{Chandra} X-ray Observatory Center, which is
operated by the Smithsonian Astrophysical Observatory for and on
behalf of NASA under contract
NAS8-03060. 
K.G.\ acknowledges support provided by the Fund for Astrophysics Research.
A.L.K.\ would like to thank the support provided by NASA through Einstein Postdoctoral Fellowship Grant No.\ PF4-150125 awarded by the \emph{Chandra} X-ray Center, operated by the Smithsonian Astrophysical Observatory for NASA under contract NAS8-03060

The National Radio Astronomy Observatory is a facility of the National Science Foundation operated under cooperative agreement by Associated Universities, Inc.

This research has made use of the NASA/IPAC Extragalactic Database, which is operated by the Jet Propulsion Laboratory, California
Institute of Technology, under contract with the National Aeronautics
and Space Administration.  
This research has made use of NASA's Astrophysics Data System.

\bibliographystyle{aasjournal}
\hypertarget{refbkmk}{}%
\bookmark[level=0,dest=refbkmk]{References}
\bibliography{gultekin}

\appendix

\section{New Radio Data}
\label{newradio}
{
We obtained new VLA observations of the nuclei of 12 galaxies harboring massive black holes with direct dynamical mass measurement (Project ID SB0514).  Our observations were centered at 8.46 GHz with a total bandwidth of 256 MHz while the array was in its most extended A configuration, leading to a typical angular resolution of $0\farcs3$.  With a total time of 60 minutes for each source, the 
time-on-source integration varied but ranged from 25--33 minutes for theoretical sensitivities in the range $15$--$22\,\mu\mathrm{Jy\,beam^{-1}}$.  Each scheduling block began with scans of the corresponding flux calibrator source given in Table \ref{t:newradio} to set the flux density scale to an accuracy of 5\% and calibrate the bandpass \citep{2013ApJS..204...19P}.  We performed phase-referencing using a nearby complex gain calibrator within 10 degrees.  Standard VLA calibration and imaging procedures were followed using CASA version 5.3.0.

After inspecting for radio frequency interference, data were averaged in 30 s temporal bins and 8- or 10-channel frequency bins.   We made images and processed with the CLEAN algorithm for imaging.  For CLEAN we halted processing at a value of 2.5 times the dirty map rms, which was typically very close to the theoretical noise, using a gain of 0.1 and robust weighting with a robust parameter of 0.5.  Our images (Fig.\ \ref{f:vlamaps}) used a cell size of 0\farcs05 with a total image size of $8192 \times 8192$, compared to our field of view of $5\farcm3$.  For NGC 4486A we used an image size of $32768 \times 32768$ in order to avoid side lobes coming from NGC 4486, but in the end we were unable to detect emission from NGC 4486A.   For the choices above, we tried several different variations but found that it made very little difference.

For each of the processed images, we looked for emission at the location of the galaxy nucleus.  Of the 12 galaxies, 6 had unambiguous point sources at the expected location.  { For the detections, we found that the radio flux was spatially coincident with the X-ray point source found.}  We attribute all of this emission to the central black hole.  We calculated the flux density from these sources by fitting a two-dimensional { elliptical} Gaussian to the point source in a $20 \times 20$ pixel region and use the total flux returned by the CASA imfit tool.  We list in Table \ref{t:newradio} flux densities and their uncertainties, calculated as the quadrature sum of fit uncertainty, image rms noise, and a 3\% uncertainty for absolute flux calibration \citep{2013ApJS..204...19P}.   Undetected sources are reported as upper limits at 3 times the flux uncertainty.
}
\kgfigstarbeg{vlamaps}
\iftoggle{lowresversion}{%
\includegraphics[width=0.33\textwidth]{./n1300_vlamap.png}
\includegraphics[width=0.33\textwidth]{./n2748_vlamap.png}
\includegraphics[width=0.33\textwidth]{./n2778_vlamap.png}\\
\includegraphics[width=0.33\textwidth]{./n3384_vlamap.png}
\includegraphics[width=0.33\textwidth]{./n4291_vlamap.png}
\includegraphics[width=0.33\textwidth]{./n4459_vlamap.png}\\
\includegraphics[width=0.33\textwidth]{./n4486A_vlamap.png}
\includegraphics[width=0.33\textwidth]{./n4596_vlamap.png}
\includegraphics[width=0.33\textwidth]{./n4742_vlamap.png}\\
\includegraphics[width=0.33\textwidth]{./n5077_vlamap.png}
\includegraphics[width=0.33\textwidth]{./n5576_vlamap.png}
\includegraphics[width=0.33\textwidth]{./n7457_vlamap.png}%
}{%
\includegraphics[width=0.33\textwidth]{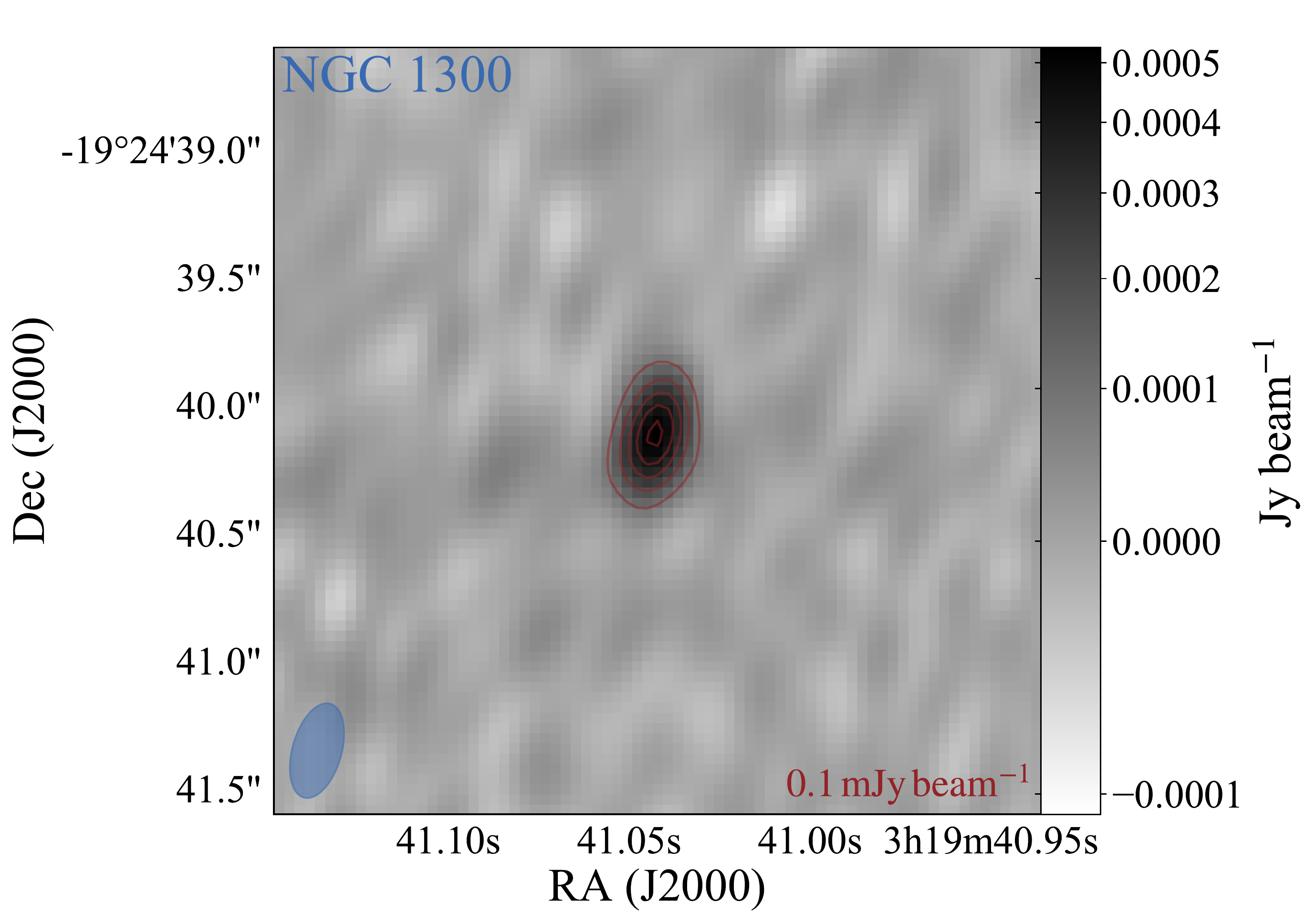}
\includegraphics[width=0.33\textwidth]{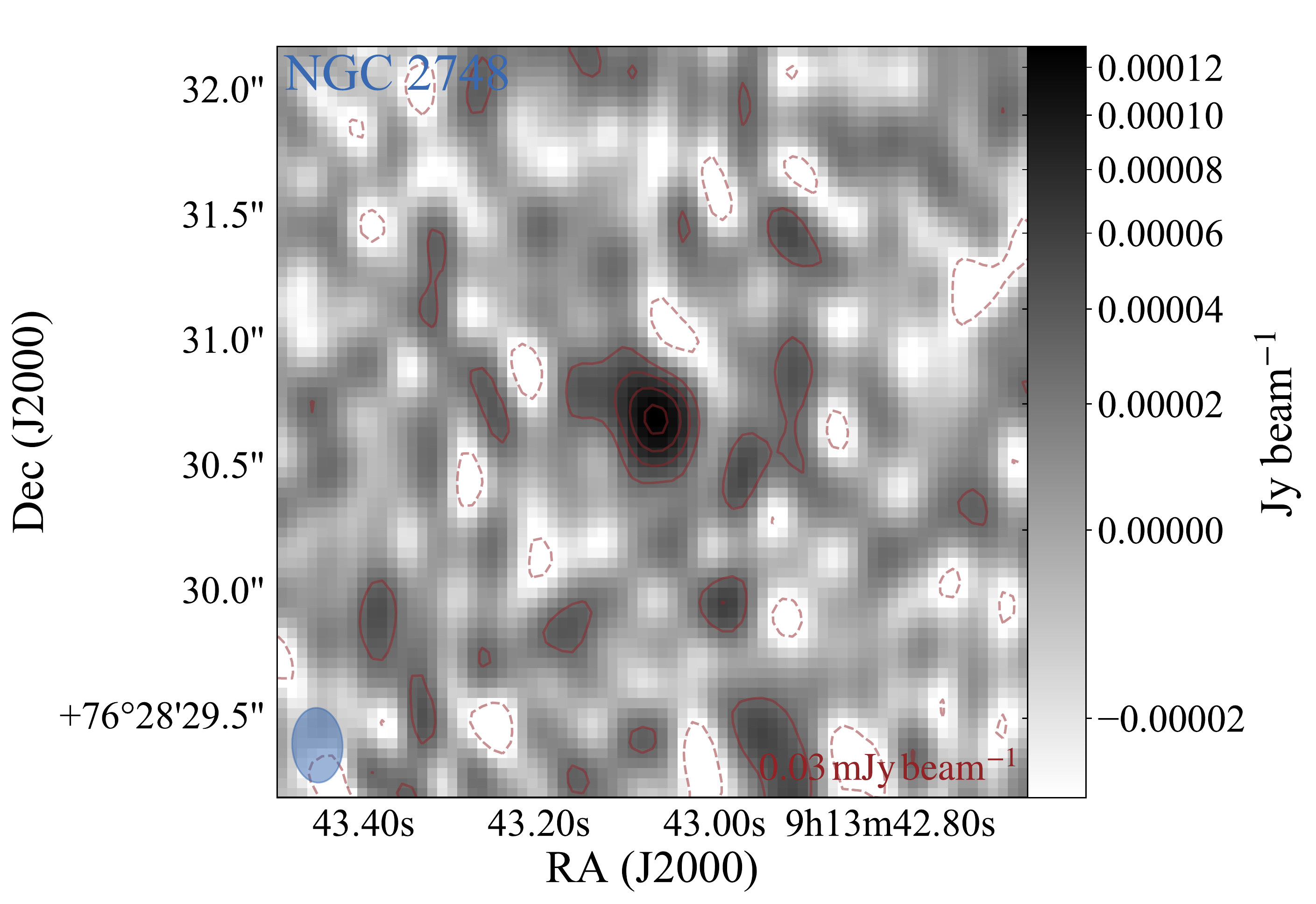}
\includegraphics[width=0.33\textwidth]{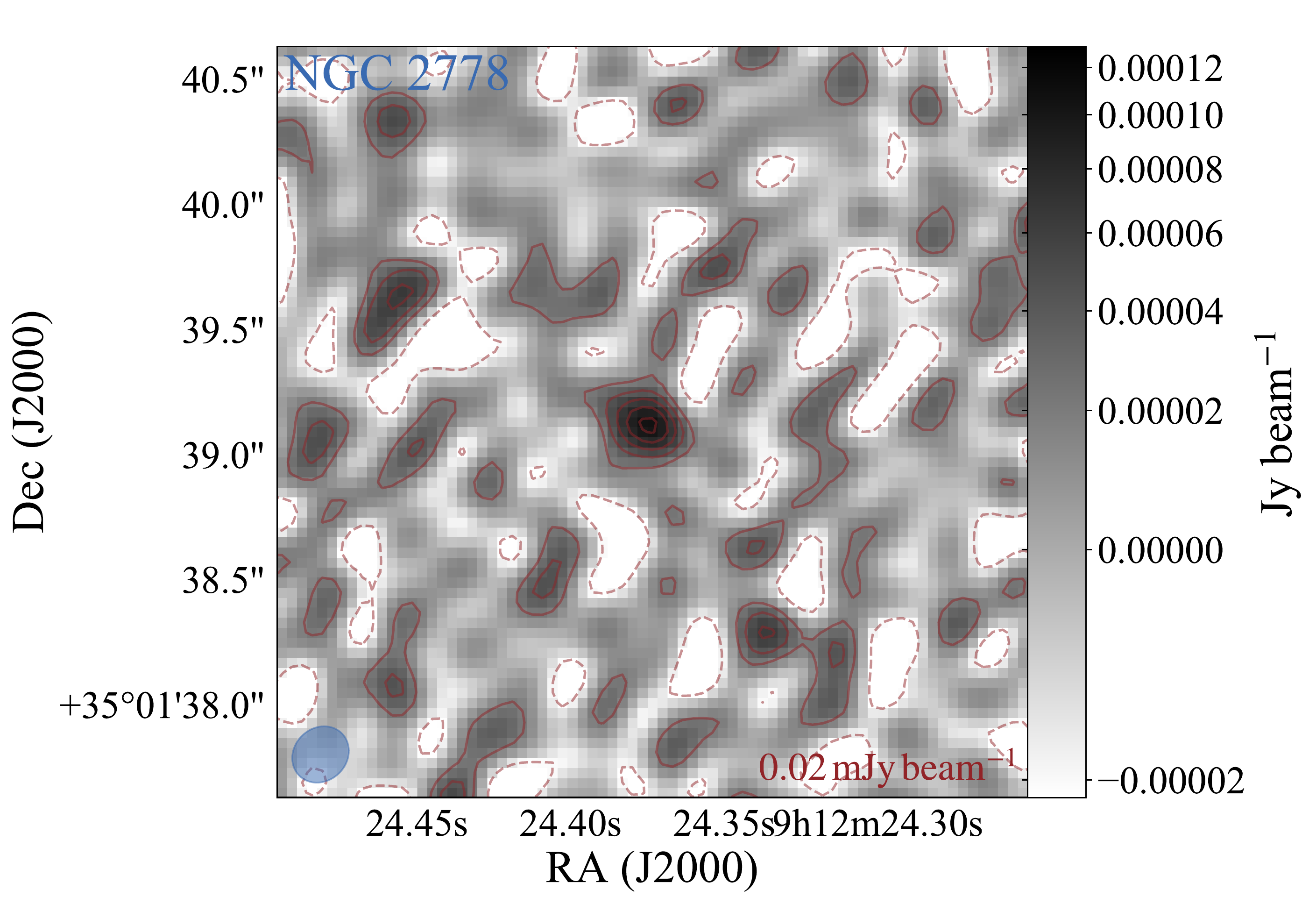}\\
\includegraphics[width=0.33\textwidth]{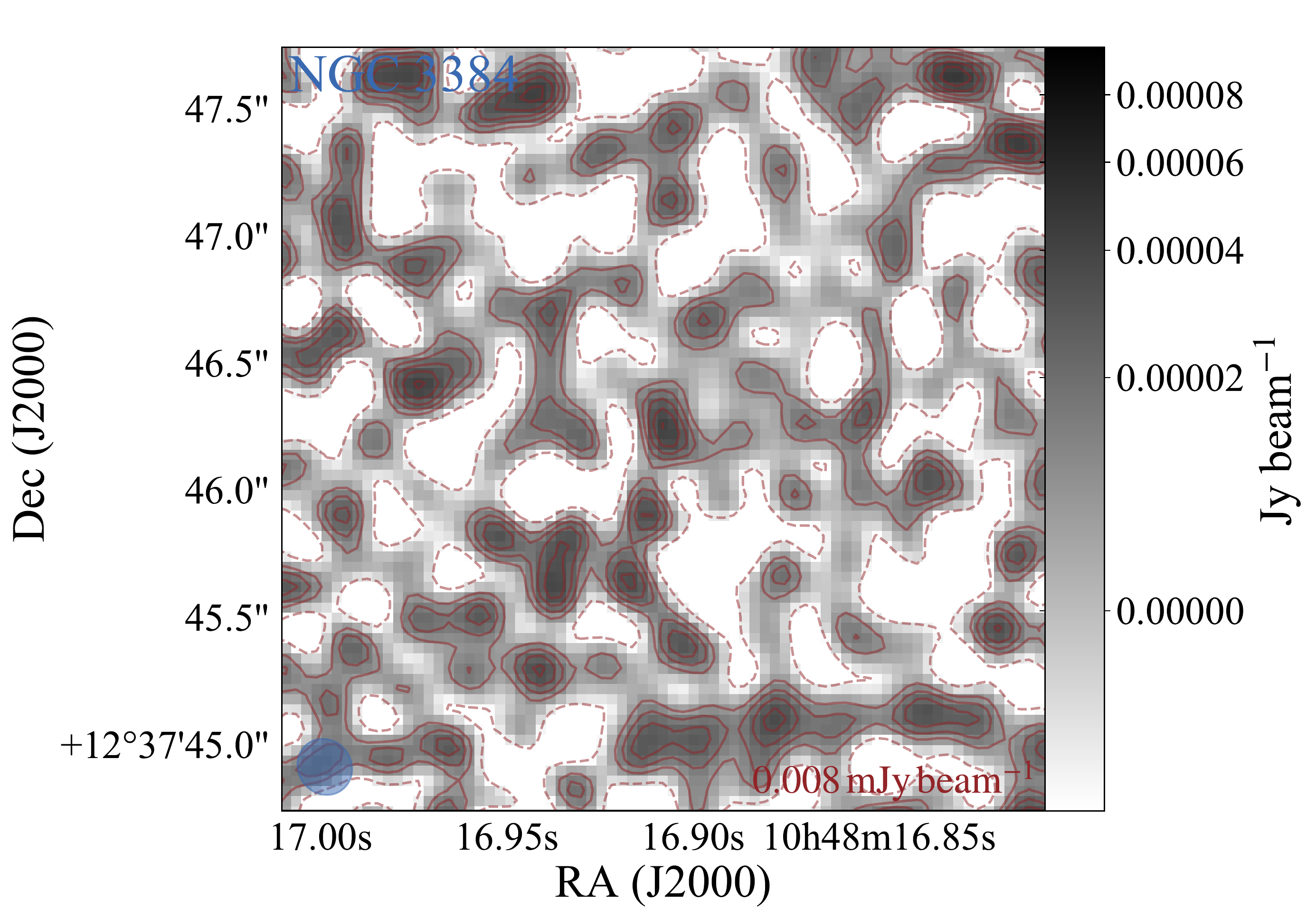}
\includegraphics[width=0.33\textwidth]{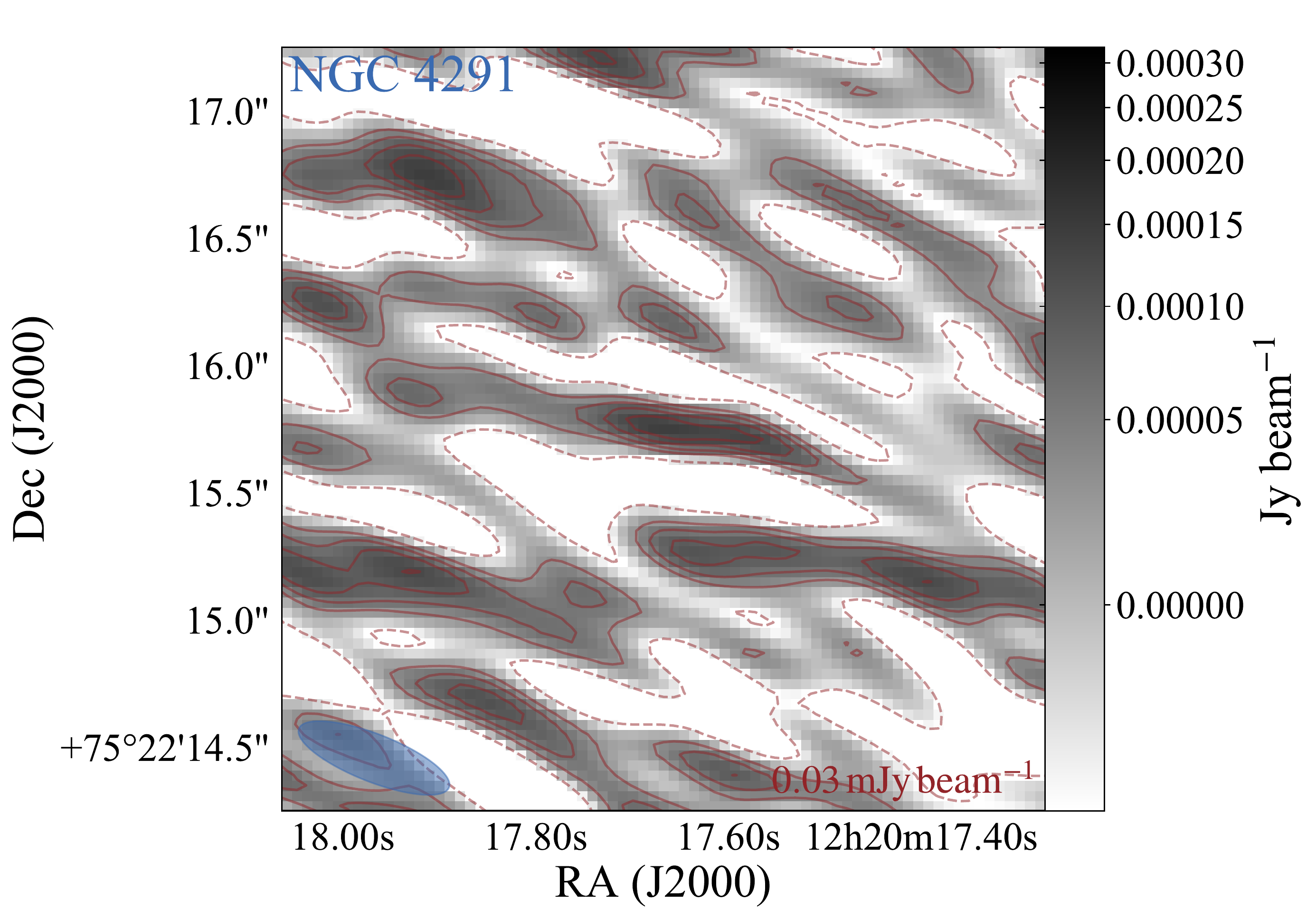}
\includegraphics[width=0.33\textwidth]{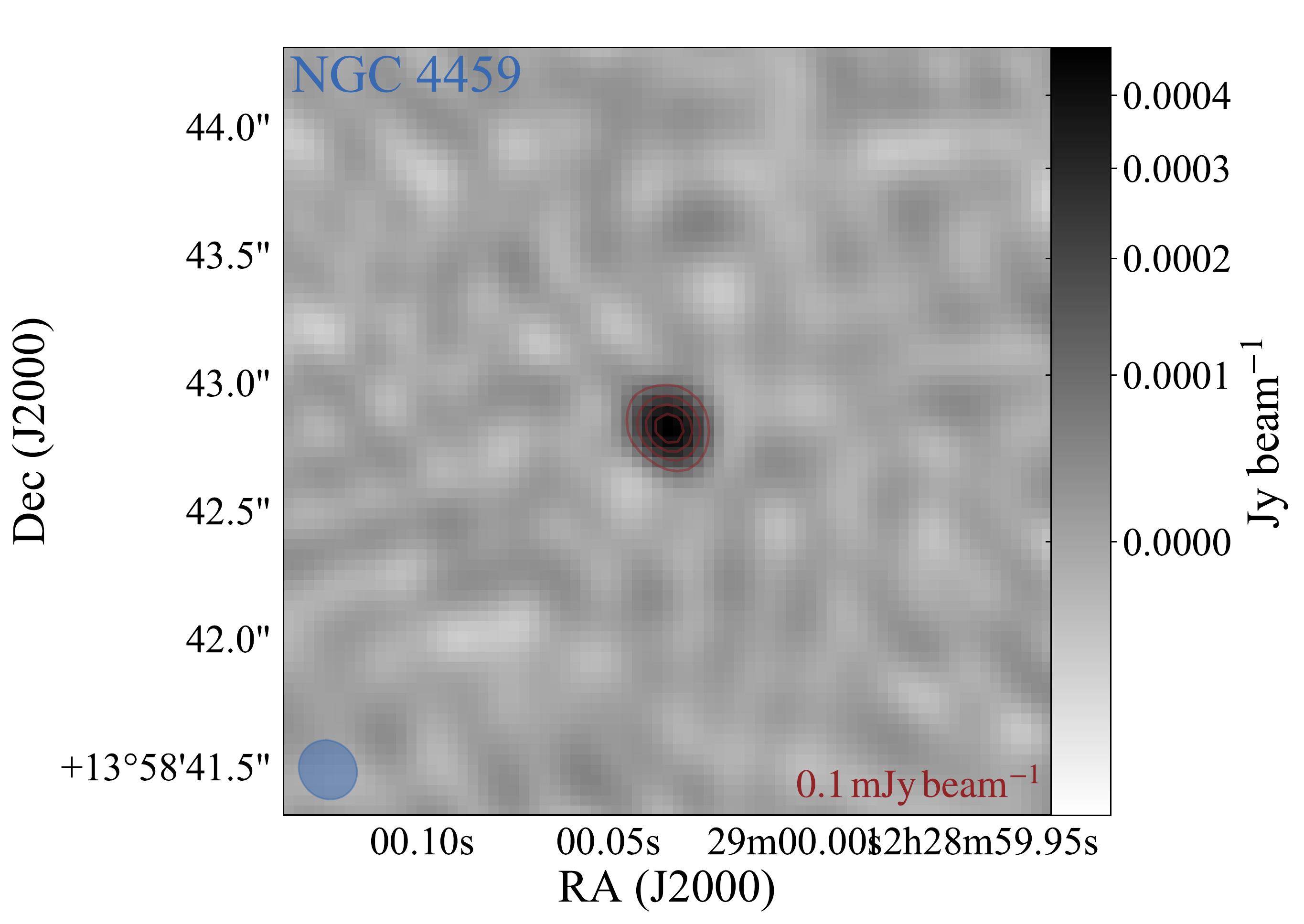}\\
\includegraphics[width=0.33\textwidth]{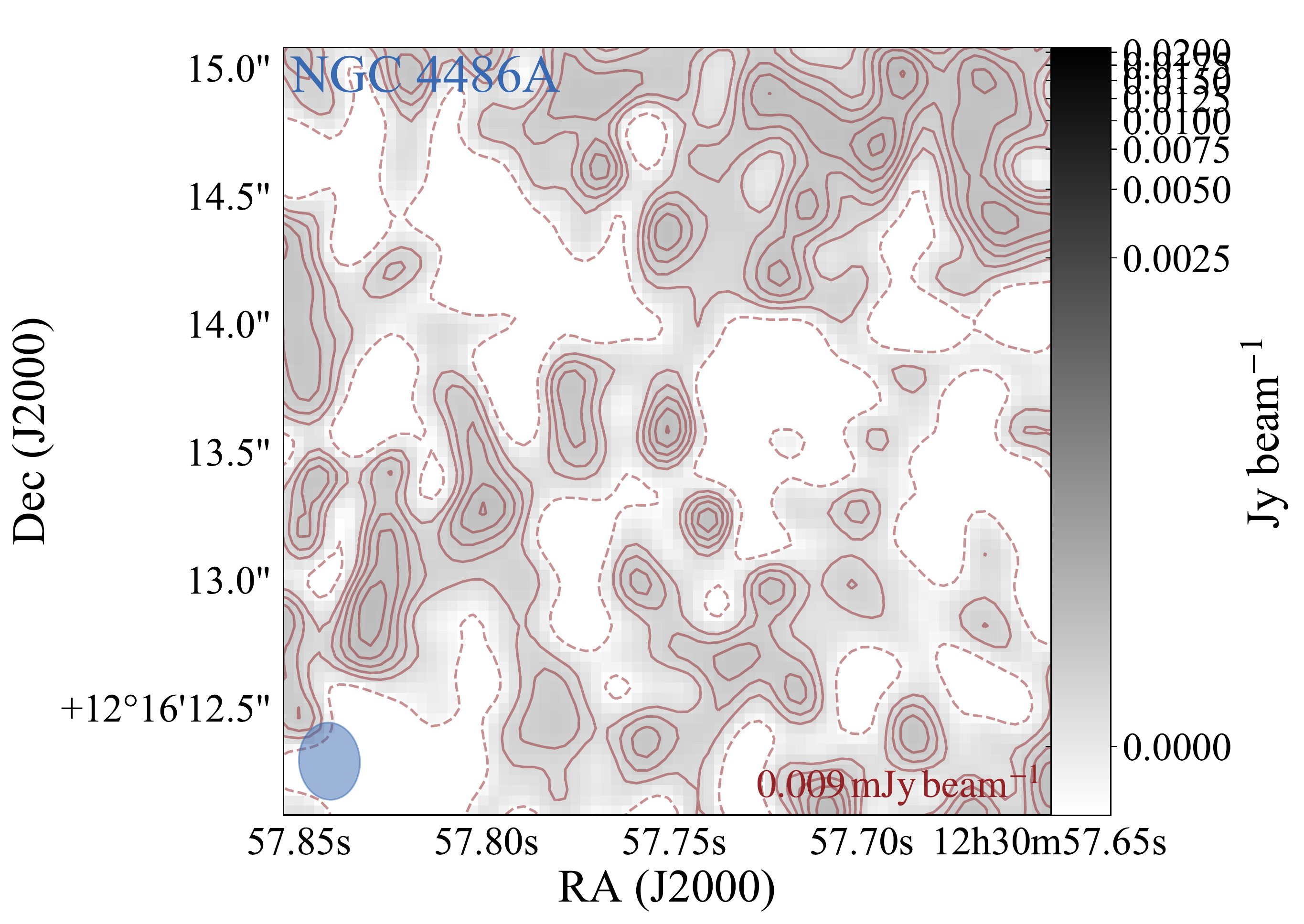}
\includegraphics[width=0.33\textwidth]{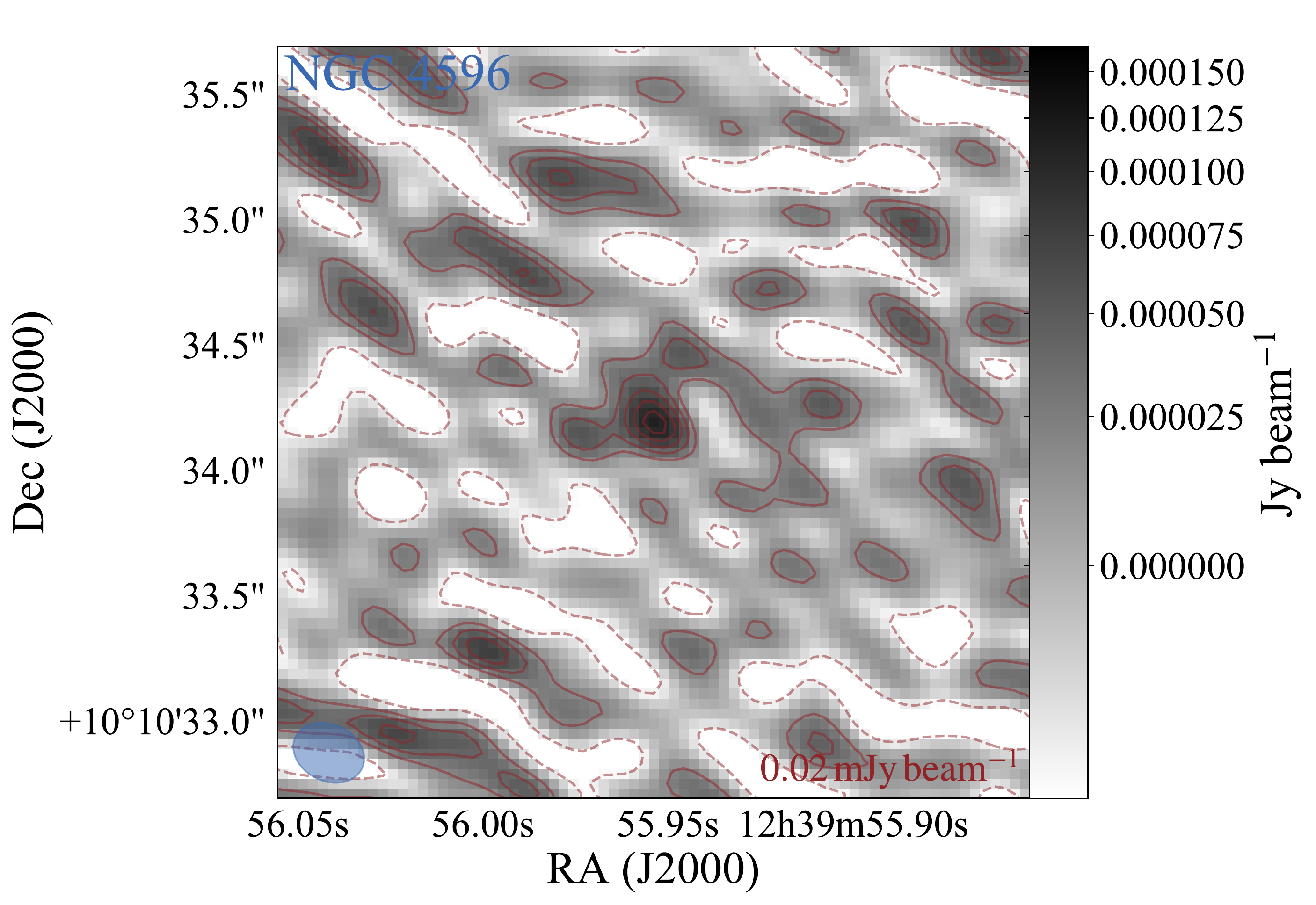}
\includegraphics[width=0.33\textwidth]{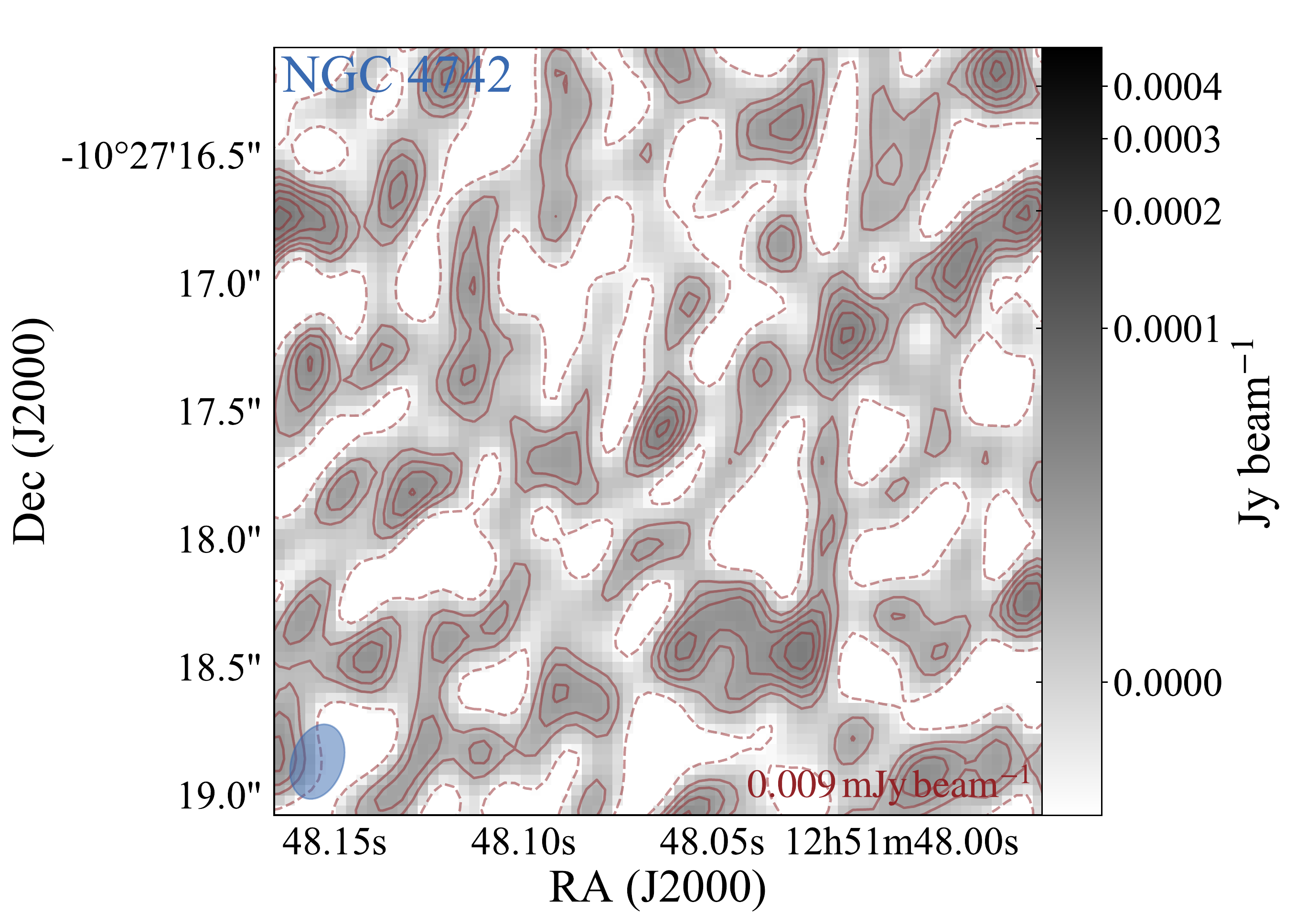}\\
\includegraphics[width=0.33\textwidth]{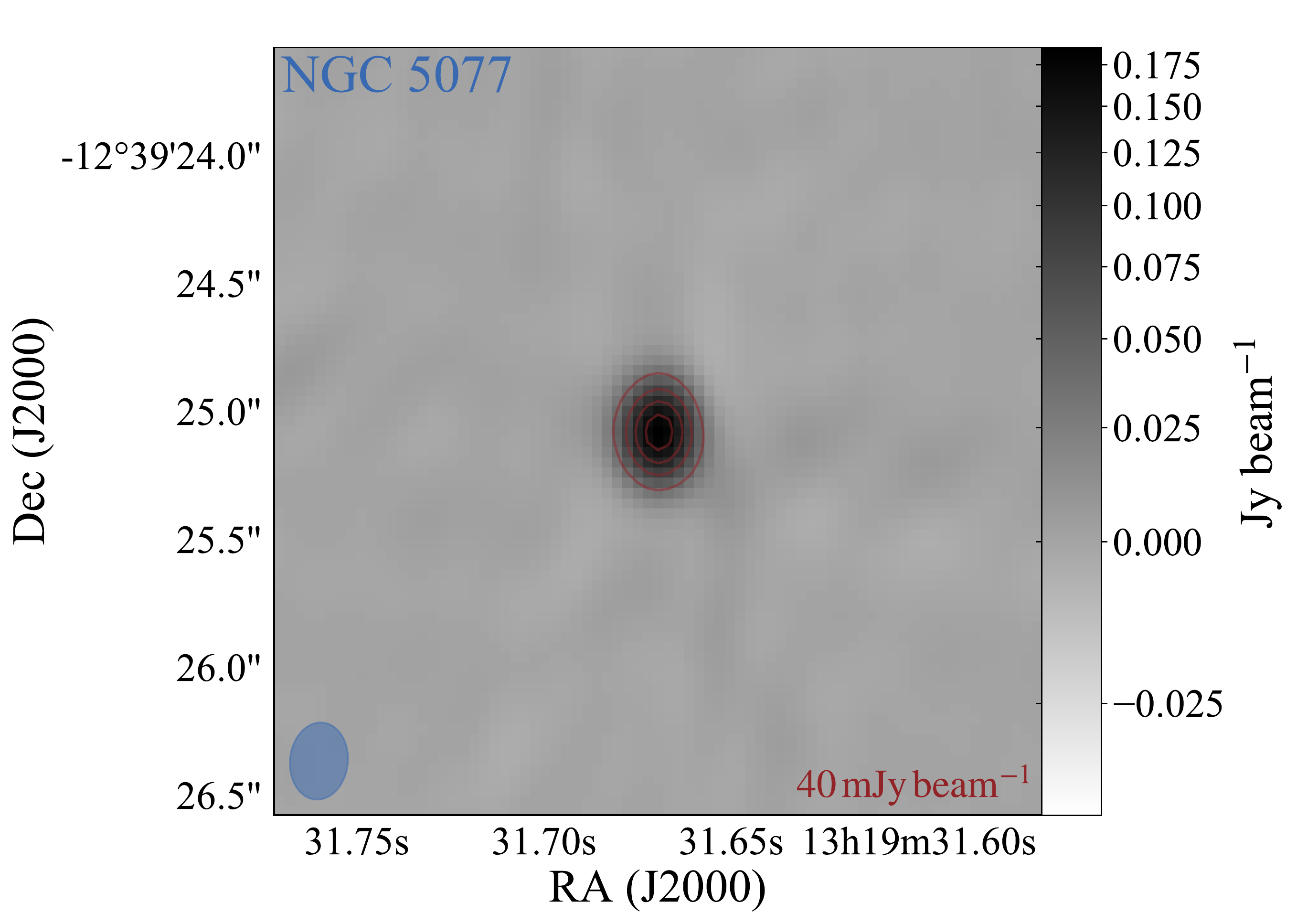}
\includegraphics[width=0.33\textwidth]{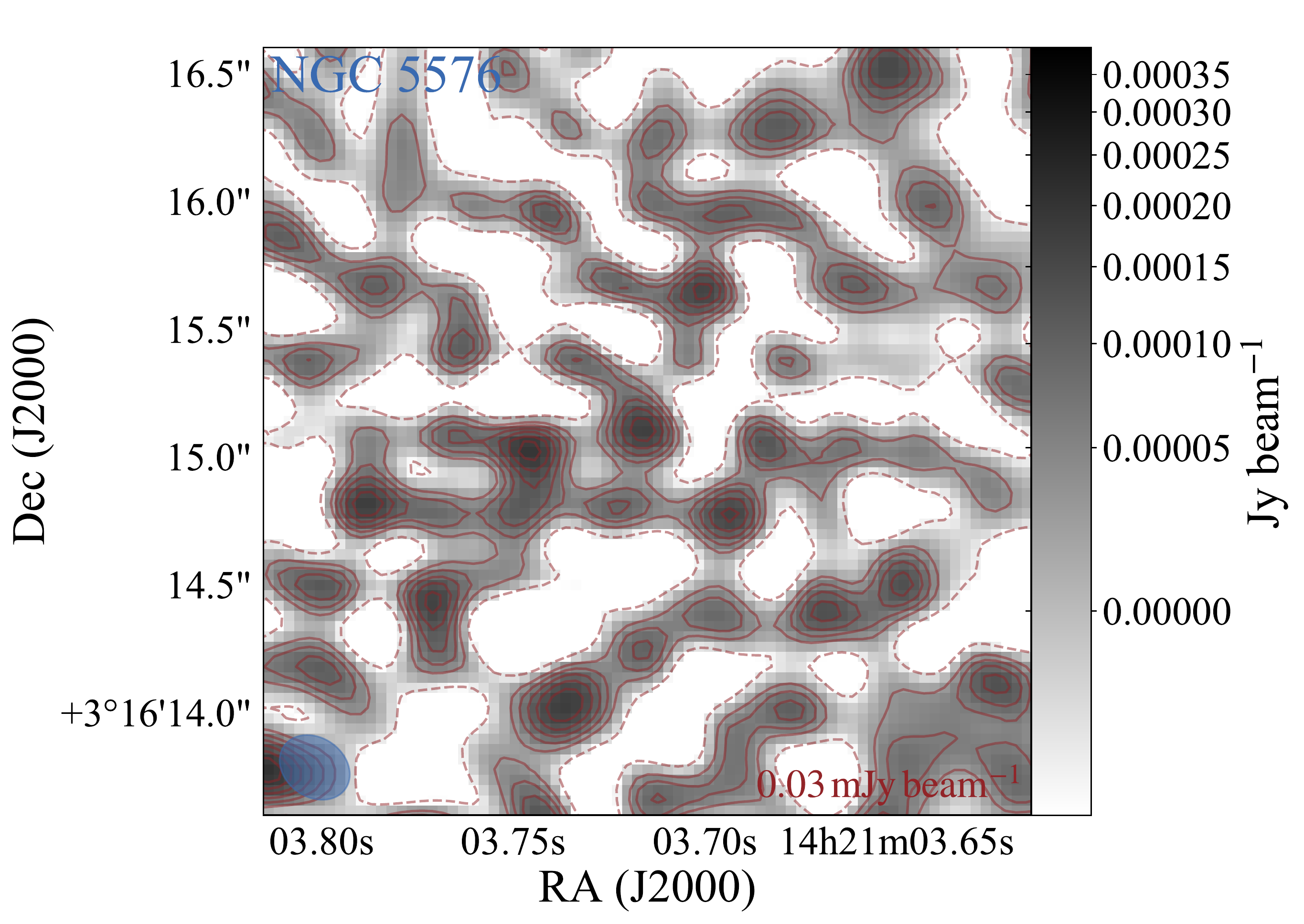}
\includegraphics[width=0.33\textwidth]{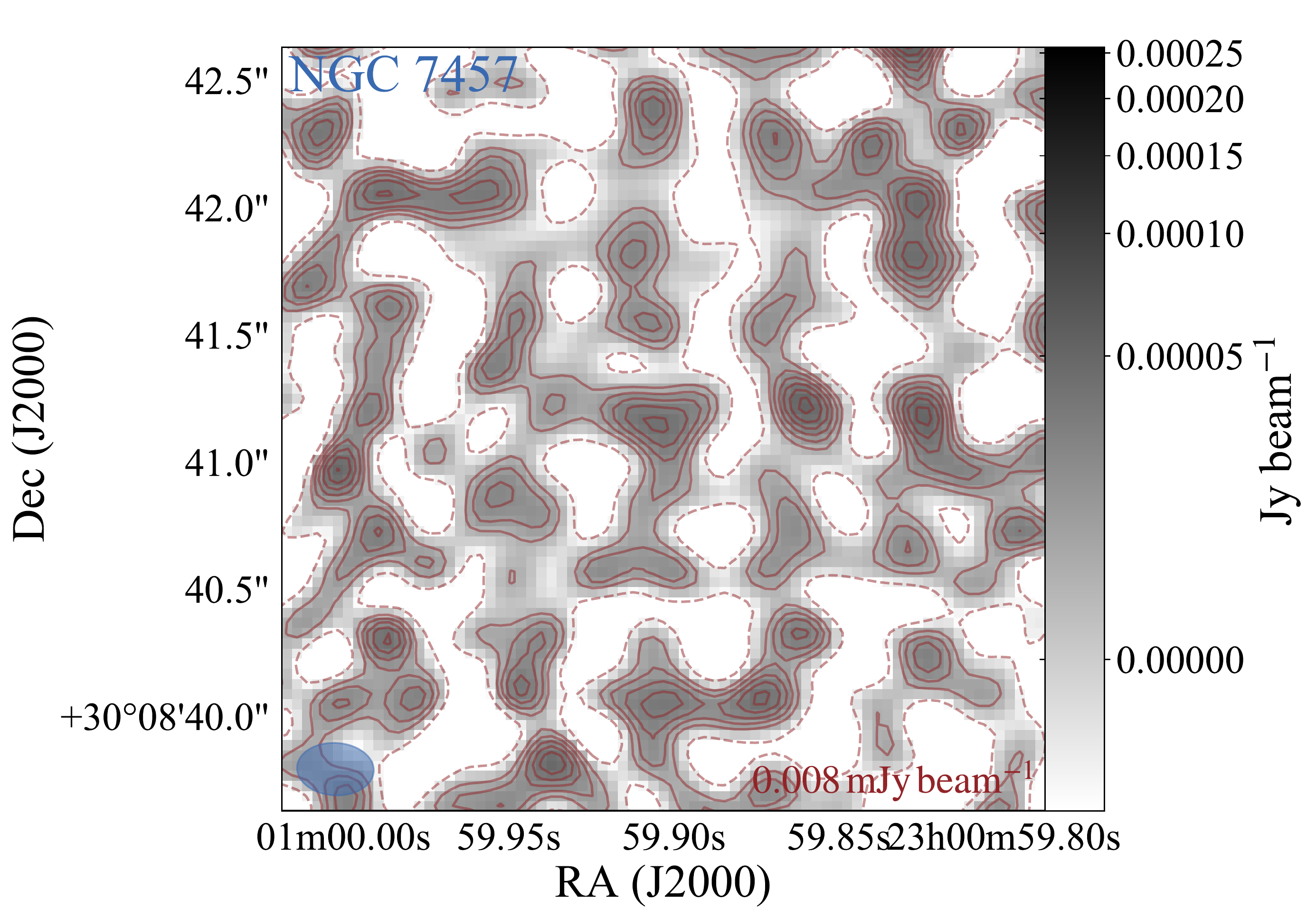}%
}
\caption{VLA maps of our new 8.4 GHz $X$-band observations of 12 sources.  Grayscale is indicated by the bar to the right of each panel, and the contours are in constant steps of the value indicated in the lower-right corner of each panel.  The blue ellipse in the lower-right corner of each panel shows the size and position angle of the synthesized beam.  The maps are centered on the brightest pixel within a 20 $\times$ 20 pixel region centered on the Simbad coordinates for the host galaxy.  When the source is securely detected, it is always consistent with a point source.  We list the integrated flux densities and upper limits in Table \ref{t:newradio}.}
\kgfigstarend{vlamaps}{VLA maps.}

\kgtabbeg{lrrrrrrrrr}
\footnotesize
\tablecaption{New 8.4 GHz VLA Data}
\tablewidth{0.9\columnwidth}

\tablehead{
\colhead{Galaxy} &
 \colhead{SB ID} &
 \colhead{MJD} &
 \colhead{Flux Cal.} &
 \colhead{Gain Cal.} &
 \colhead{$S_{\mathrm{peak}}$} &
 \colhead{$S_\nu$}  &
 \colhead{$\mathrm{rms}$} &
 \colhead{Size} &
 \colhead{PA} \\
 \colhead{} &
 \colhead{} &
 \colhead{} &
 \colhead{} &
 \colhead{} &
 \colhead{$\mathrm{mJy}$} &
 \colhead{$\mathrm{mJy}$} &
 \colhead{$\mathrm{mJy}$} &
 \colhead{$\arcsec \times \arcsec$} &
 \colhead{$^\circ$} 
}
\startdata
NGC 1300 & 3035251 & 55747 & 3C138 & J0340$-$2119 & $0.530$ & $ 0.582 \pm 0.042$ & $0.018$ & $0.37 \times 0.23$ & $168$ \\
NGC 2748 & 4463759 & 55744 & 3C147 & J0930+7420   & $0.134$ & $ 0.197 \pm 0.051$ & $0.019$ & $0.36 \times 0.25$ & $26$ \\
NGC 2778 & 4463837 & 55745 & 3C147 & J0916+3854   & $0.109$ & $ 0.115 \pm 0.039$ & $0.019$ & $0.26 \times 0.20$ & $55$ \\
NGC 3384 & 4463932 & 55735 & 3C288 & J1044+0655   & \dots & $< 0.047$           & $0.016$ & \dots & \dots \\
NGC 4291 & 4464121 & 55724 & 3C295 & J1243+7442   & \dots & $< 0.143$          & $0.048$ & \dots & \dots \\
NGC 4459 & 4464418 & 55723 & 3C286 & J1230+1223   & $0.496$ & $ 0.428 \pm 0.034$ & $0.017$ & $0.23 \times 0.20$ & $38$\\
NGC 4486A& 4464517 & 55737 & 3C286 & J1230+1223   & \dots & $< 0.273$          & $0.020$ & \dots & \dots \\
NGC 4596 & 4464593 & 55737 & 3C286 & J1230+1223   & $0.105$ & $ 0.102 \pm 0.046$ & $0.023$ & $0.30 \times 0.22$ & $32$\\
NGC 4742 & 4464671 & 55735 & 3C286 & J1256$-$0547 & \dots & $< 0.127$          & $0.015$ & \dots & \dots \\
NGC 5077 & 4464763 & 55742 & 3C286 & J1337$-$1257 & $185.4$ & $193.1 \pm 5.8$   & $0.222$ & $0.30 \times 0.23$ & $4$\\
NGC 5576 & 4464855 & 55740 & 3C295 & J1405+0415   & \dots & $< 0.202$          & $0.067$ & \dots & \dots \\
NGC 7457 & 4466045 & 55743 &  3C48 & J2255+4202   & \dots & $< 0.057$          & $0.019$ & \dots &  \dots

\enddata
\tablecomments{This table lists results from our new 8.4 GHz VLA observing campaign.  The columns list host galaxy name, VLA scheduling block identification number, MJD of the observation, flux calibrator, gain calibrator, and 8.4 GHz flux density.  Upper limits are listed at their 3$\sigma$ value.}
\kgtabend{newradio}{New 8.4 GHz VLA Data}

\section{Archival Radio Data}
\label{archivalradio}
In addition to our new radio data, we also analyzed 21 archival VLA radio datasets.  Some of the data in this had been previously published but had concentrated on radio emission that was not continuum core emission.  We analyzed these datasets in the same way as outlined in Appendix \ref{newradio}, and we present the results in Table \ref{t:arcradio}.

\kgtabbeg{llrrrrr}
\footnotesize
\tablecaption{Archival VLA Data}
\tablewidth{0.4\textwidth}

\tablehead{
\colhead{Galaxy} &
\colhead{Array} &
 \colhead{MJD} &
 \colhead{$\nu / \mathrm{GHz}$} &
 \colhead{$S_\nu / \mathrm{mJy}$} 
}
\startdata
IC 4296 & VLA A & 52381 & 8.46 & $ 247.7 \pm 1.7 $\\
IC 1459 & VLA A & 52301 & 4.86 & $ 1126 \pm 56 $\\
NGC 0224 & VLA B & 52485 & 8.40 & $ < 0.06 $\\
NGC 1068 & VLA A & 51429 & 8.46 & $ 364 \pm 23 $\\
NGC 1399 & VLA A & 45608 & 4.88 & $ 5.7 \pm 0.7 $\\
NGC 1550 & VLA B & 51033 & 8.48 & $ < 0.5 $\\
NGC 2787 & VLA A & 51899 & 8.46 & $ 13.5 \pm 0.9 $\\
NGC 3031 & VLA A & 51899 & 8.46 & $ 153 \pm 7.6 $\\
NGC 3245 & VLA A & 53259 & 8.46 & $ 1.7 \pm 0.15 $\\
NGC 3351 & VLA AB & 48596 & 4.88 & $ < 13.5 $\\
NGC 3393 & VLA AB & 52907 & 22.44 & $ < 13.2 $\\
NGC 3842 & VLA A & 47498 & 4.83 & $ < 1.1 $\\
NGC 4258 & VLA A & 51899 & 8.46 & $ 2.88 \pm 0.18 $\\
NGC 4477 & EVLA A & 56279 & 5.50 & $ 0.12 \pm 0.02 $\\
NGC 4486 & VLA A & 52793 & 8.44 & $ 2803 \pm 141 $\\
NGC 4594 & VLA A & 53280 & 8.46 & $ 59.9 \pm 3.7 $\\
NGC 4649 & VLA A & 50888 & 8.46 & $ 18.5 \pm 1.0 $\\
NGC 4889 & VLA AB & 47593 & 4.86 & $ < 2.7 $\\
NGC 5128 & VLA A & 53353 & 8.46 & $ 4700 \pm 250 $\\
NGC 7052 & VLA A & 50907 & 8.46 & $ 47.5 \pm 2.4 $\\
NGC 7619 & VLA A & 46049 & 4.89 & $ 1.92 \pm 0.12 $\\
UGC 3789 & VLA A & 54791 & 8.44 & $ < 5.4 $

\enddata
\tablecomments{This table lists results from our VLA archival analysis.  The columns list host galaxy name, array configuration, MJD of the observation, frequency of the observation, and flux density.  Upper limits are listed at their 3$\sigma$ value.}
\kgtabend{arcradio}{Archival VLA Data}

\section{Archival X-Ray Data}
\label{archivalxray}

The X-ray data for our AGN sample consists of our archival X-ray analysis in \citet{2009ApJ...706..404G}, analysis of new \emph{Chandra} data in \citet{2012ApJ...749..129G}, and a separate analysis of archival data in this work, including analysis of archival data of NGC 4486 at two epochs, which we average into a single X-ray flux for our fundamental plane analysis.  In this appendix we describe this most recent archival data analysis.

We use the same method of analysis as in \citet{2009ApJ...706..404G} and \citet{2012ApJ...749..129G}, both of which can be consulted for further details but which we summarize here.  All data reduction was performed with CIAO version $4.y$ and calibration database (CALDB) version $4.6.7$ using newly created level 2 event files.  Source identification was done with a combination of wavdetect and manual inspection of each \emph{Chandra} image.  Some sources, especially those in central early-type galaxies, did not register as a point source with wavdetect because of being surrounded by diffuse X-ray emission, presumably from hot gas.  { For such sources where any AGN emission was dominated by surrounding hot gas emission, the best that could be done was to estimate an upper limit on the flux by performing a background subtraction under the assumption that there was a point source at the center.  This upper limit to nuclear flux is a combination of AGN and XRBs, but it does not affect the rest of our analysis in the paper.}  All source locations were compared with optical and near infrared images of the host galaxy to ensure that we were selecting the likely central X-ray source.  In cases where there were multiple X-ray point sources consistent with the optical/near-infrared center of the galaxy, we registered the \emph{Chandra} image with the Sloan Digital Sky Survey or Deep Near Infrared Survey coordinates as best possible using common background AGNs.  We then took the source closest to the peak of the galaxy starlight.  Background regions were selected using an annulus with inner radius equal to the source region radius of between 1.5 and 3\arcsec, depending on potential contamination from surrounding X-ray emission.  The outer radius of the background annulus was chosen to include sufficient counts to generate a background.  We used the specextract tool provided by CIAO to extract spectra and to create auxiliary response files and response matrix files for the source and background spectra.

Spectral fitting was done with XSPEC version 12 \citep{1996ASPC..101...17A}.  All spectra were fit with a photoabsorbed power law with $N_\mathrm{H}$ column density set to the Galactic value toward the source \cite{2005A&A...440..775K}.  For the sources that had sufficient counts to warrant more detailed fitting, we also included a redshifted intrinsic photoabsorption component with a column density that was allowed to vary.  When required by the spectra, we also included other components to fully explain the data, especially APEC, pexmon, and blackbody models to account for soft emission and Gaussian features to account for emission lines.  In the vast majority of cases, the inclusion of the softer components did not make more than a 2$\sigma$ difference in the 2--10 keV power-law flux.  If we were unable to rule out zero flux from a power-law in the 2--10 keV band at the 3$\sigma$ level, we considered it an upper limit.  We report the results of our X-ray spectral fits in Table \ref{t:arcombhfits}.

\kgtabbeg{lrrrrrrrrr}
\footnotesize
\tablecaption{Spectral Fitting Results from Archival X-Ray Analysis}
\tablewidth{\columnwidth}

\tablehead{
\colhead{Galaxy} &
 \colhead{ObsID} &
 \colhead{MJD} &
 \colhead{Exposure Time} &
 \colhead{Net Count Rate} &
 \colhead{Galactic $N_\mathrm{H}$} &
 \colhead{Intrinsic $N_\mathrm{H}$} &
 \colhead{$\Gamma$} &
 \colhead{$\log{F_X}$} &
 \colhead{Notes}\\
\colhead{} &
 \colhead{} &
 \colhead{} &
 \colhead{ks} &
 \colhead{$10^{-3} \mathrm{count\,s^{-1}}$} &
 \colhead{$10^{22}\,\mathrm{cm}^{-2}$} &
 \colhead{$10^{22}\,\mathrm{cm}^{-2}$} &
 \colhead{} &
 \colhead{$\mathrm{erg\,s^{-1}\,cm^{-2}}$} &
 \colhead{}
}
\startdata
A1836-BCG & 11750 & 55342 & 60 & 41.5 & 0.0498 & $3.80_{-1.41}^{+1.69}$ & $2.95_{-0.66}^{+0.72}$ & $-13.15_{-0.08}^{+0.10}$   & 1 \\
       IC 2560 &  4908 & 55342 & 60 & 25.8 & 0.0722 & \dots                  & $2.21_{-0.07}^{+0.06}$ & $-12.33_{-0.02}^{+0.02}$   & 2 \\
      NGC 0524 &  6778 & 53976 & 15 & 1.19 & 0.0499 & \dots                  & $2$                    & $-14.30$                   &   \\
      NGC 1316 &  2022 & 52017 & 30 & 8.86 & 0.0240 & $0.13_{-0.06}^{+0.08}$ & $2.58_{-0.37}^{+0.44}$ & $-13.70_{-0.15}^{+0.14}$   &   \\
      NGC 1332 &  2915 & 52017 & 20 & 13.8 & 0.0222 & $0.19_{-0.11}^{+0.13}$ & $3.19_{-0.65}^{+0.77}$ & $-13.93_{-0.22}^{+0.19}$   & 3 \\
      NGC 1407 &   791 & 52017 & 50 & 8.69 & 0.0542 & $0.07_{-0.04}^{+0.04}$ & $2.43_{-0.21}^{+0.22}$ & $-13.75_{-0.09}^{+0.08}$   &   \\
      NGC 1550 &  5800 & 53666 & 45 & 5.34 & 0.1020 & \dots                  & $2$                    & $-15.84$                   &   \\
      NGC 3091 &  3215 & 53666 & 35 & 1.41 & 0.0409 & \dots                  & $2$                    & $-14.43$                   &   \\
      NGC 3393 & 12290 & 53666 & 79 & 9.03 & 0.0618 & \dots                  & $2.26_{-0.10}^{+0.08}$ & $-12.65_{-0.03}^{+0.03}$   & 4 \\
      NGC 3489 &   392 & 51485 &  2 & 6.48 & 0.0167 & \dots                  & $2$                    & $-13.68_{-0.14}^{+0.12}$   &   \\
      NGC 3607 &  2073 & 51485 & 37 & 4.03 & 0.0136 & \dots                  & $2$                    & $-13.90_{-0.06}^{+0.05}$   &   \\
      NGC 3842 &  4189 & 51485 & 50 & 1.11 & 0.0157 & \dots                  & $2$                    & $-14.61_{-\infty}^{+0.30}$ &   \\
      NGC 4382 &  2016 & 52059 & 43 & 1.28 & 0.0245 & \dots                  & $1.46_{-0.40}^{+0.45}$ & $-14.02_{-0.18}^{+0.17}$   &   \\
      NGC 4388 & 12291 & 52059 & 30 & 93.3 & 0.0258 & \dots                  & $1.10_{-0.00}^{+0.01}$ & $-11.09_{-0.02}^{+0.02}$   & 5 \\
      NGC 4472 & 12889 & 55607 &140 & 7.13 & 0.0153 & \dots                  & $2.86_{-0.78}^{+1.31}$ & $-14.71_{-0.26}^{+0.20}$   & 6 \\
      NGC 4477 & 12209 & 55318 & 10 & 3.11 & 0.0242 & \dots                  & $2.90_{-0.67}^{+0.79}$ & $-14.58_{-0.45}^{+0.39}$   &   \\
      NGC 4486 &  3980 & 52778 &  5 & 173.5 & 0.0194 & $0.07 \pm 0.02$       & $2.19_{-0.12}^{+0.13}$ & $-12.15 \pm 0.05$          & 7 \\
      NGC 4486 &  3981 & 52816 &  5 & 211.5 & 0.0194 & $0.11 \pm 0.02$       & $2.25 \pm0.12$         & $-11.97 \pm 0.05$          & 7 \\
      NGC 4526 &  3925 & 52957 & 44 & 5.33 & 0.0147 & \dots                  & $1.05_{-0.10}^{+0.18}$ & $-13.22_{-0.07}^{+0.06}$   &   \\
      NGC 4736 &   808 & 51677 & 50 & 57.9 & 0.0124 & $0.05_{-0.01}^{+0.01}$ & $2.11_{-0.06}^{+0.07}$ & $-12.79_{-0.03}^{+0.03}$   &   \\
      NGC 4751 & 12957 & 55666 &7.5 & 7.84 & 0.0755 & \dots                  & $2.50_{-0.28}^{+0.31}$ & $-13.88_{-0.19}^{+0.18}$   &   \\
      NGC 4826 &  9545 & 54889 &28.7& 7.82 & 0.0281 & \dots                  & $3.29_{-3.29}^{+4.92}$ & $-15.74_{-\infty}^{+0.84}$ & 8 \\
      NGC 4889 & 13996 & 56013 &125 & 5.16 & 0.0085 & $0.52$                 & $2$                    & $-14.58_{-\infty}^{+0.63}$ & 9 \\
      NGC 6861 & 11752 & 55057 &100 & 5.40 & 0.0388 & $0.05_{-0.05}^{+12}$   & $2.19_{-0.33}^{+0.35}$ & $-13.94_{-0.09}^{+0.09}$   & 10 \\
      NGC 7619 &  3955 & 52906 & 40 & 4.05 & 0.0477 & \dots                  & $2$                    & $-15.08_{-\infty}^{+0.95}$ &  

\enddata
\tablecomments{This table lists X-ray spectral fits for nuclear sources from archival \emph{Chandra} analysis.  Columns list galaxy name, \emph{Chandra} observation ID (ObsID), MJD of observation, Galactic absorption column density assumed toward each source, intrinsic absorption column density found from fits, power-law photon index, and logarithmic 2--10\,keV unabsorbed flux arising from the power-law component.  Uncertainties listed are 1$\sigma$, and values without uncertainties were held fixed.  Fluxes without uncertainties indicate that because of low count rates the X-ray flux was estimated using PIMMS with the net count rate, assuming a $\Gamma = 2$ power law with Galactic absorption only.  Such sources are treated as upper limits.  We do not report intrinsic absorption column density if the best fit is less than $10^{19}\,\mathrm{cm^{-2}}$.  The final column indicates notes as follows:  (1) Fit included APEC component with $kT = 0.54\,\mathrm{keV}$, a 0.5--2\,keV logarithmic flux $\log{F} = -13.81$ with an intrinsic absorption of $0.33 \times 10^{22}\,\mathrm{cm^{-2}}$.  (2) Fit included blackbody component with $kT = 0.17\,\mathrm{keV}$ and 0.5--2\,keV logarithmic flux $\log{F} = -13.16$ and the power law comes from a pexmon spectral model with solar abundances and inclination less than $16^\circ$. (3) Fit included APEC component with $kT = 0.81\,\mathrm{keV}$, a 0.5--2\,keV logarithmic flux $\log{F} = -13.38$. (4) Fit included blackbody component with $kT = 0.17\,\mathrm{keV}$ and 0.5--2\,keV logarithmic flux $\log{F} = -13.13$.  (5) Power law comes from pexmon spectral model with abundance 1.49 and Fe abundance of 0.11 of solar with inclination fixed at $60^\circ$.  The best-fit photon index is at the lower limit allowed by the pexmon model.  (6) Fit included APEC component with $kT = 0.72\,\mathrm{keV}$, a 0.5--2\,keV logarithmic flux $\log{F} = -14.24$.  (7) Used results to derive average value for fundamental plane fit of $F_X = (8.87_{-1.83}^{+2.33}) \times 10^{-13}\,\mathrm{erg\,s^{-1}\,cm^{-2}}$ on $\mathrm{MJD} = 52792$.   (8) Fit included APEC component with $kT = 0.78\,\mathrm{keV}$ and a normalization of $6.98 \times 10^{-6}$ at 1 keV.  (9) Fit included APEC component with $kT = 0.21\,\mathrm{keV}$, a 0.5--2\,keV logarithmic flux $\log{F} = -13.42$.  We were unable to constrain the uncertainties of the intrinsic absorption.  (10) Fit included APEC component with $kT = 0.97\,\mathrm{keV}$, a 0.5--2\,keV logarithmic flux $\log{F} = -13.94$.}
\kgtabend{arcombhfits}{Spectral Fitting Results from Archival X-ray Analysis}

\section{XRB Data}
\label{xrbdata}

We list our sample of XRBs with distances and masses in Table \ref{t:xrbsample}.  {A recent examination of \emph{Gaia} data has new distance estimates for 4 of our 6 XRBs \citep{2018arXiv180411349G}.  Of those 4, three of the distance estimates are based on \emph{Gaia} measurements with goodness of fit metrics greater than $+3$, which indicates a bad astrometric fit.  The final source, 4U 1543$-$47, for which we use distance of $7.5 \pm 1\ \mathrm{kpc}$ due to \citet{2004ApJ...610..378P}, has new distance estimates of $24.72 \pm 41.15$, $7.02^{+2.85}_{-1.86}$, and $10.11^{+4.33}_{-3.53}\ \mathrm{kpc}$, depending on the Bayesian priors assumed.  Given the order of magnitude range in distances acceptable for this source, it is not clear that it is an improvement on our adopted distance.  Nevertheless, we ran our analysis using distances for all four sources from \citet{2018arXiv180411349G} based on their ``$r_{\mathrm{exp}}$ prior.''  The quantitative differences in the fit results for the full sample were less than 1\% of the 68\% uncertainty range, indicating no difference to the results of this paper.}

\kgtabbeg{lr@{$\pm$}lrr@{$\pm$}lr}
\footnotesize
\tablecaption{X-Ray Binary Sample}
\tablewidth{0pt}

\tablehead{
\colhead{Source Name} &
 \multicolumn{2}{c}{Distance} &
 \colhead{References} &
 \multicolumn{2}{c}{BH Mass} &
 \colhead{References}\\
 \colhead{} &
 \multicolumn{2}{c}{kpc} &
 \colhead{} &
 \multicolumn{2}{c}{\msun} &
 \colhead{} 
}
\startdata
4U 1543$-$47 & 7.5&1.0 & 1 & 9.42&0.97& 2\\
Cygnus X-1 & 1.86& 0.12 & 3 & 14.8&1.0 & 4\\
GRO J1655$-$40 & 3.2&0.2& 5 & 6.3&0.25 & 6\\
GRS 1915+105 & 11&1 & 7 & 10.1&0.6 & 8\\
XTE J1118+480 & 1.8&0.6 & 9 & 7.1&1.3 & 10\\
XTE J1550$-$564 & 4.4&0.5& 11 & 9.1&0.6 & 12
\enddata
\tablecomments{This table lists the black hole X-ray binary sources used in our fundamental plane analysis.  The columns indicate source name, distance to source in units of kpc, a reference code for the distance measurement, the mass of the black hole in solar units, and a reference code for the mass measurement.}
\tablerefs{(1) \citealt{2004ApJ...610..378P}; (2) \citealt{2003IAUS..212..365O};
(3) \citealt{2011ApJ...742...83R}; (4) \citealt{2011ApJ...742...84O};
(5) \citealt{2004MNRAS.354..355J}; (6) \citealt{2001ApJ...554.1290G}, where we have converted the published
95\% uncertainty to a 68\% value assuming a normal distribution, which
is well justified by the derived probability distribution;
(7) \citealt{2013ApJ...768..185S}; (8) \citealt{2013ApJ...768..185S};
(9) \citealt{2001ApJ...551L.147M}; (10) \citealt{2001ApJ...551L.147M};
(11) \citealt{2011ApJ...730...75O}; (12) \citealt{2011ApJ...730...75O}.}
\kgtabend{xrbsample}{X-ray Binary Sample}

Our data for XRBs consists of literature radio data and a combination of literature X-ray data and our analysis of archival \emph{RXTE} data for sources 4U 1543$-$47, GRO J1655$-$40, XTE J1118+480, and XTE J15550$-$564.  

X-ray spectral fitting to 4U 1543$-$47 for observation on $\mathrm{MJD} = 52490.14009$ was done using a photoabsorbed power-law model ({\tt phabs(pow)}) with an absorption column fixed to $N_H = 4.0 \times 10^{21}\,\mathrm{cm^{-2}}$.  Our spectral fitting yielded a constraint on the power-law component's photon index of $\Gamma = 1.82 \pm 0.03$.  For the observation on $\mathrm{MJD} = 52445.60917$, during which 4U 1543$-$47 was in a very high state, we used a photoabsorbed accretion-disk emission-line plus accretion-disk blackbody plus power-law model ({\tt phabs(laor + bbody + pow)}) with  an absorption column fixed to $N_H = 4.0 \times 10^{21}\,\mathrm{cm^{-2}}$.  The power-law photon index was constrained to $\Gamma = 2.51 \pm 0.01$.

X-ray spectral fitting to GRO J1655$-$40 was done using a model of a photoabsorbed accretion-disk emission-line plus power-law model ({\tt phabs(laor + pow)}) with  an absorption column fixed to $N_H = 9.0 \times 10^{21}\,\mathrm{cm^{-2}}$.  The power-law varied slightly among the four observations from $\Gamma = 1.32 \pm 0.02$ to $1.47 \pm 0.01$.

X-ray spectral fitting to XTE J1118+480 was done using an absorbed power-law model in the 3--25 keV range using standard data products in the archive. We added 0.6\% systematic errors to all channels, and adopted an absorption column of $N_H = 1.4 \times 10^{20}\,\mathrm{cm^{-2}}$ with Tuebingen-Boulder interstellar medium absorption model (tbabs).  For all observations, the power-law photon index changed very little, from $\Gamma = 1.715 \pm 0.004$ to $1.721 \pm 0.004$.

X-ray spectral fitting to XTE J15550$-$564 for the observation on $\mathrm{MJD} = 51664.42194$, during which it was in a very high state, was done using a photoabsorbed accretion-disk emission-line plus accretion-disk blackbody plus power-law model ({\tt phabs(laor + bbody + pow)}) with  an absorption column fixed to $N_H = 9.0 \times 10^{21}\,\mathrm{cm^{-2}}$.  The fit constrained the power-law photon index to be $\Gamma = 2.22 \pm 0.01$.  The observation on $\mathrm{MJD} = 51696.48361$, during which the source was in a low/hard state, was fitted with a photoabsorbed power-law model ({\tt phabs(pow)}) with the same fixed absorption column.  The photon index was constrained to be $\Gamma = 1.64 \pm 0.01$.

We list the radio and X-ray observational data in Table \ref{t:xrbdata}.

\clearpage
\LongTables
\begin{landscape}
\newcommand{\kgagntabref}[1]{%
     \IfEqCase{#1}{%
       {KH13}{1}
       {Dalla Bonta+ 2009}{2}
       {PKSCat90}{3}
       {arcombh}{4}
       {ALK13}{5}
       {alpha Abell 3565 BCG : AT20G}{6}
       {combh}{7}
       {Greenhill+ 2003}{8}
       {AT20G}{9}
       {alpha Circinus : AT20G}{10}
       {Tadhunter+ 2003}{11}
       {Hirabayashi+ 2000}{12}
       {Genzel+ 2010 etc.}{13}
       {Zhao+ 2001}{14}
       {Bag+01}{15}
       {Nowak+ 2012}{16}
       {Cappellari+ 2002}{17}
       {alpha IC 1459 : CRATES}{18}
       {Hure+ 2011}{19}
       {Xanthopoulos+ 2010}{20}
       {alpha IC 1481 : Dressel+Condon 1978}{21}
       {Yamauchi+ 2012}{22}
       {Morganti+ 1999}{23}
       {alpha IC 2560 : NVSS}{24}
       {G09}{25}
       {Valluri+ 2005}{26}
       {Heckman+ 1980}{27}
       {van den Bosch+ 2010}{28}
       {Yang+ 2015}{29}
       {Bender+ 2005}{30}
       {Krajnovic+ 2009}{31}
       {Filho+ 2004}{32}
       {alpha NGC 0524 : Nagar+ 2005}{33}
       {Gebhardt+ 2001; Merritt+ 2001}{34}
       {Israel+ 1992}{35}
       {Schulze+ 2011}{36}
       {Nyland+ 2014}{37}
       {Bower+ 2001}{38}
       {Lodato+ 2003; Hure+ 2002}{39}
       {HU01}{40}
       {Kuo+ 2011}{41}
       {Schmidt+ 2001}{42}
       {alpha NGC 1194 : Gallimore+ 2006}{43}
       {van den Bosch+ 2012}{44}
       {Atkinson+ 2005}{45}
       {KG13}{46}
       {alpha NGC 1300 : Beck+ 2002}{47}
       {xcombh}{48}
       {Nowak+ 2008}{49}
       {Tingay+ 2003}{50}
       {alpha NGC 1316 : Tingay+ 2003}{51}
       {Rusli+ 2011}{52}
       {Rusli+ 2013}{53}
       {Nagar+ 2005}{54}
       {alpha NGC 2778 : KG13}{55}
       {Sarzi+ 2001}{56}
       {Emsellem+ 1999}{57}
       {Fabbiano+ 1989}{58}
       {Davies+ 2006}{59}
       {Barth+ 2001}{60}
       {Pastorini+ 2007}{61}
       {Saikia+ 1994}{62}
       {Sarzi+ 2002}{63}
       {Nowak+ 2010}{64}
       {Nagar+ 2002}{65}
       {Kondratko+ 2008; Hure+ 2011}{66}
       {Gultekin+ 09b}{67}
       {alpha NGC 3607 : Nagar+ 2005}{68}
       {McConnell+ 2012}{69}
       {alpha NGC 3945 : Nagar+ 2005}{70}
       {Walsh+ 2012}{71}
       {Wrobel+ 1984}{72}
       {alpha NGC 3998 : Richards+ 2011}{73}
       {Marconi+ 2003}{74}
       {Sramek 1975}{75}
       {alpha NGC 4143 : Nagar+ 2002}{76}
       {Ferrarese+ 1996}{77}
       {Laurent-Muehleisen+ 1997}{78}
       {alpha NGC 4261 : Jones+Wehrle 1997}{79}
       {Filho+ 2006}{80}
       {Cretton+ 1999b}{81}
       {Wrobel+ 1991}{82}
       {Walsh+ 2010}{83}
       {alpha NGC 4374 : ALK13}{84}
       {Gultekin+ 11}{85}
       {Coccato+ 2006}{86}
       {alpha NGC 4459 : KG13}{87}
       {Gebhardt+ 2011}{88}
       {alpha NGC 4486 : Nagar+ 2001}{89}
       {combh2}{90}
       {Nowak+ 2007}{91}
       {Kormendy+ 1997}{92}
       {Cappetti+ 2009}{93}
       {Davis+ 2013}{94}
       {Jardel+ 2011}{95}
       {Shen+Gebhardt 2010}{96}
       {alpha NGC 4649 : Shurkin+ 2008}{97}
       {Gebhardt+ 2013}{98}
       {alpha NGC 4736 : Nagar+ 2005}{99}
       {Tonry+ 2001}{100}
       {Gultekin+ inprep}{101}
       {Greenhill+ 1997}{102}
       {Elmouttie+ 1997}{103}
       {De Francesco+ 2008}{104}
       {alpha NGC 5077 : Murphy+ 2010}{105}
       {Cappellari+ 2009}{106}
       {alpha NGC 5128 : Muller+ 2011}{107}
       {McConnell+ 2011b}{108}
       {Liuzzo+ 2010}{109}
       {Ferrarese+ 1999}{110}
       {Evans+ 2005}{111}
       {alpha NGC 6251 : Evans+ 2005}{112}
       {van der Marel + 1998b}{113}
       {Wold+ 2006}{114}
       {UW84}{115}
       {alpha NGC 7582 : UW84}{116}
       {alpha NGC 7619 : FIRST}{117}
       {Kuo+ 2011; Hure+ 2011}{118}
       {Venturi+ 2004}{119}
       {alpha UGC 9799 : Venturi+ 2004}{120}
        {}{}
       }[$X$]
 }%

 \kgtabbeg{lrcr@{}lcrrr@{}lclcrr@{}lcc}
 \tablecaption{SMBH Sample and Radio and X-ray Observational Data}
 \tablewidth{0pt}

 \tablehead{
  \colhead{Source} &
  \colhead{Dist.} &
  \colhead{Ref.} &
  \multicolumn{2}{c}{$M_\mathrm{BH}$} &
  \colhead{Ref.} &
  \colhead{MJD} &
  \colhead{$\nu$} &
  \multicolumn{2}{c}{$S_\nu$} &
  \colhead{Ref.} &
  \colhead{$\alpha_R$} &
  \colhead{Ref.} &
  \colhead{MJD} &
  \multicolumn{2}{c}{${F_X}$} &
  \colhead{Ref.} &
  \colhead{Notes}\\
  \colhead{} &
  \colhead{Mpc} &
  \colhead{} &
  \multicolumn{2}{c}{\msun} &
  \colhead{} &
  \colhead{} &
  \colhead{GHz} &
  \multicolumn{2}{c}{mJy} &
  \colhead{} &
  \colhead{} &
  \colhead{} &
  \colhead{} &
  \multicolumn{2}{c}{$\mathrm{erg\,s^{-1}\,cm^{-2}}$} &
  \colhead{} &
  \colhead{}
 }
 \startdata
 \scriptsize

MGC $-$02-36-002&152.4&\kgagntabref{KH13}&&$3.74^{+0.42}_{-0.52} \times 10^{9}$&\kgagntabref{Dalla Bonta+ 2009}&44239&5&&$440.00\pm0.00$&\kgagntabref{PKSCat90}&\dots&&55342&&$7.14^{+1.93}_{-1.19} \times 10^{-14}$&\kgagntabref{arcombh}&ac\\
IC 4296&49.2&\kgagntabref{KH13}&&$1.30^{+0.20}_{-0.19} \times 10^{9}$&\kgagntabref{Dalla Bonta+ 2009}&53280&8.46&&$247.70\pm1.70$&\kgagntabref{ALK13}&$0.10\pm0.24$&\kgagntabref{alpha Abell 3565 BCG : AT20G}&52258&&$5.85^{+0.46}_{-0.70} \times 10^{-13}$&\kgagntabref{combh}&\\
Circinus&2.82&\kgagntabref{KH13}&&$1.14\pm{0.20} \times 10^{6}$&\kgagntabref{Greenhill+ 2003}&53907&4.8&&$304.00\pm15.00$&\kgagntabref{AT20G}&$1.13\pm0.11$&\kgagntabref{alpha Circinus : AT20G}&51617&&$1.61^{+0.05}_{-0.19} \times 10^{-10}$&\kgagntabref{combh}&bc\\
Cygnus A&242.7&\kgagntabref{KH13}&&$2.66^{+0.74}_{-0.75} \times 10^{9}$&\kgagntabref{Tadhunter+ 2003}&50249&5&&$160.00\pm10.00$&\kgagntabref{Hirabayashi+ 2000}&\dots&&51690&&$2.21^{+0.11}_{-0.67} \times 10^{-11}$&\kgagntabref{combh}&ab\\
Galaxy&0.00828&\kgagntabref{KH13}&&$4.30\pm{0.36} \times 10^{6}$&\kgagntabref{Genzel+ 2010 etc.}&48182&5&&$710.00\pm72.00$&\kgagntabref{Zhao+ 2001}&$1.00\pm0.10$&\kgagntabref{Zhao+ 2001}&51843&&$2.87^{+0.53}_{-0.39} \times 10^{-13}$&\kgagntabref{Bag+01}&\\
Galaxy&0.00828&\kgagntabref{KH13}&&$4.30\pm{0.36} \times 10^{6}$&\kgagntabref{Genzel+ 2010 etc.}&48182&5&&$710.00\pm72.00$&\kgagntabref{Zhao+ 2001}&$1.00\pm0.10$&\kgagntabref{Zhao+ 2001}&51843&&$1.31\pm{0.13} \times 10^{-11}$&\kgagntabref{Bag+01}&\\
Galaxy&0.00828&\kgagntabref{KH13}&&$4.30\pm{0.36} \times 10^{6}$&\kgagntabref{Genzel+ 2010 etc.}&48182&5&&$710.00\pm72.00$&\kgagntabref{Zhao+ 2001}&$1.00\pm0.10$&\kgagntabref{Zhao+ 2001}&55966&&$2.51^{+0.94}_{-0.49} \times 10^{-11}$&\kgagntabref{Nowak+ 2012}&\\
IC 1459&28.92&\kgagntabref{KH13}&&$2.48^{+0.48}_{-0.19} \times 10^{9}$&\kgagntabref{Cappellari+ 2002}&52301&4.86&&$1125.70\pm1.10$&\kgagntabref{ALK13}&$0.21\pm0.02$&\kgagntabref{alpha IC 1459 : CRATES}&52133&&$6.45^{+0.19}_{-0.21} \times 10^{-13}$&\kgagntabref{combh}&\\
IC 1481&89.9&\kgagntabref{KH13}&&$1.49^{+0.44}_{-0.45} \times 10^{7}$&\kgagntabref{Hure+ 2011}&51160&4.994&&$1.29\pm0.15$&\kgagntabref{Xanthopoulos+ 2010}&$3.83\pm0.24$&\kgagntabref{alpha IC 1481 : Dressel+Condon 1978}&52133&&\dots&&a\\
IC 2560&37.2&\kgagntabref{KH13}&&$5.01^{+0.71}_{-0.72} \times 10^{6}$&\kgagntabref{Yamauchi+ 2012}&49914&8.6&&$6.20\pm0.26$&\kgagntabref{Morganti+ 1999}&$0.90\pm0.04$&\kgagntabref{alpha IC 2560 : NVSS}&53052&&$4.72\pm{0.24} \times 10^{-13}$&\kgagntabref{arcombh}&b\\
NGC 0205&0.74&\kgagntabref{G09}&$<$&$3.80 \times 10^{4}$&\kgagntabref{Valluri+ 2005}&43431&4.885&$<$&2.00&\kgagntabref{Heckman+ 1980}&\dots&&53062&$<$&$3.90 \times 10^{-13}$&\kgagntabref{combh}&ac\\
NGC 0221&0.805&\kgagntabref{KH13}&&$2.45^{+1.01}_{-1.02} \times 10^{6}$&\kgagntabref{van den Bosch+ 2010}&56137&6.6&&$0.05\pm0.01$&\kgagntabref{Yang+ 2015}&$-2.00\pm1.70$&\kgagntabref{Yang+ 2015}&53517&&$1.71^{+0.17}_{-0.28} \times 10^{-14}$&\kgagntabref{combh}&\\
NGC 0224&0.774&\kgagntabref{KH13}&&$1.43^{+0.91}_{-0.31} \times 10^{8}$&\kgagntabref{Bender+ 2005}&52485&8.4&$<$&0.07&\kgagntabref{ALK13}&\dots&&51696&&$2.18^{+0.37}_{-0.65} \times 10^{-13}$&\kgagntabref{combh}&a\\
NGC 0524&24.22&\kgagntabref{KH13}&&$8.67^{+0.94}_{-0.46} \times 10^{8}$&\kgagntabref{Krajnovic+ 2009}&52153&5&&$1.50\pm0.09$&\kgagntabref{Filho+ 2004}&$0.00\pm0.13$&\kgagntabref{alpha NGC 0524 : Nagar+ 2005}&53976&$<$&$1.58 \times 10^{-14}$&\kgagntabref{arcombh}&\\
NGC 0598&0.8&\kgagntabref{G09}&$<$&$3.00 \times 10^{3}$&\kgagntabref{Gebhardt+ 2001; Merritt+ 2001}&42199&5&$<$&1000.00&\kgagntabref{Israel+ 1992}&\dots&&51786&&$3.61^{+0.12}_{-0.98} \times 10^{-12}$&\kgagntabref{combh}&ac\\
NGC 0821&23.44&\kgagntabref{KH13}&&$1.65^{+0.74}_{-0.73} \times 10^{8}$&\kgagntabref{Schulze+ 2011}&56204&5&$<$&0.09&\kgagntabref{Nyland+ 2014}&\dots&&53543&&$3.58^{+0.85}_{-3.35} \times 10^{-15}$&\kgagntabref{combh}&a\\
NGC 1023&10.81&\kgagntabref{KH13}&&$4.13^{+0.43}_{-0.42} \times 10^{7}$&\kgagntabref{Bower+ 2001}&56204&5&$<$&0.07&\kgagntabref{Nyland+ 2014}&\dots&&54276&&$3.71^{+0.42}_{-0.67} \times 10^{-14}$&\kgagntabref{combh}&a\\
NGC 1068&15.9&\kgagntabref{KH13}&&$8.39\pm{0.44} \times 10^{6}$&\kgagntabref{Lodato+ 2003; Hure+ 2002}&51429&8.46&&$364.00\pm23.00$&\kgagntabref{ALK13}&$-0.76\pm0.08$&\kgagntabref{HU01}&51595&&$1.25^{+0.06}_{-0.08} \times 10^{-13}$&\kgagntabref{combh}&b\\
NGC 1194&57.98&\kgagntabref{KH13}&&$7.08^{+0.33}_{-0.32} \times 10^{7}$&\kgagntabref{Kuo+ 2011}&50916&8.46&&$0.40\pm0.03$&\kgagntabref{Schmidt+ 2001}&$2.51\pm0.18$&\kgagntabref{alpha NGC 1194 : Gallimore+ 2006}&51595&&\dots&&a\\
NGC 1277&73&\kgagntabref{KH13}&&$1.70\pm{0.30} \times 10^{10}$&\kgagntabref{van den Bosch+ 2012}&&&&\dots&&\dots&&51595&&\dots&&a\\
NGC 1300&21.5&\kgagntabref{KH13}&&$7.55^{+7.25}_{-3.66} \times 10^{7}$&\kgagntabref{Atkinson+ 2005}&55747&8.46&&$0.58\pm0.04$&\kgagntabref{KG13}&$5.74\pm0.37$&\kgagntabref{alpha NGC 1300 : Beck+ 2002}&55115&&$1.10^{+0.40}_{-0.30} \times 10^{-13}$&\kgagntabref{xcombh}&\\
NGC 1316&20.95&\kgagntabref{KH13}&&$1.69^{+0.28}_{-0.30} \times 10^{8}$&\kgagntabref{Nowak+ 2008}&50430&4.8&&$40.00\pm0.10$&\kgagntabref{Tingay+ 2003}&$1.19\pm0.01$&\kgagntabref{alpha NGC 1316 : Tingay+ 2003}&52017&&$2.01^{+0.69}_{-0.60} \times 10^{-14}$&\kgagntabref{arcombh}&\\
NGC 1332&22.66&\kgagntabref{KH13}&&$1.47^{+0.21}_{-0.20} \times 10^{9}$&\kgagntabref{Rusli+ 2011}&&&&\dots&&\dots&&52535&&$1.19^{+0.66}_{-0.47} \times 10^{-14}$&\kgagntabref{arcombh}&a\\
NGC 1374&19.57&\kgagntabref{KH13}&&$5.90^{+0.61}_{-0.51} \times 10^{8}$&\kgagntabref{Rusli+ 2013}&&&&\dots&&\dots&&52535&&\dots&&a\\
NGC 1399&20.85&\kgagntabref{KH13}&&$5.04\pm{0.69} \times 10^{8}$&\kgagntabref{KH13}&45608&4.885&&$5.74\pm0.72$&\kgagntabref{ALK13}&\dots&&51561&$<$&$9.38 \times 10^{-16}$&\kgagntabref{combh}&a\\
NGC 1407&29&\kgagntabref{KH13}&&$4.65^{+0.73}_{-0.41} \times 10^{9}$&\kgagntabref{Rusli+ 2013}&&&&\dots&&\dots&&51772&&$1.78^{+0.36}_{-0.32} \times 10^{-14}$&\kgagntabref{arcombh}&a\\
NGC 1550&52.5&\kgagntabref{KH13}&&$3.87^{+0.61}_{-0.71} \times 10^{9}$&\kgagntabref{Rusli+ 2013}&51033&8.485&$<$&0.49&\kgagntabref{ALK13}&\dots&&53666&$<$&$8.58 \times 10^{-16}$&\kgagntabref{arcombh}&a\\
NGC 2273&29.5&\kgagntabref{KH13}&&$8.61\pm{0.46} \times 10^{6}$&\kgagntabref{Kuo+ 2011}&52260&5&&$2.40\pm0.20$&\kgagntabref{Nagar+ 2005}&$-0.54\pm0.05$&\kgagntabref{HU01}&53666&&\dots&&ab\\
NGC 2549&12.7&\kgagntabref{KH13}&&$1.45^{+0.20}_{-1.14} \times 10^{7}$&\kgagntabref{Krajnovic+ 2009}&56302&5&$<$&0.07&\kgagntabref{Nyland+ 2014}&\dots&&53666&&\dots&&a\\
NGC 2748&23.4&\kgagntabref{KH13}&&$4.44^{+1.76}_{-1.82} \times 10^{7}$&\kgagntabref{Atkinson+ 2005}&55744&8.46&&$0.20\pm0.05$&\kgagntabref{KG13}&\dots&&55377&&$7.30^{+27.50}_{-6.00} \times 10^{-16}$&\kgagntabref{xcombh}&\\
NGC 2778&23.44&\kgagntabref{KH13}&$<$&$1.45 \times 10^{7}$&\kgagntabref{Schulze+ 2011}&56254&5&&$0.11\pm0.01$&\kgagntabref{Nyland+ 2014}&$-0.11\pm0.78$&\kgagntabref{alpha NGC 2778 : KG13}&55197&&$1.90^{+68.90}_{-1.80} \times 10^{-16}$&\kgagntabref{xcombh}&a\\
NGC 2787&7.45&\kgagntabref{KH13}&&$4.07^{+0.40}_{-0.52} \times 10^{7}$&\kgagntabref{Sarzi+ 2001}&51899&8.46&&$13.51\pm0.54$&\kgagntabref{ALK13}&\dots&&53143&&$6.80^{+0.72}_{-1.06} \times 10^{-14}$&\kgagntabref{combh}&\\
NGC 2960&67.1&\kgagntabref{KH13}&&$1.08^{+0.04}_{-0.05} \times 10^{7}$&\kgagntabref{Kuo+ 2011}&&&&\dots&&\dots&&53143&&\dots&&a\\
NGC 3031&3.604&\kgagntabref{KH13}&&$6.50^{+2.50}_{-1.50} \times 10^{7}$&\kgagntabref{KG13}&51899&8.46&&$153.29\pm0.09$&\kgagntabref{ALK13}&$0.14\pm0.03$&\kgagntabref{HU01}&53869&$<$&$2.52 \times 10^{-10}$&\kgagntabref{combh}&ab\\
NGC 3091&53.02&\kgagntabref{KH13}&&$3.72^{+0.11}_{-0.51} \times 10^{9}$&\kgagntabref{Rusli+ 2013}&&&&\dots&&\dots&&52359&$<$&$9.83 \times 10^{-15}$&\kgagntabref{arcombh}&a\\
NGC 3115&9.54&\kgagntabref{KH13}&&$8.97^{+0.57}_{-2.77} \times 10^{8}$&\kgagntabref{Emsellem+ 1999}&46840&4.86&$<$&0.33&\kgagntabref{Fabbiano+ 1989}&\dots&&52074&&$8.92^{+1.28}_{-6.49} \times 10^{-15}$&\kgagntabref{combh}&ac\\
NGC 3227&23.75&\kgagntabref{KH13}&&$2.10^{+0.69}_{-1.12} \times 10^{7}$&\kgagntabref{Davies+ 2006}&51922&5&&$4.70\pm0.30$&\kgagntabref{Nagar+ 2005}&$-0.90\pm0.09$&\kgagntabref{HU01}&51542&&$1.05^{+0.09}_{-0.13} \times 10^{-11}$&\kgagntabref{combh}&b\\
NGC 3245&21.38&\kgagntabref{KH13}&&$2.39^{+0.27}_{-0.76} \times 10^{8}$&\kgagntabref{Barth+ 2001}&53259&8.46&&$1.72\pm0.15$&\kgagntabref{ALK13}&\dots&&52303&&$3.67^{+0.41}_{-3.08} \times 10^{-14}$&\kgagntabref{combh}&\\
NGC 3310&17.4&\kgagntabref{G09}&$<$&$4.20 \times 10^{7}$&\kgagntabref{Pastorini+ 2007}&46658&4.86&&$2.10\pm0.21$&\kgagntabref{Saikia+ 1994}&\dots&&52664&&$3.72^{+0.26}_{-0.90} \times 10^{-13}$&\kgagntabref{combh}&a\\
NGC 3351&8.7&\kgagntabref{G09}&$<$&$8.60 \times 10^{6}$&\kgagntabref{Sarzi+ 2002}&48596&4.885&$<$&13.50&\kgagntabref{ALK13}&\dots&&53409&$<$&$1.67 \times 10^{-10}$&\kgagntabref{combh}&a\\
NGC 3368&10.62&\kgagntabref{KH13}&&$7.66\pm{1.53} \times 10^{6}$&\kgagntabref{Nowak+ 2010}&50877&15&$<$&1.00&\kgagntabref{Nagar+ 2002}&\dots&&51868&&$5.79^{+17.41}_{-3.02} \times 10^{-16}$&\kgagntabref{combh}&a\\
NGC 3377&10.99&\kgagntabref{KH13}&&$1.78^{+0.94}_{-0.93} \times 10^{8}$&\kgagntabref{Schulze+ 2011}&56304&5&&$0.19\pm0.01$&\kgagntabref{Nyland+ 2014}&\dots&&52645&&$6.21^{+0.74}_{-4.63} \times 10^{-15}$&\kgagntabref{combh}&\\
NGC 3379&10.7&\kgagntabref{KH13}&&$4.16\pm{1.04} \times 10^{8}$&\kgagntabref{van den Bosch+ 2010}&56304&5&&$0.71\pm0.01$&\kgagntabref{Nyland+ 2014}&\dots&&54110&&$9.17^{+1.43}_{-5.05} \times 10^{-15}$&\kgagntabref{combh}&\\
NGC 3384&11.49&\kgagntabref{KH13}&&$1.08\pm{0.49} \times 10^{7}$&\kgagntabref{Schulze+ 2011}&56304&5&$<$&0.07&\kgagntabref{Nyland+ 2014}&\dots&&55215&&$1.70^{+1.50}_{-0.80} \times 10^{-14}$&\kgagntabref{xcombh}&a\\
NGC 3393&49.2&\kgagntabref{KH13}&&$1.57^{+0.98}_{-0.99} \times 10^{7}$&\kgagntabref{Kondratko+ 2008; Hure+ 2011}&52907&22.44&$<$&13.20&\kgagntabref{ALK13}&\dots&&55633&&$2.26^{+0.16}_{-0.15} \times 10^{-13}$&\kgagntabref{arcombh}&ab\\
NGC 3489&11.98&\kgagntabref{KH13}&&$5.94^{+0.84}_{-0.83} \times 10^{6}$&\kgagntabref{Nowak+ 2010}&56266&5&$<$&0.21&\kgagntabref{Nyland+ 2014}&\dots&&51485&&$3.71^{+3.75}_{-2.14} \times 10^{-14}$&\kgagntabref{arcombh}&a\\
NGC 3585&20.51&\kgagntabref{KH13}&&$3.29^{+1.45}_{-0.58} \times 10^{8}$&\kgagntabref{Gultekin+ 09b}&&&&\dots&&\dots&&52063&&$1.82^{+0.27}_{-0.82} \times 10^{-14}$&\kgagntabref{combh}&a\\
NGC 3607&22.65&\kgagntabref{KH13}&&$1.37^{+0.45}_{-0.47} \times 10^{8}$&\kgagntabref{Gultekin+ 09b}&56240&5&&$2.05\pm0.07$&\kgagntabref{Nyland+ 2014}&$0.35\pm0.10$&\kgagntabref{alpha NGC 3607 : Nagar+ 2005}&52072&&$1.26^{+0.16}_{-0.15} \times 10^{-14}$&\kgagntabref{arcombh}&\\
NGC 3608&22.75&\kgagntabref{KH13}&&$4.65\pm{0.99} \times 10^{8}$&\kgagntabref{Schulze+ 2011}&56240&5&&$0.24\pm0.01$&\kgagntabref{Nyland+ 2014}&\dots&&52072&$<$&$1.21 \times 10^{-15}$&\kgagntabref{combh}&a\\
NGC 3842&92.2&\kgagntabref{KH13}&&$9.09^{+2.31}_{-2.81} \times 10^{9}$&\kgagntabref{McConnell+ 2012}&47498&4.835&$<$&1.10&\kgagntabref{ALK13}&\dots&&52663&$<$&$8.90 \times 10^{-15}$&\kgagntabref{arcombh}&a\\
NGC 3945&19.5&\kgagntabref{KH13}&$<$&$8.80 \times 10^{6}$&\kgagntabref{Gultekin+ 09b}&56305&5&&$2.87\pm0.09$&\kgagntabref{Nyland+ 2014}&$0.28\pm0.07$&\kgagntabref{alpha NGC 3945 : Nagar+ 2005}&53841&&$7.76^{+0.89}_{-1.48} \times 10^{-14}$&\kgagntabref{combh}&a\\
NGC 3982&18.2&\kgagntabref{G09}&$<$&$8.00 \times 10^{7}$&\kgagntabref{Sarzi+ 2002}&51482&4.86&&$1.79\pm0.05$&\kgagntabref{HU01}&$-0.56\pm0.06$&\kgagntabref{HU01}&53007&$<$&$2.80 \times 10^{-14}$&\kgagntabref{combh}&ab\\
NGC 3998&14.3&\kgagntabref{KH13}&&$8.45^{+0.70}_{-0.66} \times 10^{8}$&\kgagntabref{Walsh+ 2012}&44216&4.885&&$83.00\pm3.00$&\kgagntabref{Wrobel+ 1984}&$-0.77\pm0.03$&\kgagntabref{alpha NGC 3998 : Richards+ 2011}&53917&&$1.05\pm{0.02} \times 10^{-11}$&\kgagntabref{combh}&\\
NGC 4026&13.35&\kgagntabref{KH13}&&$1.80^{+0.60}_{-0.35} \times 10^{8}$&\kgagntabref{Gultekin+ 09b}&&&&\dots&&\dots&&53886&$<$&$2.78 \times 10^{-15}$&\kgagntabref{combh}&a\\
NGC 4041&20.9&\kgagntabref{G09}&$<$&$6.40 \times 10^{6}$&\kgagntabref{Marconi+ 2003}&41469&5&$<$&41.00&\kgagntabref{Sramek 1975}&\dots&&53940&$<$&$4.18 \times 10^{-13}$&\kgagntabref{combh}&ac\\
NGC 4143&16.8&\kgagntabref{G09}&$<$&$1.40 \times 10^{8}$&\kgagntabref{Sarzi+ 2002}&51269&4.9&&$8.70\pm0.20$&\kgagntabref{Nagar+ 2002}&$-0.12\pm0.03$&\kgagntabref{alpha NGC 4143 : Nagar+ 2002}&51994&&$2.68^{+238.32}_{-1.95} \times 10^{-13}$&\kgagntabref{combh}&a\\
NGC 4203&16&\kgagntabref{G09}&$<$&$3.80 \times 10^{7}$&\kgagntabref{Sarzi+ 2002}&50877&15&&$9.00\pm0.23$&\kgagntabref{Nagar+ 2002}&$0.44\pm0.04$&\kgagntabref{HU01}&51486&&$8.17^{+0.96}_{-1.61} \times 10^{-13}$&\kgagntabref{combh}&a\\
NGC 4258&7.27&\kgagntabref{KH13}&&$3.78\pm{0.04} \times 10^{7}$&\kgagntabref{KG13}&51899&8.46&&$2.88\pm0.10$&\kgagntabref{ALK13}&$-0.18\pm0.03$&\kgagntabref{HU01}&52058&&$1.12^{+0.09}_{-0.35} \times 10^{-11}$&\kgagntabref{combh}&b\\
NGC 4261&32.36&\kgagntabref{KH13}&&$5.29^{+1.07}_{-1.08} \times 10^{8}$&\kgagntabref{Ferrarese+ 1996}&49479&4.885&&$285.00\pm2.77$&\kgagntabref{Laurent-Muehleisen+ 1997}&$1.92\pm0.03$&\kgagntabref{alpha NGC 4261 : Jones+Wehrle 1997}&54508&&$7.74^{+0.42}_{-0.58} \times 10^{-13}$&\kgagntabref{combh}&\\
NGC 4291&26.58&\kgagntabref{KH13}&&$9.78^{+3.12}_{-3.08} \times 10^{8}$&\kgagntabref{Schulze+ 2011}&55724&8.46&$<$&0.14&\kgagntabref{KG13}&\dots&&55541&&$1.40^{+0.60}_{-0.40} \times 10^{-14}$&\kgagntabref{xcombh}&a\\
NGC 4321&18&\kgagntabref{G09}&$<$&$2.70 \times 10^{7}$&\kgagntabref{Sarzi+ 2002}&52201&5&&$0.53\pm0.06$&\kgagntabref{Filho+ 2006}&\dots&&51488&$<$&$4.34 \times 10^{-15}$&\kgagntabref{combh}&a\\
NGC 4342&22.91&\kgagntabref{KH13}&&$4.53^{+2.65}_{-1.48} \times 10^{8}$&\kgagntabref{Cretton+ 1999b}&46778&4.86&$<$&0.50&\kgagntabref{Wrobel+ 1991}&\dots&&53412&&$3.54^{+0.59}_{-1.48} \times 10^{-14}$&\kgagntabref{combh}&ac\\
NGC 4374&18.51&\kgagntabref{KH13}&&$9.25^{+0.95}_{-0.87} \times 10^{8}$&\kgagntabref{Walsh+ 2010}&51269&4.9&&$160.00\pm0.20$&\kgagntabref{Nagar+ 2002}&$0.17\pm0.02$&\kgagntabref{alpha NGC 4374 : ALK13}&51683&$<$&$6.78 \times 10^{-11}$&\kgagntabref{combh}&a\\
NGC 4382&17.88&\kgagntabref{KH13}&$<$&$1.30 \times 10^{7}$&\kgagntabref{Gultekin+ 11}&&&&\dots&&\dots&&52059&&$9.45^{+4.45}_{-3.14} \times 10^{-15}$&\kgagntabref{arcombh}&a\\
NGC 4388&16.53&\kgagntabref{KH13}&&$7.31^{+0.17}_{-0.18} \times 10^{6}$&\kgagntabref{Kuo+ 2011}&50877&15&&$3.70\pm0.09$&\kgagntabref{Nagar+ 2002}&$-0.80\pm0.08$&\kgagntabref{HU01}&55902&&$8.21^{+0.35}_{-0.33} \times 10^{-12}$&\kgagntabref{arcombh}&b\\
NGC 4435&17&\kgagntabref{G09}&$<$&$8.00 \times 10^{6}$&\kgagntabref{Coccato+ 2006}&56303&5&&$0.12\pm0.01$&\kgagntabref{Nyland+ 2014}&\dots&&54507&$<$&$7.53 \times 10^{-11}$&\kgagntabref{combh}&a\\
NGC 4459&16.01&\kgagntabref{KH13}&&$6.96^{+1.33}_{-1.34} \times 10^{7}$&\kgagntabref{Sarzi+ 2001}&56303&5&&$0.37\pm0.03$&\kgagntabref{Nyland+ 2014}&$-0.21\pm0.20$&\kgagntabref{alpha NGC 4459 : KG13}&55310&&$2.30^{+-0.30}_{-0.70} \times 10^{-14}$&\kgagntabref{xcombh}&\\
NGC 4472&16.72&\kgagntabref{KH13}&&$2.54^{+0.58}_{-0.10} \times 10^{9}$&\kgagntabref{Rusli+ 2013}&51269&4.9&&$2.30\pm0.20$&\kgagntabref{Nagar+ 2002}&$-0.45\pm0.05$&\kgagntabref{HU01}&55607&&$1.95^{+1.11}_{-0.89} \times 10^{-15}$&\kgagntabref{arcombh}&\\
NGC 4473&15.25&\kgagntabref{KH13}&&$9.00\pm{4.50} \times 10^{7}$&\kgagntabref{Schulze+ 2011}&56303&5&$<$&0.15&\kgagntabref{Nyland+ 2014}&\dots&&53427&$<$&$7.09 \times 10^{-15}$&\kgagntabref{combh}&a\\
NGC 4477&18&\kgagntabref{G09}&$<$&$8.40 \times 10^{7}$&\kgagntabref{Sarzi+ 2002}&56279&5.5&&$0.12\pm0.02$&\kgagntabref{ALK13}&\dots&&55318&&$2.65^{+3.87}_{-1.72} \times 10^{-15}$&\kgagntabref{arcombh}&ab\\
NGC 4486&16.68&\kgagntabref{KH13}&&$6.15^{+0.38}_{-0.37} \times 10^{9}$&\kgagntabref{Gebhardt+ 2011}&52793&8.435&&$2803.00\pm15.00$&\kgagntabref{ALK13}&$0.19\pm0.01$&\kgagntabref{alpha NGC 4486 : Nagar+ 2001}&52797&&$8.87^{+2.33}_{-1.83} \times 10^{-13}$&\kgagntabref{combh2}&\\
NGC 4486A&18.36&\kgagntabref{KH13}&&$1.44^{+0.53}_{-0.52} \times 10^{7}$&\kgagntabref{Nowak+ 2007}&55737&8.394&$<$&0.00&\kgagntabref{KG13}&\dots&&55299&&$4.00^{+4.30}_{-2.40} \times 10^{-15}$&\kgagntabref{xcombh}&a\\
NGC 4486B&16.26&\kgagntabref{KH13}&&$6.00^{+3.00}_{-2.00} \times 10^{8}$&\kgagntabref{Kormendy+ 1997}&53842&8.4&$<$&0.15&\kgagntabref{Cappetti+ 2009}&\dots&&52964&&$2.96^{+0.66}_{-1.71} \times 10^{-15}$&\kgagntabref{combh}&a\\
NGC 4501&18&\kgagntabref{G09}&$<$&$7.90 \times 10^{7}$&\kgagntabref{Sarzi+ 2002}&51482&4.86&&$1.14\pm0.06$&\kgagntabref{HU01}&$-0.48\pm0.05$&\kgagntabref{HU01}&52617&&$3.93^{+0.70}_{-1.62} \times 10^{-14}$&\kgagntabref{combh}&ab\\
NGC 4526&16.44&\kgagntabref{KH13}&&$4.51^{+1.40}_{-1.03} \times 10^{8}$&\kgagntabref{Davis+ 2013}&56302&5&&$1.48\pm0.05$&\kgagntabref{Nyland+ 2014}&\dots&&52957&&$6.08^{+0.85}_{-0.88} \times 10^{-14}$&\kgagntabref{arcombh}&\\
NGC 4548&20.3&\kgagntabref{G09}&$<$&$3.40 \times 10^{7}$&\kgagntabref{Sarzi+ 2002}&50877&15&&$1.60\pm0.04$&\kgagntabref{Nagar+ 2002}&\dots&&51992&$<$&$3.03 \times 10^{-13}$&\kgagntabref{combh}&a\\
NGC 4564&15.94&\kgagntabref{KH13}&&$8.81^{+2.49}_{-2.43} \times 10^{7}$&\kgagntabref{Schulze+ 2011}&56303&5&$<$&0.09&\kgagntabref{Nyland+ 2014}&\dots&&52964&&$1.86^{+0.39}_{-1.48} \times 10^{-14}$&\kgagntabref{combh}&a\\
NGC 4594&9.87&\kgagntabref{KH13}&&$6.65^{+0.40}_{-0.41} \times 10^{8}$&\kgagntabref{Jardel+ 2011}&53280&8.46&&$59.90\pm2.20$&\kgagntabref{ALK13}&\dots&&52060&&$1.23^{+0.15}_{-0.89} \times 10^{-12}$&\kgagntabref{combh}&\\
NGC 4596&16.53&\kgagntabref{KH13}&&$7.67^{+3.73}_{-3.24} \times 10^{7}$&\kgagntabref{Sarzi+ 2001}&56302&5&$<$&0.10&\kgagntabref{Nyland+ 2014}&\dots&&55151&&$3.90^{+4.30}_{-2.20} \times 10^{-15}$&\kgagntabref{xcombh}&a\\
NGC 4649&16.46&\kgagntabref{KH13}&&$4.72^{+1.04}_{-1.05} \times 10^{9}$&\kgagntabref{Shen+Gebhardt 2010}&50888&8.46&&$18.45\pm0.32$&\kgagntabref{ALK13}&$1.52\pm0.01$&\kgagntabref{alpha NGC 4649 : Shurkin+ 2008}&54130&$<$&$1.72 \times 10^{-14}$&\kgagntabref{combh}&a\\
NGC 4697&12.54&\kgagntabref{KH13}&&$2.02^{+0.51}_{-0.50} \times 10^{8}$&\kgagntabref{Schulze+ 2011}&56306&5&$<$&0.07&\kgagntabref{Nyland+ 2014}&\dots&&51558&&$9.74^{+2.56}_{-9.34} \times 10^{-15}$&\kgagntabref{combh}&a\\
NGC 4698&18&\kgagntabref{G09}&$<$&$7.60 \times 10^{7}$&\kgagntabref{Sarzi+ 2002}&51482&4.86&&$0.23\pm0.06$&\kgagntabref{HU01}&\dots&&52441&&$1.71^{+0.24}_{-0.61} \times 10^{-14}$&\kgagntabref{combh}&ab\\
NGC 4736&5&\kgagntabref{KH13}&&$6.77\pm{1.56} \times 10^{6}$&\kgagntabref{Gebhardt+ 2013}&49479&4.885&&$4.00\pm0.22$&\kgagntabref{Laurent-Muehleisen+ 1997}&$0.76\pm0.09$&\kgagntabref{alpha NGC 4736 : Nagar+ 2005}&51677&&$1.62^{+0.11}_{-0.10} \times 10^{-13}$&\kgagntabref{arcombh}&\\
NGC 4742&16.4&\kgagntabref{Tonry+ 2001}&&$1.40^{+0.40}_{-0.50} \times 10^{7}$&\kgagntabref{Gultekin+ inprep}&55735&8.46&$<$&0.00&\kgagntabref{KG13}&\dots&&55160&&$2.50^{+0.80}_{-0.60} \times 10^{-14}$&\kgagntabref{xcombh}&a\\
NGC 4751&32.81&\kgagntabref{KH13}&&$2.44^{+0.12}_{-0.37} \times 10^{9}$&\kgagntabref{Rusli+ 2013}&&&&\dots&&\dots&&55666&&$1.30^{+0.66}_{-0.47} \times 10^{-14}$&\kgagntabref{arcombh}&a\\
NGC 4826&7.27&\kgagntabref{KH13}&&$1.56\pm{0.39} \times 10^{6}$&\kgagntabref{Gebhardt+ 2013}&50877&15&$<$&0.90&\kgagntabref{Nagar+ 2002}&\dots&&54889&$<$&$7.85 \times 10^{-15}$&\kgagntabref{arcombh}&a\\
NGC 4889&102&\kgagntabref{KH13}&&$2.08^{+1.58}_{-1.59} \times 10^{10}$&\kgagntabref{McConnell+ 2012}&47593&4.86&$<$&2.70&\kgagntabref{ALK13}&\dots&&56013&$<$&$2.62 \times 10^{-14}$&\kgagntabref{arcombh}&a\\
NGC 4945&3.58&\kgagntabref{KH13}&&$1.35^{+0.68}_{-0.48} \times 10^{6}$&\kgagntabref{Greenhill+ 1997}&49133&4.998&&$297.00\pm0.03$&\kgagntabref{Elmouttie+ 1997}&$-0.60\pm0.06$&\kgagntabref{Elmouttie+ 1997}&51570&$<$&$2.25 \times 10^{-12}$&\kgagntabref{combh}&ab\\
NGC 5077&38.7&\kgagntabref{KH13}&&$8.55^{+4.35}_{-4.48} \times 10^{8}$&\kgagntabref{De Francesco+ 2008}&55742&8.46&&$193.10\pm5.80$&\kgagntabref{KG13}&$0.32\pm0.07$&\kgagntabref{alpha NGC 5077 : Murphy+ 2010}&55325&&$2.80^{+0.90}_{-0.70} \times 10^{-14}$&\kgagntabref{xcombh}&\\
NGC 5128&3.62&\kgagntabref{KH13}&&$5.69\pm{1.04} \times 10^{7}$&\kgagntabref{Cappellari+ 2009}&53353&8.46&&$4700.00\pm250.00$&\kgagntabref{ALK13}&$0.33\pm0.06$&\kgagntabref{alpha NGC 5128 : Muller+ 2011}&52896&&$7.15^{+0.74}_{-1.81} \times 10^{-12}$&\kgagntabref{combh}&\\
NGC 5516&55.3&\kgagntabref{KH13}&&$3.69^{+0.10}_{-1.04} \times 10^{9}$&\kgagntabref{Rusli+ 2013}&&&&\dots&&\dots&&52862&&\dots&&a\\
NGC 5576&25.68&\kgagntabref{KH13}&&$2.73^{+0.68}_{-0.79} \times 10^{8}$&\kgagntabref{Gultekin+ 09b}&56302&5&$<$&0.09&\kgagntabref{Nyland+ 2014}&\dots&&55342&&$2.70^{+5.60}_{-1.80} \times 10^{-15}$&\kgagntabref{xcombh}&a\\
NGC 5845&25.87&\kgagntabref{KH13}&&$4.87\pm{1.53} \times 10^{8}$&\kgagntabref{Schulze+ 2011}&56302&5&$<$&0.07&\kgagntabref{Nyland+ 2014}&\dots&&52642&&$1.21^{+0.13}_{-1.16} \times 10^{-14}$&\kgagntabref{combh}&a\\
NGC 6086&138&\kgagntabref{KH13}&&$3.74^{+1.76}_{-1.15} \times 10^{9}$&\kgagntabref{McConnell+ 2011b}&54542&5&$<$&0.55&\kgagntabref{Liuzzo+ 2010}&\dots&&52642&&\dots&&a\\
NGC 6251&108.4&\kgagntabref{KH13}&&$6.14^{+2.04}_{-2.05} \times 10^{8}$&\kgagntabref{Ferrarese+ 1999}&51679&4.9891&&$380.00\pm0.40$&\kgagntabref{Evans+ 2005}&$-0.02\pm0.09$&\kgagntabref{alpha NGC 6251 : Evans+ 2005}&52954&$<$&$6.06 \times 10^{-12}$&\kgagntabref{combh}&a\\
NGC 6264&147.6&\kgagntabref{KH13}&&$3.08\pm{0.04} \times 10^{7}$&\kgagntabref{Kuo+ 2011}&&&&\dots&&\dots&&52954&&\dots&&a\\
NGC 6323&113.4&\kgagntabref{KH13}&&$1.01^{+0.02}_{-0.01} \times 10^{7}$&\kgagntabref{Kuo+ 2011}&&&&\dots&&\dots&&52954&&\dots&&a\\
NGC 6861&28.71&\kgagntabref{KH13}&&$2.10^{+0.63}_{-0.10} \times 10^{9}$&\kgagntabref{Rusli+ 2013}&&&&\dots&&\dots&&55057&&$1.15^{+0.26}_{-0.22} \times 10^{-14}$&\kgagntabref{arcombh}&a\\
NGC 7052&70.4&\kgagntabref{KH13}&&$3.96^{+2.76}_{-1.56} \times 10^{8}$&\kgagntabref{van der Marel + 1998b}&50907&8.46&&$47.53\pm0.56$&\kgagntabref{ALK13}&\dots&&52538&$<$&$1.45 \times 10^{-14}$&\kgagntabref{combh}&a\\
NGC 7457&12.53&\kgagntabref{KH13}&&$9.00^{+5.30}_{-5.40} \times 10^{6}$&\kgagntabref{Schulze+ 2011}&56234&5&$<$&0.07&\kgagntabref{Nyland+ 2014}&\dots&&55085&&$4.30^{+16.00}_{-3.40} \times 10^{-15}$&\kgagntabref{xcombh}&a\\
NGC 7582&22.3&\kgagntabref{KH13}&&$5.51^{+1.30}_{-0.95} \times 10^{7}$&\kgagntabref{Wold+ 2006}&45149&4.86&&$69.00\pm1.25$&\kgagntabref{UW84}&$0.71\pm0.02$&\kgagntabref{alpha NGC 7582 : UW84}&51831&&$8.21^{+1.04}_{-4.66} \times 10^{-12}$&\kgagntabref{combh}&\\
NGC 7619&53.85&\kgagntabref{KH13}&&$2.30^{+1.15}_{-0.11} \times 10^{9}$&\kgagntabref{Rusli+ 2013}&46049&4.89&&$1.92\pm0.12$&\kgagntabref{ALK13}&$1.66\pm0.05$&\kgagntabref{alpha NGC 7619 : FIRST}&52906&$<$&$5.07 \times 10^{-14}$&\kgagntabref{arcombh}&a\\
NGC 7768&116&\kgagntabref{KH13}&&$1.34^{+0.51}_{-0.41} \times 10^{9}$&\kgagntabref{McConnell+ 2012}&54512&5&&$0.72\pm0.09$&\kgagntabref{Liuzzo+ 2010}&\dots&&52906&&\dots&&a\\
UGC 3789&49.9&\kgagntabref{KH13}&&$9.65\pm{1.55} \times 10^{6}$&\kgagntabref{Kuo+ 2011; Hure+ 2011}&54791&8.435&$<$&5.40&\kgagntabref{ALK13}&\dots&&52906&&\dots&&a\\
UGC 9799&151.1&\kgagntabref{G09}&$<$&$4.90 \times 10^{9}$&\kgagntabref{Dalla Bonta+ 2009}&51242&4.9&&$345.00\pm14.00$&\kgagntabref{Venturi+ 2004}&$0.88\pm0.11$&\kgagntabref{alpha UGC 9799 : Venturi+ 2004}&51790&$<$&$1.24 \times 10^{-13}$&\kgagntabref{combh}&a

\enddata
\tablenotetext{a}{Did not use in fundamental plane analysis.}
\tablenotetext{b}{Seyfert.}
\tablenotetext{c}{Radio data comes from observation with large beam and therefore may contain emission that is not solely core AGN emission.}
\tablerefs{
 (1) \citealt{2013ARA&A..51..511K}; 
 (2) \citealt{2009ApJ...690..537D}; 
 (3) \citealt{1990PKS...C......0W}; 
 (4) our analysis of archival \emph{Chandra} data (section \ref{archivalxray});
 (5) our analysis of archival VLA data (section \ref{archivalradio});
 (6) from $S_{5} = 262 \pm 34\ \mathrm{mJy}$ \citep{2010MNRAS.402.2403M};
 (7) \citealt{2009ApJ...706..404G}; 
 (8) \citealt{2003ApJ...590..162G}; 
 (9) \citealt{2010MNRAS.402.2403M};
 (10) from $S_{8.64} = 156 \pm 7\ \mathrm{mJy}$ \citep{2010MNRAS.402.2403M}; 
 (11) \citealt{2003MNRAS.342..861T}; 
 (12) \citealt{2000PASJ...52..997H}; 
 (13) \citealt{2010RvMP...82.3121G}; 
 (14) \citealt{2001ApJ...547L..29Z}; 
 (15) \citealt{2001Natur.413...45B}; 
 (16) \citealt{2012ApJ...759...95N}; 
 (17) \citealt{2002ApJ...578..787C}; 
 (18) from $S_{8.4} = 1001.5 \pm 13\ \mathrm{mJy}$ \citep{2007ApJS..171...61H};
 (19) \citealt{2002A&A...395L..21H}; 
 (20) \citealt{2010MNRAS.404.1966X}; 
 (21) from $S_{2.38} = 22 \pm 3\ \mathrm{mJy}$ \citep{1978ApJS...36...53D};
 (22) \citealt{2012PASJ...64..103Y}; 
 (23) \citealt{1999A&AS..137..457M}; 
 (24) from $S_{1.4} = 32 \pm 1.7\ \mathrm{mJy}$ \citep{1998AJ....115.1693C};
 (25) \citealt{2009ApJ...698..198G}; 
 (26) \citealt{2005ApJ...628..137V}; 
 (27) \citealt{1980A&AS...40..295H}; 
 (28) \citealt{2010MNRAS.401.1770V}; 
 (29) \citealt{2015ApJ...807L..19Y}; 
 (30) \citealt{2005ApJ...631..280B}; 
 (31) \citealt{2009MNRAS.399.1839K}; 
 (32) \citealt{2004A&A...418..429F}; 
 (33) from $S_{15} = 1.5 \pm 0.2\ \mathrm{mJy}$ \citep{2005A&A...435..521N};
 (34) \citealt{2001AJ....122.2469G, 2001Sci...293.1116M};
 (35) \citealt{1992A&A...261...47I}; 
 (36) \citealt{2011ApJ...729...21S}; 
 (37) \citealt{2016MNRAS.458.2221N}; 
 (38) \citealt{2001ApJ...550...75B}; 
 (39) \citealt{2003A&A...398..517L, 2002A&A...395L..21H};
 (40) \citealt{2001ApJS..133...77H}; 
 (41) \citealt{2011ApJ...727...20K}; 
 (42) \citealt{2001ApJS..132..199S}; 
 (43) from $S_{5} = 1.5 \pm 0.083\ \mathrm{mJy}$ \citep{2006AJ....132..546G};
 (44) \citealt{2012Natur.491..729V}; 
 (45) \citealt{2005MNRAS.359..504A}; 
 (46) our analsysis of new EVLA data (section \ref{newradio});
 (47) from $S_{5} = 11 \pm 2\ \mathrm{mJy}$ \citep{2002A&A...391...83B};
 (48) \citealt{2012ApJ...749..129G}; 
 (49) \citealt{2008MNRAS.391.1629N}; 
 (50) \citealt{2003PASJ...55..351T}; 
 (51) from $S_{8.6} = 20 \pm 0.1\ \mathrm{mJy}$ \citep{2003PASJ...55..351T};
 (52) \citealt{2011MNRAS.410.1223R}; 
 (53) \citealt{2013AJ....146...45R}; 
 (54) \citealt{2005A&A...435..521N}; 
 (56) \citealt{2001ApJ...550...65S}; 
 (57) \citealt{1999MNRAS.303..495E}; 
 (58) \citealt{1989ApJ...347..127F}; 
 (59) \citealt{2006ApJ...646..754D}; 
 (60) \citealt{2001ApJ...555..685B}; 
 (61) \citealt{2007A&A...469..405P}; 
 (62) \citealt{1994MNRAS.270...46S}; 
 (63) \citealt{2002ApJ...567..237S}; 
 (64) \citealt{2010MNRAS.403..646N}; 
 (65) \citealt{2002A&A...392...53N}; 
 (66) \citealt{2008ApJ...678...87K, 2011A&A...530A.145H};
 (67) \citealt{2009ApJ...695.1577G}; 
 (68) from $S_{15} = 1.4 \pm 0.15\ \mathrm{mJy}$ \citep{2005A&A...435..521N};
 (69) \citealt{2012ApJ...756..179M}; 
 (70) from $S_{15} = 2.1 \pm 0.15\ \mathrm{mJy}$ \citep{2005A&A...435..521N};
 (71) \citealt{2012ApJ...753...79W}; 
 (72) \citealt{1984ApJ...287...41W}; 
 (73) from $S_{15} = 196 \pm 3\ \mathrm{mJy}$ \citep{2011ApJS..194...29R};
 (74) \citealt{2003ApJ...586..868M}; 
 (75) \citealt{1975AJ.....80..771S}; 
 (76) from $S_{15} = 10 \pm 0.25\ \mathrm{mJy}$ \citep{2002A&A...392...53N};
 (77) \citealt{1996ApJ...470..444F}; 
 (78) \citealt{1997A&AS..122..235L}; 
 (79) from $S_{8.387} = 101 \pm 1\ \mathrm{mJy}$ \citep{1997ApJ...484..186J};
 (80) \citealt{2006A&A...451...71F}; 
 (81) \citealt{1999ApJ...514..704C}; 
 (82) \citealt{1991AJ....101..148W}; 
 (83) \citealt{2010ApJ...721..762W}; 
 (84) from $S_{8.46} = 146 \pm 1.3\ \mathrm{mJy}$ from our analysis of archival VLA data (section \ref{archivalradio});
 (85) \citealt{2011ApJ...741...38G}; 
 (86) \citealt{2006MNRAS.366.1050C}; 
 (87) from $S_{8.394} = 0.452 \pm 0.028\ \mathrm{mJy}$ from our analysis of archival VLA data (section \ref{archivalradio});
 (88) \citealt{2011ApJ...729..119G}; 
 (89) from $S_{5} = 3097.1 \pm 0.1\ \mathrm{mJy}$ \citep{2001ApJ...559L..87N};
 (91) \citealt{2007MNRAS.379..909N}; 
 (92) \citealt{1997ApJ...482L.139K}; 
 (93) \citealt{2009AJ....138.1990C}; 
 (94) \citealt{2013Natur.494..328D}; 
 (95) \citealt{2011ApJ...739...21J}; 
 (96) \citealt{2010ApJ...711..484S}; 
 (97) from $S_{1.4} = 282 \pm 0.03\ \mathrm{mJy}$ \citep{2008MNRAS.383..923S};
 (98) K.\ Gebhardt et al.\ in prep;
 (99) from $S_{15} = 1.7 \pm 0.15\ \mathrm{mJy}$ \citep{2005A&A...435..521N};
 (100) \citealt{2001ApJ...546..681T}; 
 (101) K.\ G{\"u}ltekin et al.\ in prep;
 (102) \citealt{1997ApJ...481L..23G}; 
 (103) \citealt{1997MNRAS.284..830E}; 
 (104) \citealt{2008A&A...479..355D}; 
 (105) from $S_{20} = 149 \pm 8\ \mathrm{mJy}$ \citep{2010MNRAS.402.2403M};
 (106) \citealt{2009MNRAS.394..660C}; 
 (107) from $S_{22.3} = 3400 \pm 100\ \mathrm{mJy}$ \citep{2011A&A...530L..11M};
 (108) \citealt{2011ApJ...728..100M}; 
 (109) \citealt{2010A&A...516A...1L}; 
 (110) \citealt{1999ApJ...515..583F}; 
 (111) \citealt{2005MNRAS.359..363E}; 
 (112) from $S_{15} = 390 \pm 40\ \mathrm{mJy}$ \citep{2005MNRAS.359..363E};
 (113) \citealt{1998AJ....116.2220V}; 
 (114) \citealt{2006A&A...460..449W}; 
 (115) \citealt{1984ApJ...285..439U}; 
 (116) from $S_{1.4} = 166 \pm 2\ \mathrm{mJy}$ \citep{1984ApJ...285..439U};
 (117) from $S_{1.4} = 15.4 \pm 0.123\ \mathrm{mJy}$ \citep{1997ApJ...475..479W};
 (118) \citealt{2011ApJ...727...20K, 2011A&A...530A.145H};
 (119) \citealt{2004A&A...422..515V}; 
 (120) from $S_{8.3} = 217 \pm 9\ \mathrm{mJy}$ \citep{2004A&A...422..515V}.
}
\tablecomments{This table shows the SMBH data we use in
our fundamental plane analysis. The columns provide source name,
distance in units of Mpc, a reference code for the distance
measurement, the mass of the black hole in solar units, a reference
code for the mass measurement, the MJD of the radio observation, the
frequency of the radio observation, the core radio flux density, a
reference code for the radio observation, the MJD of the X-ray
observation, the 2--10 keV X-ray flux, a reference code for the X-ray
measurement.}
\kgtabend{agndata}{AGN Data}

\clearpage
\end{landscape}

\kgtabbeg{lcccr@{$\pm$}lcccr@{$\pm$}lclcc}
\footnotesize
\tablecaption{X-Ray Binary Radio and X-Ray Observational Data}
\tablewidth{0pt}

\tablehead{
 \colhead{Source} &
 \colhead{Radio Obs.} &
 \colhead{Radio MJD} &
 \colhead{$\nu$} &
 \multicolumn{2}{c}{$S_\nu$} &
 \colhead{Ref.} &
 \colhead{X-Ray Obs.} &
 \colhead{X-ray MJD} &
 \multicolumn{2}{c}{${F_X}$} &
 \colhead{Ref.} &
 \colhead{State} &
 \colhead{Ref.} &
 \colhead{Notes}\\
 \colhead{} &
 \colhead{} &
 \colhead{} &
 \colhead{GHz} &
 \multicolumn{2}{c}{mJy} &
 \colhead{} &
 \colhead{} &
 \colhead{} &
 \multicolumn{2}{c}{$10^{-10}\,\mathrm{erg\,s^{-1}\,cm^{-2}}$} &
 \colhead{} &
 \colhead{} &
 \colhead{} &
 \colhead{}
}
\startdata
4U 1543$-$47 & ATCA & 52445 & 4.8 & $3.18$&$0.19$ & 1 & \textit{RXTE} & 52445.6 & $58.30$&$1.20$ & 2 & Very high & 1 & \tablenotemark{a}\\
4U 1543$-$47 & ATCA & 52490 & 4.8 & $4.00$&$0.05$ & 1 & \textit{RXTE} & 52490.1 & $0.88$&$0.01$ & 2 & Low/hard & 1 & \tablenotemark{a}\\
Cyg X-1 & AMI & 54928 & 15 & $9.30$&$0.20$ & 3 & \textit{Suzaku} & 54925.4 & $100.00$&$2.60$ & 3 & Low/hard & 3 & \tablenotemark{b} \\
Cyg X-1 & AMI & 54933 & 15 & $11.60$&$0.10$ & 3 & \textit{Suzaku} & 54930.4 & $79.60$&$2.40$ & 3 & Low/hard & 3 & \tablenotemark{b} \\
Cyg X-1 & AMI & 54940 & 15 & $8.80$&$0.10$ & 3 & \textit{Suzaku} & 54936.4 & $67.30$&$2.00$ & 3 & Low/hard & 3 & \tablenotemark{b} \\
Cyg X-1 & AMI & 54945 & 15 & $10.80$&$0.10$ & 3 & \textit{Suzaku} & 54945.4 & $70.00$&$2.10$ & 3 & Low/hard & 3 & \tablenotemark{b} \\
Cyg X-1 & AMI & 54953 & 15 & $10.20$&$0.20$ & 3 & \textit{Suzaku} & 54950.4 & $74.10$&$2.20$ & 3 & Low/hard & 3 & \tablenotemark{b} \\
Cyg X-1 & AMI & 54958 & 15 & $9.60$&$0.10$ & 3 & \textit{Suzaku} & 54958.4 & $89.10$&$2.70$ & 3 & Low/hard & 3 & \tablenotemark{b} \\
Cyg X-1 & AMI & 54971 & 15 & $13.50$&$0.10$ & 3 & \textit{Suzaku} & 54972.4 & $85.30$&$2.50$ & 3 & Low/hard & 3 & \tablenotemark{b} \\
Cyg X-1 & AMI & 54983 & 15 & $14.70$&$0.20$ & 3 & \textit{Suzaku} & 54977.4 & $111.00$&$3.00$ & 3 & Low/hard & 3 & \tablenotemark{b} \\
Cyg X-1 & AMI & 54983 & 15 & $14.70$&$0.20$ & 3 & \textit{Suzaku} & 54981.4 & $116.00$&$3.00$ & 3 & Low/hard & 3 & \tablenotemark{b} \\
Cyg X-1 & AMI & 54987 & 15 & $19.60$&$0.10$ & 3 & \textit{Suzaku} & 54985.4 & $121.00$&$4.00$ & 3 & Low/hard & 3 & \tablenotemark{b} \\
Cyg X-1 & AMI & 54987 & 15 & $19.60$&$0.10$ & 3 & \textit{Suzaku} & 54987.4 & $115.00$&$3.00$ & 3 & Low/hard & 3 & \tablenotemark{b} \\
Cyg X-1 & AMI & 55126 & 15 & $9.40$&$0.80$ & 3 & \textit{Suzaku} & 55126.4 & $68.90$&$2.10$ & 3 & Low/hard & 3 & \tablenotemark{b} \\
Cyg X-1 & AMI & 55131 & 15 & $6.50$&$0.10$ & 3 & \textit{Suzaku} & 55131.4 & $68.60$&$2.00$ & 3 & Low/hard & 3 & \tablenotemark{b} \\
Cyg X-1 & AMI & 55142 & 15 & $7.20$&$0.30$ & 3 & \textit{Suzaku} & 55139.4 & $52.70$&$1.50$ & 3 & Low/hard & 3 & \tablenotemark{b} \\
Cyg X-1 & AMI & 55147 & 15 & $6.10$&$0.20$ & 3 & \textit{Suzaku} & 55146.4 & $46.10$&$1.30$ & 3 & Low/hard & 3 & \tablenotemark{b} \\
Cyg X-1 & AMI & 55153 & 15 & $6.00$&$0.10$ & 3 & \textit{Suzaku} & 55153.4 & $103.00$&$3.00$ & 3 & Low/hard & 3 & \tablenotemark{b} \\
Cyg X-1 & AMI & 55161 & 15 & $13.30$&$0.20$ & 3 & \textit{Suzaku} & 55160.4 & $81.80$&$2.40$ & 3 & Low/hard & 3 & \tablenotemark{b} \\
Cyg X-1 & AMI & 55167 & 15 & $13.60$&$0.10$ & 3 & \textit{Suzaku} & 55167.4 & $76.60$&$2.30$ & 3 & Low/hard & 3 & \tablenotemark{b} \\
Cyg X-1 & AMI & 55176 & 15 & $16.20$&$0.20$ & 3 & \textit{Suzaku} & 55174.4 & $79.10$&$2.40$ & 3 & Low/hard & 3 & \tablenotemark{b} \\
Cyg X-1 & AMI & 55188 & 15 & $15.30$&$0.50$ & 3 & \textit{Suzaku} & 55183.4 & $143.00$&$5.00$ & 3 & Low/hard & 3 & \tablenotemark{b} \\
GRO J1655$-$40 & VLA B & 53425 & 4.86 & $1.46$&$0.07$ & 4 & \textit{RXTE} & 53424.0 & $2.07$&$0.04$ & 2 & Low/hard & 4 &  \\
GRO J1655$-$40 & VLA B & 53426 & 4.86 & $1.52$&$0.11$ & 4 & \textit{RXTE} & 53425.1 & $2.56$&$0.05$ & 2 & Low/hard & 4 &  \\
GRO J1655$-$40 & VLA B & 53429 & 4.86 & $1.86$&$0.06$ & 4 & \textit{RXTE} & 53429.0 & $4.02$&$0.05$ & 2 & Low/hard & 4 &  \\
GRO J1655$-$40 & VLA B & 53434 & 4.86 & $2.01$&$0.10$ & 4 & \textit{RXTE} & 53434.0 & $5.69$&$0.13$ & 2 & Low/hard & 4 &  \\
GRS 1915+105 & GBI & 50584 & 5 & $28.56$&$4.14$ & 5 & \textit{RXTE} & 50583.5 & $181.00$&$18.00$ & 5 & Hard-steady & 5 & \tablenotemark{c} \\
GRS 1915+105 & GBI & 50725 & 5 & $156.11$&$6.55$ & 5 & \textit{RXTE} & 50724.9 & $593.00$&$60.00$ & 5 & Hard-steady & 5 & \tablenotemark{c} \\
GRS 1915+105 & GBI & 50729 & 5 & $26.00$&$4.45$ & 5 & \textit{RXTE} & 50729.3 & $247.00$&$25.00$ & 5 & Hard-steady & 5 & \tablenotemark{c} \\
GRS 1915+105 & GBI & 50730 & 5 & $38.25$&$4.20$ & 5 & \textit{RXTE} & 50730.4 & $230.00$&$23.00$ & 5 & Hard-steady & 5 & \tablenotemark{c} \\
GRS 1915+105 & GBI & 50736 & 5 & $58.72$&$3.88$ & 5 & \textit{RXTE} & 50735.6 & $130.00$&$13.00$ & 5 & Hard-steady & 5 & \tablenotemark{c} \\
GRS 1915+105 & GBI & 50737 & 5 & $63.45$&$3.58$ & 5 & \textit{RXTE} & 50737.4 & $139.00$&$14.00$ & 5 & Hard-steady & 5 & \tablenotemark{c} \\
GRS 1915+105 & GBI & 50743 & 5 & $56.37$&$3.77$ & 5 & \textit{RXTE} & 50743.3 & $144.00$&$15.00$ & 5 & Hard-steady & 5 & \tablenotemark{c} \\
GRS 1915+105 & GBI & 50746 & 5 & $42.74$&$4.09$ & 5 & \textit{RXTE} & 50746.3 & $153.00$&$15.00$ & 5 & Hard-steady & 5 & \tablenotemark{c} \\
GRS 1915+105 & GBI & 50910 & 5 & $85.44$&$3.91$ & 5 & \textit{RXTE} & 50909.9 & $210.00$&$21.00$ & 5 & Hard-steady & 5 & \tablenotemark{c} \\
GRS 1915+105 & GBI & 50913 & 5 & $85.55$&$3.96$ & 5 & \textit{RXTE} & 50912.9 & $230.00$&$23.00$ & 5 & Hard-steady & 5 & \tablenotemark{c} \\
GRS 1915+105 & GBI & 50913 & 5 & $85.67$&$3.98$ & 5 & \textit{RXTE} & 50913.0 & $234.00$&$23.00$ & 5 & Hard-steady & 5 & \tablenotemark{c} \\
GRS 1915+105 & GBI & 50923 & 5 & $111.09$&$5.29$ & 5 & \textit{RXTE} & 50923.3 & $440.00$&$44.00$ & 5 & Hard-steady & 5 & \tablenotemark{c} \\
GRS 1915+105 & GBI & 50926 & 5 & $111.60$&$4.28$ & 5 & \textit{RXTE} & 50925.9 & $207.00$&$21.00$ & 5 & Hard-steady & 5 & \tablenotemark{c} \\
GRS 1915+105 & GBI & 50932 & 5 & $98.64$&$4.32$ & 5 & \textit{RXTE} & 50931.7 & $183.00$&$19.00$ & 5 & Hard-steady & 5 & \tablenotemark{c} \\
GRS 1915+105 & GBI & 50939 & 5 & $126.56$&$4.86$ & 5 & \textit{RXTE} & 50938.9 & $163.00$&$16.00$ & 5 & Hard-steady & 5 & \tablenotemark{c} \\
GRS 1915+105 & GBI & 50945 & 5 & $79.89$&$3.88$ & 5 & \textit{RXTE} & 50944.9 & $198.00$&$20.00$ & 5 & Hard-steady & 5 & \tablenotemark{c} \\
GRS 1915+105 & GBI & 50945 & 5 & $80.01$&$3.79$ & 5 & \textit{RXTE} & 50945.0 & $190.00$&$19.00$ & 5 & Hard-steady & 5 & \tablenotemark{c} \\
GRS 1915+105 & GBI & 50945 & 5 & $79.78$&$3.82$ & 5 & \textit{RXTE} & 50945.1 & $189.00$&$19.00$ & 5 & Hard-steady & 5 & \tablenotemark{c} \\
GRS 1915+105 & GBI & 50945 & 5 & $80.17$&$3.82$ & 5 & \textit{RXTE} & 50945.2 & $187.00$&$19.00$ & 5 & Hard-steady & 5 & \tablenotemark{c} \\
GRS 1915+105 & GBI & 50953 & 5 & $96.61$&$4.05$ & 5 & \textit{RXTE} & 50952.6 & $152.00$&$15.00$ & 5 & Hard-steady & 5 & \tablenotemark{c} \\
GRS 1915+105 & GBI & 50958 & 5 & $51.19$&$4.08$ & 5 & \textit{RXTE} & 50957.8 & $136.00$&$14.00$ & 5 & Hard-steady & 5 & \tablenotemark{c} \\
GRS 1915+105 & GBI & 50965 & 5 & $48.11$&$4.23$ & 5 & \textit{RXTE} & 50964.8 & $191.00$&$19.00$ & 5 & Hard-steady & 5 & \tablenotemark{c} \\
GRS 1915+105 & GBI & 50975 & 5 & $68.81$&$5.01$ & 5 & \textit{RXTE} & 50975.3 & $186.00$&$19.00$ & 5 & Hard-steady & 5 & \tablenotemark{c} \\
GRS 1915+105 & GBI & 50981 & 5 & $57.47$&$5.07$ & 5 & \textit{RXTE} & 50980.8 & $246.00$&$25.00$ & 5 & Hard-steady & 5 & \tablenotemark{c} \\
GRS 1915+105 & GBI & 50992 & 5 & $32.77$&$4.86$ & 5 & \textit{RXTE} & 50991.6 & $168.00$&$17.00$ & 5 & Hard-steady & 5 & \tablenotemark{c} \\
GRS 1915+105 & GBI & 50992 & 5 & $32.93$&$4.63$ & 5 & \textit{RXTE} & 50991.7 & $167.00$&$17.00$ & 5 & Hard-steady & 5 & \tablenotemark{c} \\
GRS 1915+105 & GBI & 51002 & 5 & $47.68$&$5.64$ & 5 & \textit{RXTE} & 51002.2 & $138.00$&$14.00$ & 5 & Hard-steady & 5 & \tablenotemark{c} \\
GRS 1915+105 & GBI & 51002 & 5 & $47.82$&$5.50$ & 5 & \textit{RXTE} & 51002.2 & $148.00$&$15.00$ & 5 & Hard-steady & 5 & \tablenotemark{c} \\
GRS 1915+105 & GBI & 51003 & 5 & $128.39$&$5.06$ & 5 & \textit{RXTE} & 51003.2 & $250.00$&$25.00$ & 5 & Hard-steady & 5 & \tablenotemark{c} \\
GRS 1915+105 & GBI & 51004 & 5 & $198.77$&$7.88$ & 5 & \textit{RXTE} & 51004.4 & $486.00$&$48.00$ & 5 & Hard-steady & 5 & \tablenotemark{c} \\
GRS 1915+105 & GBI & 51005 & 5 & $72.55$&$5.82$ & 5 & \textit{RXTE} & 51005.2 & $303.00$&$30.00$ & 5 & Hard-steady & 5 & \tablenotemark{c} \\
GRS 1915+105 & GBI & 51006 & 5 & $53.88$&$5.56$ & 5 & \textit{RXTE} & 51006.2 & $202.00$&$20.00$ & 5 & Hard-steady & 5 & \tablenotemark{c} \\
GRS 1915+105 & GBI & 51006 & 5 & $53.64$&$5.36$ & 5 & \textit{RXTE} & 51006.2 & $183.00$&$18.00$ & 5 & Hard-steady & 5 & \tablenotemark{c} \\
GRS 1915+105 & GBI & 51057 & 5 & $32.12$&$4.53$ & 5 & \textit{RXTE} & 51056.8 & $206.00$&$20.00$ & 5 & Hard-steady & 5 & \tablenotemark{c} \\
GRS 1915+105 & GBI & 51194 & 5 & $25.75$&$3.82$ & 5 & \textit{RXTE} & 51194.0 & $125.00$&$13.00$ & 5 & Hard-steady & 5 & \tablenotemark{c} \\
GRS 1915+105 & GBI & 51221 & 5 & $65.87$&$4.08$ & 5 & \textit{RXTE} & 51221.1 & $219.00$&$22.00$ & 5 & Hard-steady & 5 & \tablenotemark{c} \\
XTE J1118+480 & Ryle & 51634 & 15 & $6.20$&$0.50$ & 6 & \textit{RXTE} & 51634.0 & $7.55$&$0.08$ & 2 & Low/hard & 7 & \tablenotemark{d} \\
XTE J1118+480 & Ryle & 51637 & 15 & $7.50$&$0.30$ & 8 & \textit{RXTE} & 51636.2 & $7.97$&$0.08$ & 2 & Low/hard & 7 & \tablenotemark{d} \\
XTE J1118+480 & VLA C & 51637 & 8.3 & $6.00$&$0.10$ & 8 & \textit{RXTE} & 51637.0 & $7.86$&$0.08$ & 2 & Low/hard & 7 & \tablenotemark{d} \\
XTE J1550$-$564 & ATCA & 51665 & 4.8 & $7.45$&$0.12$ & 9 & \textit{RXTE} & 51664.4 & $57.80$&$1.00$ & 2 & Int./very high & 9 & \tablenotemark{e} \\
XTE J1550$-$564 & ATCA & 51697 & 4.8 & $0.88$&$0.08$ & 9 & \textit{RXTE} & 51696.5 & $3.86$&$0.03$ & 2 & Low/hard & 9 & \tablenotemark{e}
\enddata
\tablenotetext{a}{Radio spectral index measurements of $\alpha = -0.24$ and $0.08$ for the first and second data, respectively \citep{2005ApJ...622..508K}.}
\tablenotetext{b}{X-ray data were converted from published 0.8--10 keV band to 2--10 keV band using PIMMS.}
\tablenotetext{c}{Radio data interpolated between 2.25 and 8.3 GHz.}
\tablenotetext{d}{Radio spectral index measurement of $\alpha = 0.5$ for the first datum only \citep{2001MNRAS.322L..23F}.}
\tablenotetext{e}{Radio spectral index measurements of $\alpha = -0.46$ and $0.37$ for the first and second data, respectively \citep{2001ApJ...554...43C}.}
\tablerefs{
(1) \citealt{2005ApJ...622..508K}; 
(2) our analysis of archival X-ray data;
(3) \citealt{2012ApJ...757...11M};
(4) \citealt{2007ApJ...655..434S};
(5) \citealt{2001ApJ...556..515M};
(6) \citealt{2000IAUC.7390....2P};
(7) \citealt{2001MNRAS.322L..23F};
(8) \citealt{2000IAUC.7395....2D};
(9) \citealt{2001ApJ...554...43C}.
}
\tablecomments{This table shows the X-ray binary data we use in
our fundamental plane analysis. Columns indicate the source, the radio observatory
used, MJD of the radio observation, frequency of the radio
observation in GHz, flux density of the radio observation in mJy, the
reference code for the radio observation, the X-ray observatory used,
the MJD of the X-ray observation, the 2--10$\,$keV flux, the reference
code for the X-ray data, the state of the source, and the reference
code for the state identification.}
\kgtabend{xrbdata}{X-ray Binary Data}

\label{lastpage}
\end{document}